\documentclass[a4paper,12pt]{article}
\pdfoutput=1
\usepackage[english]{babel}
\usepackage{graphicx}
\usepackage{colordvi}
\usepackage{fancybox}
\usepackage{cancel}
\usepackage{amsmath,amsfonts, amssymb}
\usepackage{comment}
\usepackage{longtable}
\usepackage{booktabs}
\usepackage{hyperref}
\usepackage{rotating}
\usepackage[hypcap]{caption} 
\usepackage{cite}

\usepackage{perpage} 
\MakePerPage{footnote}

\def\abar{{\bar a}}

\def\Lama{\Lambda_1}
\def\Lamb{\Lambda_2}
\def\Lamc{\Lambda_3}
\def\Lamd{\Lambda_4}
\def\Lame{\Lambda_5}
\def\Lamf{\Lambda_6}
\def\Lamg{\Lambda_7}

\def\lam{\lambda}

\def\ben{\begin{enumerate}}
\def\een{\end{enumerate}}
\def\beq{\begin{equation}}
\def\eeq{\end{equation}}
\def\beqa{\begin{eqnarray}}
\def\eeqa{\end{eqnarray}}

\def\ifmath#1{\relax\ifmmode #1\else $#1$\fi}
\def\lsim{\mathrel{\raise.3ex\hbox{$<$\kern-.75em\lower1ex\hbox{$\sim$}}}}
\def\gsim{\mathrel{\raise.3ex\hbox{$>$\kern-.75em\lower1ex\hbox{$\sim$}}}}

\def\ctwob{c_{2\beta}}
\def\stwob{s_{2\beta}}
\def\sthreeb{s_{3\beta}}

\def\cthreeb{c_{3\beta}}

\def\lamtil{\lam\ls{345}}

\def\beq{\begin{equation}}
\def\eeq{\end{equation}}
\def\ifmath#1{\relax\ifmmode #1\else $#1$\fi}
\def\calm{\mathcal{M}}

\def\sb  {s_{\beta}}
\def\cb  {c_{\beta}}
\def\stwob  {s_{2\beta}}
\def\ctwob  {c_{2\beta}}

\def\tanb{\tan\beta}

\def\sinb{\sin\beta}
\def\cosb{\cos\beta}

\def\hl{h}
\def\ha{A}
\def\hh{H}
\def\hpm{{H^\pm}}

\def\lamtil{\lam_{345}}

\def\mha{m_{\ha}}
\def\mhl{m_{\hl}}
\def\mhh{m_{\hh}}
\def\mhpm{m_{\hpm}}

\def\ls#1{\ifmath{_{\lower1.5pt\hbox{$\scriptstyle #1$}}}}
\def\lss#1{\ifmath{^{\,\lower2.5pt\hbox{$\scriptstyle #1$}}}}

\def\half{\ifmath{{\textstyle{1 \over 2}}}}

\def\quarter{\ifmath{{\textstyle{1 \over 4}}}}

\renewcommand{\Re}{{\rm Re}}
\renewcommand{\Im}{{\rm Im}}
\usepackage{color}
\definecolor{gray}{rgb}{0.5,0.5,0.5}

\renewcommand{\arraystretch}{1.2}

\usepackage{adjustbox}
\usepackage{refstyle}
\usepackage[toc,page]{appendix}
\usepackage{booktabs,siunitx}
\usepackage{siunitx}
\usepackage{mathrsfs}
\usepackage{mathtools}
\usepackage{pstricks}
\usepackage{slashed}
\usepackage{mathbbol}
\usepackage{multirow}

\begin{document}

\baselineskip=18pt

\begin{flushright}
OSU-HEP-18-07
\end{flushright}

\vspace*{0.8in}

\bigskip

\begin{center}

{\Large\bf Enhanced Di-Higgs Production in the\\[0.1in]
Two Higgs Doublet Model}

\vspace{1cm}

\centerline{
\bf K.S. Babu\footnote{babu@okstate.edu} and
\bf Sudip Jana\footnote{sudip.jana@okstate.edu}
}
\vspace{0.5cm}
\centerline{\it\small Department of Physics, Oklahoma State University, Stillwater, OK, 74078, USA }

\end{center}

\bigskip
\begin{abstract}
We show that the rate for di-Higgs production at the LHC can be enhanced by a factor as large as  25 compared to the Standard Model value in the two Higgs doublet model, while being consistent with the known properties of the observed Higgs boson $h$.  There are correlated modifications in $t\overline{t}h$ and resonant $Zh$ production rates, which can serve as tests of this model.  Our framework treats  both Higgs doublets on equal footing, each with comparable Yukawa couplings to fermions. The Cheng-Sher ansatz for multi-Higgs doublet model is shown to be strongly disfavored by current experiments. We propose a new ansatz for the Yukawa couplings of the Higgs doublets $\Phi_a$ is proposed, where $Y_{ij}^{(a)} =  C_{ij}^{(a)}\cdot {\rm min}\{m_i, \, m_j\}/v$, with $C_{ij}^{(a)}$ being order one coefficients, $m_i$ the mass of fermion $i$ and $v$ the electroweak vacuum expectation value.  Such a pattern of couplings can explain the observed features of fermion masses and mixings and satisfies all flavor violation constraints arising from the exchange of neutral Higgs bosons. The rate for $\mu \rightarrow e \gamma$ decay and new contributions to CP violation in $B_s-\overline{B}_s$ mixing are predicted to be close to the experimental limits.

\end{abstract}

\clearpage

\tableofcontents

\clearpage

\section{Introduction}\label{sec:intro}
The observation of a new particle with properties matching that of the Higgs boson predicted by the Standard Model (SM) by the ATLAS and CMS experiments \cite{Aad:2012tfa,Chatrchyan:2012xdj} has been an important step towards understanding the mechanism of electroweak (EW) symmetry breaking. With this discovery, attention has now shifted to testing whether the Higgs boson exhibits any property that deviates from the Standard Model expectation, and to searching for additional Higgs bosons that may take part in EW symmetry breaking. Experimental results to date, including Run$-$II data from the LHC, have shown no deviations from the SM.  Furthermore, no signals of new resonances which could take part in symmetry breaking have been observed.  However, as detailed in this paper, there is plenty of room for new physics in the symmetry breaking sector.  Di-Higgs production has not been measured to date, with an upper limit set to about 25 times the SM prediction. $t\overline{t}h$ production measurement allows its rate to be as large as 1.9 times the SM value (or as small as 0.5 of the SM value), and $Zh$ production rate is allowed to be as large as 2 times its SM value.  The purpose of this paper is to study these processes and their correlations in the context of the two Higgs doublet model (2HDM).

There are a variety of motivations for extending the SM with the addition of a second Higgs doublet.  Supersymmetric models require a second Higgs doublet to generate fermion masses; electroweak baryogenesis can be consistently realized with a second Higgs doublet \cite{baryo}; TeV scale dark matter can be realized in such extensions \cite{inert}; vacuum stability can be maintained all the way to the Planck scale with a second doublet \cite{hill} unlike in the SM \cite{vacuum}, and small neutrinos masses may by generated as radiative corrections with a second doublet (along with a singlet scalar so that lepton number is broken) \cite{zee}, to name a few.  A second Higgs doublet appears naturally in models with extended symmetries such as left-right symmetric models \cite{lr}, axion models \cite{dfsz}, and grand unified theories.  While the second doublet may have a mass of order the scale of higher symmetry breaking, it may also survive down to the TeV scale, in which case its signatures can be observed experimentally.  It is to be noted that a second Higgs doublet which participates in the EW symmetry breaking can easily be made consistent with EW precision measurements, as the $\rho$ parameter maintains its tree-level value of 1.  The 2HDM also provides a foil to test the properties of the SM Higgs boson.

The phenomenology of 2HDM has been extensively studied over the years \cite{report}. However, most studies restrict the form of the Lagrangian by assuming additional discrete symmetries.  The type-II 2HDM, for example, allows only one doublet to couple to up-type quarks, with the second doublet coupling to the down-type quarks and charged leptons. While this is natural in the supersymmetric extension, a discrete $Z_2$ symmetry has to be assumed to achieve this restriction in other cases.  One motivation for such a discrete symmetry is to suppress Higgs-mediated flavor changing neutral currents (FCNC)  \cite{gw}.  However, it has been recognized that there is no need to completely suppress such FCNC \cite{Cheng}, an appropriate hierarchy in the Yukawa couplings can achieve the necessary suppression. We present a modified ansatz for the Yukawa couplings of Higgs doublet $\Phi_a$, where $Y_{ij}^{(a)} = \sqrt{2} C_{ij}^{(a)} {\rm min}\{m_i,\,m_j\}/v$ which is valid for the couplings of each of the Higgs doublet.  Here $C_{ij}^{(a)}$ are order one coefficients, $m_i$ stands for the mass of fermion $i$ and $v \simeq 246$ GeV is the electroweak vacuum expectation value (VEV). Our modified ansatz can be realized in the context of unification \cite{ansatz:unification}.   We refer to the 2HDM with no additional symmetry as simply the two Higgs doublet model (2HDM), with no qualifier, as opposed to type-I or type-II models, which require additional assumptions.

The modified Yukawa coupling ansatz that we propose, viz., $Y_{ij}^{(a)} =\sqrt{2} C_{ij}^{(a)} {\rm min}\{m_i,\,m_j\}/v$, improves on the Cheng-Sher (CS) ansatz \cite{Cheng} which assumes  $Y_{ij}^{(a)} = \sqrt{2} C_{ij}^{(a)} \sqrt{m_i m_j}/v$.  (A factor $\sqrt{2}$ has been inserted in both ansatze as we normalize $v \simeq 246$ GeV, rather than $174$ GeV used in Ref. \cite{Cheng}.) The pattern of quark mixings is compatible with both these ansatze, as we shall illustrate.  The order one coefficients in the CS ansatz will have to be somewhat smaller than unity for explaining the quark mixings, while in the modified ansatz we present some $C_{ij}^{(a)}$ are slightly larger than unity. We find that the overall goodness to fit is similar in the two cases as regards the CKM mixings.  We compare the two ansatze for their consistency with Higgs mediated flavor violation and show that in  processes such as $K^0-\overline{K^0}$ mixing, $B_{d,s}-\overline{B}_{d,s}$ mixing, and especially $\mu \rightarrow e \gamma$, the modified ansatz gives a better description of current data.  We also find that the rate for $\mu \rightarrow e \gamma$ decay and new contributions to CP violation in $B_s-\overline{B}_s$ mixing in the modified ansatz are close to the experimental limits, which may therefore provide tests of the model.

In the 2HDM framework that we adopt there is no fundamental distinction between the two Higgs doublets. The top quark, bottom quark and tau lepton Yukawa couplings -- most relevant for LHC phenomenology --  with both doublets are then comparable, unlike in the case of type-I or type-II 2HDM.  We shall see that these additional couplings, especially of the top and bottom quarks, would lead to distinct signatures in the $hh$, $t\bar{t}h$ and $Zh$  production rates in a correlated manner. Here $h$ is the standard model-like Higgs boson of mass 125 GeV.  While there would be some deviations in the properties of $h$ from the SM predictions, in our analysis we ensure that such deviations are within experimental limits.  Modifications in $hh$, $t\bar{t}h$ and $Zh$  production rates will also be correlated with such deviations in the properties of $h$.  Discovery of these correlated modifications would be tests of the model.

{\bf \boldmath{$t \overline{t}h$} production:} Theoretically, $t\bar{t}h$ production process is very interesting, as its rate is proportional to $Y_t^2$, the square of the top Yukawa coupling with the Higgs boson. Within the SM, $Y_t$ is known to a good accuracy, as it is related to the top quark mass. But this proportionality relation is disrupted in the 2HDM we present.  In Run$-$I of the LHC, a not-so-small signal of $t\bar{t}h$ production was observed by the ATLAS and CMS collaborations in several channels. Assuming SM-like  branching fractions of the Higgs boson, the $t\bar{t}h$ signal strength normalized to the Standard Model prediction was found to be $\mu_{t\bar{t}h}$ =  $2.3\substack{+0.7 \\ -0.6}$ by the
combined ATLAS and CMS collaborations  with an observed significance of 4.4$\sigma$ \cite{higgs}.
With data collected in the LHC Run$-$II with 13 TeV center of mass energy, both ATLAS and CMS have presented results for $t\bar{t}h$
production, with ATLAS quoting a $\mu$ value of $1.32\substack{+0.28 \\ -0.26}$ with an observed significance of $5.8\sigma$ \cite{tthatlas}, and CMS quoting a $\mu$ value of $1.26\substack{+0.31 \\ -0.26}$ \cite{tthcms} with an observed significance of $5.2\sigma$. (In the  SM, $t\bar{t}h$ production cross-section is $\sim 0.509$ pb.) As can be seen with the errors associated with these measurements, $\mu_{t\bar{t}h}$ can be as large as 1.9 and also as low as 0.5.  Such large deviations can be achieved in the 2HDM, as shall be shown below.  If any significant deviation in $t\bar{t}h$ production rate is observed at the LHC, the 2HDM framework we present can serve as an excellent platform for explaining it.

{\bf Di-Higgs production:} ATLAS and CMS collaborations have reported new results on di-Higgs boson searches \cite{econf, mor,CMShh,CMShh2,ATLAShh} using 36 fb$^{-1}$ data from Run$-$II of LHC at center of mass energy of 13 TeV. These analyses are based on different final states  in the decay of the two Higgs bosons ($b\bar{b}\gamma\gamma, b\bar{b}\tau^{+} \tau^{-}, b\bar{b}b\bar{b}$ and  $b\bar{b}W^+W^-$). Current experimental searches exclude non-resonant $hh$ production to be less than 19 times the SM prediction \cite{econf,CMShh}, whereas for  resonant di-Higgs production, this limit is  correlated with the resonance mass and can be as large as 25  times \cite{CMShh} the SM value, or even larger.  In the SM, $hh$ production cross-section is $\sim 33.5$ fb. In the 2HDM, the signal strength $\mu_{hh}$ can be enhanced by a factor of 25,  including both resonant and non-resonant di-Higgs boson production, which gives ample room for its potential observation at the LHC. While other extensions of the SM, such as the singlet scalar extension, can enhance the di-Higgs production, the enhancement is much smaller compared to the 2HDM, owing primarily to severe constraints on the mixing of the SM Higgs and the singlet scalar from the measured properties of $h$ \cite{lewis}.  Large enhancement for di-Higgs is possible in the 2HDM, since both Higgs bosons couple to top and bottom quarks, which allows for the properties of $h$ to be within observed limits.

{\bf \boldmath{$Zh$} production:} A third di-boson channel of experimental interest is the production of $h$ in association with a $Z$.  The rate for $Zh$ production will also be modified in our 2HDM framework. Recently the ATLAS collaboration has reported a small excess in the $pp \rightarrow A \rightarrow Zh $ cross section \cite{440}, corresponding to a potential pseudoscalar mass of about 440 GeV. The statistical significance is larger if the pseudoscalar $A$ is produced in association with bottom quarks rather than through gluon fusion, but both production processes show deviations. The statistical significance of this excess is too low to conclude anything meaningful, but it does raise the question: would it be possible to account for such a possible excess arising from a pseudoscalar resonance within a self-consistent framework?
We show that this can indeed be achieved in our 2HDM framework.

We base our numerical analysis of the 2HDM on data set which takes into account the light Higgs boson properties as well as searches for heavy Higgs bosons. We also consider theoretical constraints arising from boundedness, stability and perturbativity of the scalar potential, and ensure that the data respects bounds from $B$-physics and electroweak precision measurements.

The paper is organized as follows: In Sec. \ref{sec:2}, we briefly review the Higgs sector and the Yukawa couplings of the 2HDM.
In Sec. \ref{ansatz}, we present our modified ansatz for flavor, and in Sec. \ref{sec:3} we show the consistency of the ansatz with FCNC constraints. 
In Sec. \ref{sec:collider}, we perform numerical simulations for collider signatures of the 2HDM.  Here we discuss enhanced di-Higgs production, as well as modifications in $t \overline{t}h$ and $hZ$ productions. We show correlations among these as well as other modified properties of the 125 GeV Higgs boson $h$. In Sec. \ref{ewpt} we discuss EW precision constraints, boundedness of the potential and unitarity constraints. Finally in Sec. \ref{con} we conclude.

\section{  Brief Review of the Two Higgs Doublet Model}\label{sec:2}
We denote the two SU(2)$\ls{L}$ doublet scalar fields with hypercharge $Y=\frac{1}{2}$ as $\Phi_1$ and $\Phi_2$. The most general gauge invariant scalar potential of this 2HDM is given in Appendix \ref{appendix1}, Eq. (\ref{pot}). Here we shall choose a particularly convenient rotated basis in which only one neutral Higgs has a nonzero vacuum expectation value. The two Higgs doublets in the new basis are denoted as $H_1$ and $H_2$, with $\left\langle H_2^0 \right \rangle = 0$., and $\left\langle H_1^0 \right \rangle = v/\sqrt{2}$. These new states are related to $\Phi_1$ and $\Phi_2$ by
\beqa \label{higgsbasis}
\,H_1&=&\Phi_1\cosb+e^{-i\xi}\,\Phi_2\sinb\,,\nonumber \\
\,H_2&=&-e^{i\xi}\,\Phi_1\sinb+\Phi_2\cosb\,,.
\eeqa
The VEVs of the neutral components of $\Phi_{1,2}$ are denoted as
\beq \label{potmin}
\langle \Phi_1 \rangle={1\over\sqrt{2}} v_1, \qquad \langle
\Phi_2\rangle=
{1\over\sqrt{2}} v_2\, e^{i\xi}~
\eeq
and the mixing angle is given by
\beq
\tanb\equiv\frac{v_2}{v_1}\,.
\label{tanbdef}
\eeq
One then obtains
\beq \label{abbasis}
\,H_1=\left(\begin{array}{c}
G^+ \\ {1\over\sqrt{2}}\left(v+\varphi_1^0+iG^0\right)\end{array}
\right)\,,\qquad
\,H_2=\left(\begin{array}{c}
H^+ \\ {1\over\sqrt{2}}\left(\varphi_2^0+i\ha\right)\end{array}
\right)\,,
\eeq
where $\varphi_1^0$, $\varphi_2^0$ are CP-even neutral Higgs fields (which are not the mass eigenstates),
$\ha$ is a CP-odd neutral Higgs field, $H^+$ is the physical
charged Higgs boson, and $G^+$ and $G^0$ are unphysical Goldstone bosons . The vavuum expectation value is $v= \sqrt{v_1^2+v_2^2} \simeq 246$ GeV.  Here, for simplicity, we have set the phase $\xi$ to be zero.

\subsection{Higgs Potential in the rotated basis}

Since the Higgs potential given in Eq. (\ref{pot}) of Appendix \ref{appendix1} is the most general, once the Higgs fields are rotated as in Eq. (\ref{higgsbasis}), the form of the potential would remain the same, but with a redefinition of parameters.  In the new rotated basis the Higgs potential is:
\beqa  \label{pothbasis}
\mathcal{V}&=& M_{11}^2 H_1^\dagger H_1+M_{22}^2 H_2^\dagger H_2
-[M_{12}^2 H_1^\dagger H_2+{\rm h.c.}]\nonumber\\[4pt]
&&\quad\!\!\!\! +\half\Lama(H_1^\dagger H_1)^2
+\half\Lamb(H_2^\dagger H_2)^2
+\Lamc(H_1^\dagger H_1)(H_2^\dagger H_2)
+\Lamd(H_1^\dagger H_2)(H_2^\dagger H_1)
\nonumber\\[4pt]
&&\quad\!\!\!\! +\left\{\half\Lame(H_1^\dagger H_2)^2
+\big[\Lamf\,(H_1^\dagger H_1)
+\Lamg(H_2^\dagger H_2)\big]
H_1^\dagger H_2+{\rm h.c.}\right\}\,.
\eeqa
The correspondence between the old (Eq. (\ref{pot})and the new (Eq. (\ref{pothbasis}) parameters is given in Eqs. (\ref{maa})-(\ref{lamtildef}) of Appendix \ref{appendix1}. 
Now the minimization conditions read simply as:
\beq \label{hbasismincond}
M_{11}^2=-\half\Lama v^2\,,\qquad\qquad  M_{12}^2 =\half  \Lamf  v^2\,.
\eeq
The $3 \times 3$ neutral scalar mass matrix defined in the basis $\lbrace \varphi_1^0, \varphi_2^0, A \rbrace$ reads as:
\beq
\calm^2 = \left(
  \begin{array}{ccc}  \Lama v^2 & \Re(\Lamf) v^2& -\Im(\Lamf) v^2\\
                  \Re(\Lamf) v^2 & \quad M_{22}^2+\half v^2(\Lamc+\Lamd+\Re(\Lame))\ & -\frac{1}{2}\Im(\Lame) v^2 \\ -\Im(\Lamf) v^2 & -\frac{1}{2}\Im(\Lame) v^2 & M_{22}^2+\half v^2(\Lamc+\Lamd-\Re(\Lame))\, \end{array}\right)\,,
\label{massmhh}
\eeq

The mass eigenvalues and eigenstates can be readily obtained from Eq. (\ref{massmhh}). For simplicity we shall assume the Higgs sector to be 
CP-conserving, and take the VEVs as well as the couplings $\Lambda_{5,6}$ to be real.  The CP odd eigenstate will then decouple from the CP even eigenstates in Eq. (\ref{massmhh}).  The CP -even mass eigenvalues in this limit are:
\beq
m^2_{h, H}=\half\left[\mha^2+v^2(\Lama+\Lame)\mp\sqrt{
[\mha^2+(\Lame-\Lama)v^2]^2+4\Lamf^2 v^4}\right]\,,
\eeq
while the CP-odd and the charged Higgs boson masses are given by:
\beqa \label{massmha}
\mha^2=\mhpm^2-\half v^2(\Lame-\Lamd)\,,\\ 
\mhpm^2=M_{22}^2+\half v^2\Lamc\,,~.
\eeqa
Here $H^\pm=-e^{i\xi}\sinb\,\Phi_1^\pm+\cosb\,\Phi_2^\pm$
is the physical charged Higgs, which is orthogonal to the charged Goldstone boson $G^\pm$.

The CP-even neutral Higgs mass eigenstates are given by:\footnote{Since the original Higgs doublets $\Phi_{1,2}$ were rotated by an angle $\beta$ to go to the $H_{1,2}$ basis, defining the CP-even Higgs mixing angle as $\alpha-\beta$ (rather than as $\alpha$) would make it consistent with the standard notation used in the literature \cite{report}.}
\beqa
\hl&=&\varphi_1^0\cos{(\alpha-\beta)}+\varphi_2^0\sin{(\alpha-\beta)}, \label{hbasis}\\
\hh&=&\varphi_2^0\cos{(\alpha-\beta)}-\varphi_1^0\sin{(\alpha-\beta)}, \label{Hbasis}
\eeqa
where $\varphi_1^0$ and $\varphi_2^0$ are defined in Eq. (\ref{abbasis}) 
and the angle $(\alpha-\beta)$ is defined as:
\beqa \label{exactbma}
\sin\, 2(\alpha-\beta)  &=&{2\Lamf v^2\over\mhh^2-\mhl^2}\,.
\eeqa
The field $h$ is identified as the observed Higgs boson of mass 125 GeV.
We shall use these results in the discussion of flavor phenomenology as well as collider physics.

\subsection{Yukawa sector of the 2HDM}

As noted in the Introduction, we treat the two Higgs doublets on equal footing.  Thus, both doublets will couple to fermions with comparable strengths.  If the Yukawa coupling matrices have a certain hierarchy, consistent with the mass and mixing angles of fermions, then Higgs-mediated flavor changing neutral currents can be sufficiently suppressed \cite{Cheng}.  This statement will be further elaborated in the next section. The original Higgs doublets $\Phi_1$ and $\Phi_2$ will then have the following Yukawa couplings to fermions:
\beqa
\mathcal{L}_{y} &=& Y_{d}^{(1)} \bar{Q}_L  d_R \Phi_1 + {Y}_{d}^{(2)} \bar{Q}_L  d_R \Phi_2 +  Y_{u}^{(1)} \bar{Q}_L u_R \tilde{\Phi}_1 + {Y}_{u}^{(2)} \bar{Q}_L  u_R \tilde{\Phi}_2 \nonumber \\
&+& Y_{\ell}^{(1)} \bar{\psi}_L \Phi_1 \ell_R + {Y}_{\ell}^{(2)} \bar{\psi}_L \Phi_2 \ell_R +h.c. \qquad 
\label{fullyuk}
\eeqa
Here $Q_L = (u,\,d)_L^T$ and $\psi_\ell= (\nu,\,e)_L^T$ are the left-handed quark and lepton doublets, while 
$\tilde{\Phi}_a = i \tau_2 \Phi_a^*$. 

In the rotated Higgs basis (see Eq. (\ref{higgsbasis})) the Yukawa couplings can be written as
\beqa
\mathcal{L}_{y} &=& Y_{d} \bar{Q}_L  d_R H_1 + \tilde{Y}_{d} \bar{Q}_L  d_R H_2 +  Y_{u} \bar{Q}_L u_R \tilde{H_1} + \tilde{Y}_{u} \bar{Q}_L  u_R \tilde{H_2}\nonumber \\
&+& Y_{\ell} \bar{\psi}_L H_1 \psi_R + \tilde{Y}_{\ell} \bar{\psi}_L H_2 \psi_R +h.c. \qquad 
\label{yuk}
\eeqa
Here $(Y_d,\, \tilde{Y}_d)$,  $(Y_u,\, \tilde{Y}_u)$ and  $(Y_\ell,\, \tilde{Y}_\ell)$ are related to the original Yukawa coupling matrices as
\beqa
Y_d &=& \cos\beta \,Y_d^{(1)} + \sin\beta e^{i\xi}\, Y_d^{(2)},~~
\tilde{Y}_d = -\sin\beta e^{-i \xi} \,Y_d^{(1)} + \cos\beta \, Y_d^{(2)} \nonumber \\
Y_u &=& \cos\beta\, Y_u^{(1)} + \sin\beta e^{-i\xi} \,Y_u^{(2)},~~
\tilde{Y}_u = -\sin\beta e^{i \xi}\, Y_u^{(1)} + \cos\beta \, Y_u^{(2)}
\nonumber \\
Y_\ell &=& \cos\beta \, Y_\ell^{(1)} + \sin\beta e^{i\xi} \, Y_\ell^{(2)},~~
\tilde{Y}_\ell = -\sin\beta e^{-i \xi} \,Y_\ell^{(1)} + \cos\beta \, Y_\ell^{(2)}~.
\eeqa
Since the VEV of $H_2$ is zero, the up-quark, down-quark, and charged lepton mass matrices are given by
\beqa
M_u = {Y}_{u}\, v / \sqrt{2},~~~
M_d = {Y}_{d}\, v / \sqrt{2},~~~
M_l = {Y}_{\ell}\, v / \sqrt{2}.
\eeqa
We can diagonalize these mass matrices, which would simultaneously diagonalize the Yukawa coupling matrices of the neutral Higgs bosons of the doublet $H_1$. Note, however, that the $\varphi_1^0$  component of $H_1$ is not a mass eigenstates. All Higgs-induced FCNC will arise from the Yukawa coupling matrices of the neutral members of $H_2$.  Thus, in the down quark sector, such FCNC will be proportional to $\tilde{Y}_d$ of Eq. (\ref{yuk}), etc.  We shall define the matirces $\tilde{Y}_d,\, \tilde{Y}_u$ and $\tilde{Y}_\ell$ of Eq. (\ref{yuk}) in a basis where $M_{u,d,\ell}$ have been made diagonal.

In the type-II 2HDM one assumes $Y_d^{(2)} = Y_\ell^{(2)} = 
Y_u^{(1)} =  0$ in Eq. (\ref{fullyuk}), invoking a $Z_2$ symmetry. This would lead to $Y_d$ and $\tilde{Y}_d$ being proportional in Eq. (\ref{yuk}) (and similarly $Y_u \propto \tilde{Y}_u$, $Y_\ell \propto \tilde{Y}_\ell$). Thus, in a basis where $Y_d$ is diagonal, $\tilde{Y}_d$ will also be diagonal. As a result, type-II 2HDM  would have no Higgs mediated flavor violation at the tree level \cite{gw}. In a similar fashion, type-I 2HDM, where one assumes $Y_{d,u,\ell}^{(2)} = 0$, $Y_f$ and $\tilde{Y}_f$ are diagonal simultaneously for $f=d,u,\ell$, resulting in neutral flavor conservation.  The idea of alignment  \cite{pich,report} assumes $Y_f^{(1)}$ and $Y_f^{(2)}$ are proportional, again leading to neutral flavor conservation.

In our framework, the coupling matrices $\tilde{Y}_{u}$, $\tilde{Y}_{d}$ and $\tilde{Y}_{\ell}$ are a priori arbitrary matrices. To be consistent with flavor violation mediated by neutral Higgs bosons, we assume a hierarchy in $\tilde{Y}_{u}$ similar to the hierarchy in $Y_{u}$, and so forth. This ansatz will be elaborated in the next section. For collider studies, only $\tilde{Y}_{t}$, $\tilde{Y}_{b}$ and $\tilde{Y}_{\tau}$ couplings, defined as the $\left(3, 3\right)$ elements of $\tilde{Y}_{u}$, $\tilde{Y}_{d}$ and $\tilde{Y}_{l}$  in a basis where $M_{u,d,\ell}$ are diagonal, will play a role.


\section{A Modified Ansatz for the Yukawa Couplings}\label{ansatz}

The 2HDM can potentially lead to flavor changing neutral currents (FCNC) mediated by the neutral Higgs bosons at an unacceptable level.  As already noted, one could completely suppress tree level FCNC by assuming a discrete $Z_2$ symmetry \cite{gw}. One can also assume alignment of the two sets of Yukawa couplings to alleviate this problem \cite{pich,report}.  However, compete suppression of Higgs mediated FCNC is not necessary, as emphasized in Ref. \cite{Cheng}. A hierarchical pattern of Yukawa couplings that can generate realistic fermion masses and mixings may suppress excessive FCNC.  

In Ref. \cite{Cheng}, a particular pattern of the Higgs Yukawa couplings was suggested, referred to as the Cheng-Sher (CS) ansatz.  The Yukawa couplings of the two Higgs doublets $\Phi_a$ are taken to be of the form
\begin{equation}
(Y^{(a)}_f)_{ij} = \frac{\sqrt{2} C_{ij} \sqrt{m_i m_j}}{v}, ~~~f=(u,\,d,\,\ell)~.
\label{cs}
\end{equation}
Here $m_i$ is the mass of the $i$th fermion, $v = 246$ GeV is the electroweak VEV, and $C_{ij}$ are order one coefficients.  We have inserted a factor $\sqrt{2}$ since $v$ is normalized to 246 GeV in our analysis (rather than 174 GeV). This is a well motivated ansatz, as this form can explain qualitatively several features of the CKM mixing angles. With this form of the Yukawa coupling matrices, the mass matrices will also take a similar form. The CKM mixing angles will be given by
\begin{eqnarray}
V_{ij} = K_{ij} \sqrt{\frac{m_i}{m_j}},~~~i < j~.
\label{kij}
\end{eqnarray}
Here $K_{ij}$ are order one coefficients, expressible in terms of $C_{ij}$ of Eq. (\ref{cs}).  Furthermore, in this symmetric form of the Yukawa couplings, the form of Eq. (\ref{cs}) is the maximal allowed off-diagonal couplings consistent with the masses of fermions, assuming no cancellations among various contributions.

Take for example the case where the quark mixings arise entirely from the down quark matrix.  Using  quark masses evaluated at the top quark mass scale, $m_b \approx 2.75$ GeV, $m_s \approx 52$ MeV, $m_d \approx 2.76$ MeV \cite{xing}, and with the central values of the CKM angles, $|V_{us}| = 0.2243$, $|V_{cb}| = 0.0422$, $|V_{ub}| = 0.00394$, one would obtain for the coefficients $K_{ij}$ of Eq. (\ref{kij}) the following values:
\begin{equation}
K_{ds} \approx 0.974,~~~K_{db} \approx 0.124,~~~K_{sb} \approx 0.307~.
\label{Kcs}
\end{equation}
These coefficients are roughly of order one, which provides justification to the ansatz of Eq. (\ref{cs}).  Note, however, that $K_{db} \sim 1/8$ in particular, is significantly smaller than order one.

We shall present an updated analysis of flavor violation constraints for the CS ansatz in the next section.  There we show that with the current data, various order one coefficients $C_{ij}$ appearing in Eq. (cs) will have to be smaller than one, provided that the masses of the additional Higgs bosons of the model are below a TeV.  In particular, the constraint from $\mu \to e \gamma$ decay sets  $|C_{e\mu}| \leq 0.12$.  Constraints from $K^0-\overline{K^0}$ mixing, $B_{d,s}-\overline{B}_{d,s}$ mixing and $D^0-\overline{D^0}$ mixings on the relevant $C_{ij}$ are also of similar order.  While these constraints do not exclude the scenario, they do make the ansatz somewhat less motivated.

In view of the strains faced by the CS ansatz, we propose a modified ansatz, which fares equally well in explaining the pattern of CKM mixing angles, but causes acceptable FCNC mediated by the neutral Higgs bosons.  We take the Yukawa couplings of the two Higgs doublets $\Phi_a$ to fermions to be of the form
\begin{equation}
Y_{ij}^{(a)} =\sqrt{2} C_{ij}^{(a)} {\rm min}\{m_i,\,m_j\}/v,
\label{bj}
\end{equation}
where $C_{ij}$ are order one coefficients, $m_i$ is the mass of the $i$th fermion and $v \simeq 246$ GeV.  The main difference of this ansatz, compared to the CS ansatz of Eq. (\ref{cs}) is that the Yukawa couplings scale linearly with the lighter fermion masses, rather than  as the geometric means.  We first show that this form of the couplings can generate reasonable CKM mixings.  With the form of Eq. (\ref{bj}) for the Yukawa couplings, the CKM mixing angles would be given by
\begin{equation}
V_{ij} = K_{ij} \frac{m_i}{m_j},~~i < j~.
\label{Kbj}
\end{equation}
Since the mass hierarchies are stronger in the up-quark sector, this would imply that the CKM angles scale as the mass ratios in the down-quark sector.  Using the values of the masses of $(d,\,s,\,b)$ quarks evaluated at the top quark mass scale, and using central values of the CKM mixing angles, this leads to the following $K_{ij}$ values:
\begin{eqnarray}
K_{ds} \approx 4.23,~~~K_{db} \approx 3.93,~~~K_{sb} \approx 2.23~.
\end{eqnarray}
These values are roughly of order one, and the overall goodness to fit to the CKM angles is comparable to that of the CS ansatz (compare Eq. (\ref{Kcs}) with Eq. (\ref{Kbj})).  While in the CS
ansatz the $K_{ij}$ are somewhat smaller than one, in the modified ansatz they are slightly larger than one.

We turn to the flavor phenomenology of the modified ansatz in the next section.  There we derive limits on the $C_{ij}$ parameters in the modified ansatz and compare them with the limits on $C_{ij}$ of the CS ansatz.  We find that the modified ansatz allows for all $C_{ij}$, at least in the CP conserving sector, to be of order one.

\section{ Higgs Mediated Flavor Phenomenology}\label{sec:3}
In this section we derive constraints on the $C_{ij}$ parameters for the modified ansatz as well as the CS ansatz arising from FCNC mediated by the neutral Higgs bosons of 2HDM.  We write down the Yukawa couplings of the two Higgs doublets in a rotated Higgs basis $H_1$ and $H_2$ such that $\langle H_2\rangle =0$.  The Yukawa Lagrangian is given in Eq. (\ref{yuk}).  We define the matrices $\tilde{Y}_{u,d,\ell}$ in a basis where the matrices $Y_{u,d,\ell}$ are diagonal. This is the physical mass eigenbasis.  Since both Higgs doublets are treated on equal footing, the rotation that is needed to go to the $(H_1,\,H_2)$ basis should not change the form or the hierarchy of the original Yukawa couplings of Eq. (\ref{fullyuk}).  This is true for the CS ansatz as well as the modified ansatz.  Thus the form of the $\tilde{Y}_{ij}$ should be the same as that of $Y^{(a)}_{ij}$.

The Yukawa couplings of $H_2$ will then take, in the modified ansatz,
the following form:
\begin{equation}
\tilde{Y}_{u}= \frac{\sqrt{2}}{v}
\left( \begin{array}{ccc} 
 C_{uu} m_u & C_{uc} m_u & C_{ut} m_u \\
C_{cu} m_u & C_{cc} m_c & C_{ct} m_c \\
C_{tu} m_u & C_{tc} m_c & C_{tt} m_t
\end{array}\right), \qquad
\tilde{Y}_{d}= \frac{\sqrt{2}}{v}
\left( \begin{array}{ccc}
C_{dd} m_d & C_{ds} m_d & C_{db} m_d \\
C_{sd} m_d & C_{ss} m_s & C_{sb} m_s \\
C_{bd} m_d & C_{bs} m_s & C_{bb} m_b
\end{array}\right),  \nonumber
\end{equation}
\begin{equation}
\tilde{Y}_{\ell}= \frac{\sqrt{2}}{v}
\left( \begin{array}{ccc}
C_{ee} m_e & C_{e\mu} m_e & C_{e\tau} m_e \\
C_{\mu e} m_e & C_{\mu\mu} m_{\mu} & C_{\mu \tau} m_{\mu} \\
C_{\tau e} m_e & C_{\tau \mu} m_{\mu} & C_{\tau \tau} m_{\tau}
\end{array}\right)
\label{full}
\end{equation}
We shall show that Higgs mediated FCNC allows all the $C_{ij}$ appearing in Eq. (\ref{full}) to be of order one. The main constraints come from $K^0-\overline{K^0}$ mixing, $B^0_{d,s}-\overline{B^0}_{d,s}$ mixing and $D^0-\overline{D^0}$ mixing, mediated by neutral Higgs bosons of 2HDM at the tree-level. The $\mu \rightarrow e\gamma$ decay, although it arises only at the loop level, is also found to provide important constraints.  We now turn to derivations of these constraints.

\subsection{Constarints from tree level Higgs induced FCNC processes}

 \begin{figure}[htb!]
$$
 \includegraphics[height=4cm,width=0.35\textwidth]{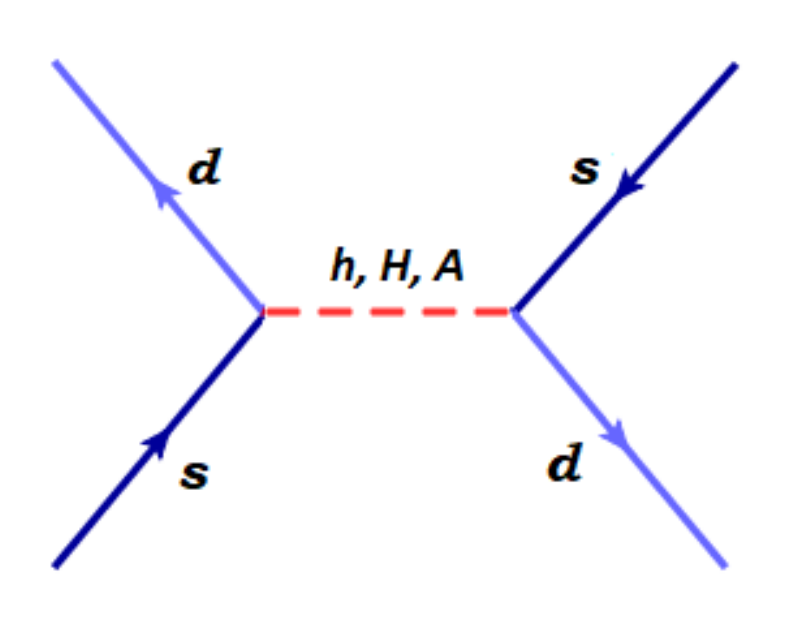} \hspace{0.3 in}
  \includegraphics[height=4cm,width=0.35\textwidth]{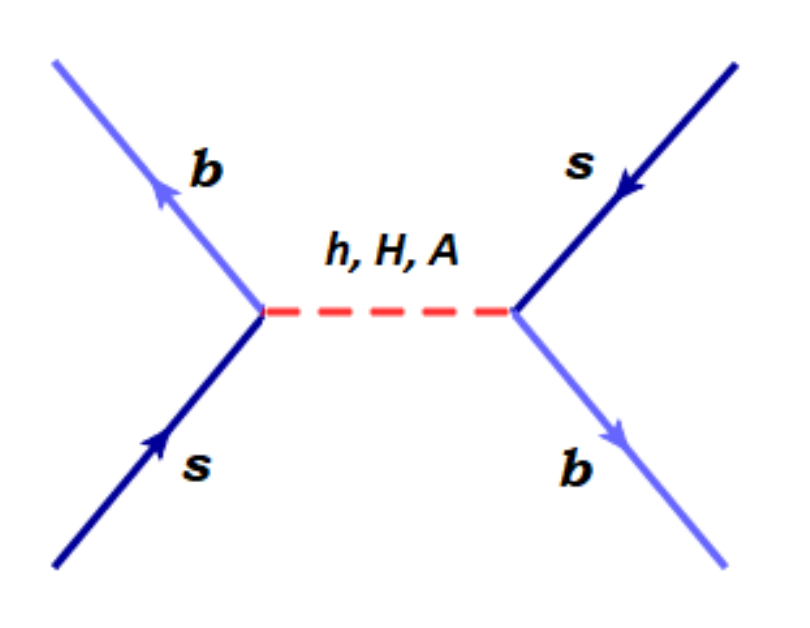}
 $$
 $$
\includegraphics[height=4cm,width=0.35\textwidth]{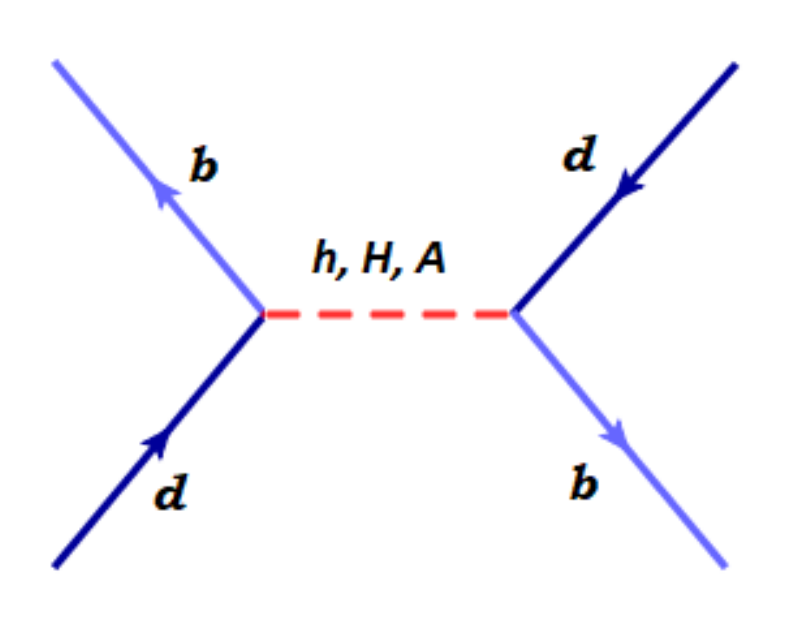}\hspace{0.3 in}
\includegraphics[height=4cm,width=0.35\textwidth]{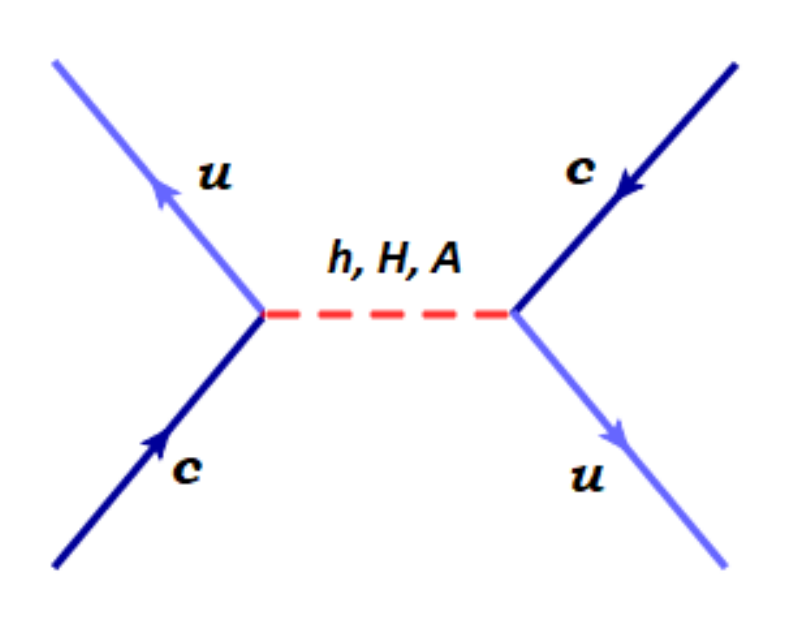}
 $$
 \caption{Feynman diagrams for various FCNC processes mediated by tree--level neutral Higgs boson exchange: top left: $K^0-\overline{K^0}$ mixing, top right:  $B_s^0 - \overline{B_s^0}$ mixing, bottom left:  $B^0_d - \overline{B_d^0}$ mixing and bottom right:  $D^0 - \overline{D^0}$ mixing.  }
\label{flav1}
\end{figure}

 There are accurate  experimental measurements \cite{charged1} of neutral meson--antimeson mixings in the $K^0-\overline{K^0}$, $B_d^0-\overline{B_d^0}$, $B_s^0-\overline{B_s^0}$ and in $D^0-\overline{D^0}$ sectors, all of which are in agreement with SM predictions. In the 2HDM there are new contributions to these mixings arising through tree level Higgs boson exchange diagrams shown in Fig. \ref{flav1}. The SM predictions will be modified with these new contributions; here we derive constraints on the 2HDM parameters from these processes.

We can write down the neutral Higgs boson mediated contributions to  $\Delta F=2$ Hamiltonian responsible for the neutral meson--antimeson mixings as \cite{des,q6}:
\begin{equation}
H_{\rm eff}=-\frac{1}{2{M_k}^2} \left( \bar{q}_i\left[Y^k_{ij}\frac{1+\gamma_5}{2}+{Y^k_{ji}}^*\frac{1-\gamma_5}{2}\right]q_j\right)^2.
\end{equation}
Here $Y^k_{ij}$ denote the Yukawa couplings of $q_i$, $q_j$ with Higgs mass eigenstate $H^k$, with $k$ taking values $(h,\, H,\, A)$,  
and $q_{i,j}$ represent the relevant quark fields contained in the meson.

The transition matrix element for meson mixing can be expressed as
\begin{eqnarray}
&&\lefteqn{M^{\phi}_{12}=\langle\phi|H_{{\rm eff}}|\bar{\phi}\rangle
=-\frac{{f_{\phi}}^2m_{\phi}}{2{M_k}^2}\Big[-\frac{5}{24}
\frac{m_{\phi}^2}{(m_{q_i}+m_{q_j})^2}\left({Y^k_{ij}}^2+{{Y^k_{ji}}^*}^2\right)\cdot B_{2}\cdot \eta_2(\mu)}\hspace{14 cm}\nonumber\\
&&\lefteqn{\hspace{5.5cm}+Y^k_{ij}{Y^k_{ji}}^*\left(\frac{1}{12}+\frac{1}{2}\frac{{m_{\phi}}^2}{(m_{q_i}+m_{q_j})^2}\right)
\cdot B_{4}\cdot \eta_4(\mu)\Big].}\label{m12}
\end{eqnarray}
Here the neutral mesons $(K^0,~B_d^0, ~B_s^0,~D^0)$ are denoted as $\phi$. We adopt the modified vacuum saturation and factorization to parametrize the matrix elements, but use lattice evaluations of the matrix elements for our numerical study:
$$
\langle\phi|\bar{f}_i(1\pm\gamma_5)f_j\bar{f}_i(1\mp\gamma_5)f_j|\bar{\phi}\rangle={f_{\phi}}^2m_{\phi}
\left(\frac{1}{6}+\frac{{m_{\phi}}^2}{(m_{q_i}+m_{q_j})^2}\right)\cdot B_{4},
$$
\begin{equation}
\langle\phi|\bar{f}_i(1\pm\gamma_5)f_j\bar{f}_i(1\pm\gamma_5)f_j|\bar{\phi}\rangle
=-\frac{5}{6}{f_{\phi}}^2m_{\phi}\frac{{m_{\phi}}^2}{(m_{q_i}+m_{q_j})^2}\cdot B_{2}.
\end{equation}
We use  the values: $(B_2,~B_4) = (0.66,~1.03)$ for the $K^0$ system, $(0.82,~1.16)$ for the $B_d^0$  and $B_s^0$ systems, and $(0.82,~1.08)$ for the $D^0$ system \cite{q6,neubert}. The QCD correction factors of the Wilson coefficients $C_2$ and $C_4$ of the effective $\Delta F = 2$ Hamiltonian in going from the heavy Higgs mass scale $M_H$ to the hadronic scale $\mu$ are denoted by $\eta_2(\mu)$ and $\eta_4(\mu)$ in Eq. (\ref{m12}).
These correction factors are computed as follows.  We can write the $\Delta F = 2$ effective Hamiltonian in the general form as
\begin{equation}
\mathcal{H}_{\rm eff}^{\Delta F = 2}=\sum^5_{i=1}C_i~Q_i+\sum^3_{i=1}\tilde{C}_i~\tilde{Q},
\end{equation}
where
\begin{eqnarray}
Q_1={\bar{q_i}}^{\alpha}_L\gamma_{\mu}{q_j}^{\alpha}_L\bar{q_i}^{\beta}_L\gamma^{\nu}{q_j}^{\beta}_L, ~~~~
Q_2={\bar{q_i}}^{\alpha}_R {q_j}^{\alpha}_L \bar{q_i}^{\beta}_R {q_j}^{\beta}_R,~~~~
Q_3={\bar{q_i}}^{\alpha}_R {q_j}^{\beta}_L \bar{q_i}^{\beta}_R {q_j}^{\alpha}_L,\nonumber\\
Q_4={\bar{q_i}}^{\alpha}_R {q_j}^{\alpha}_L \bar{q_i}^{\beta}_L {q_j}^{\beta}_R,~~~~~~~~~~~
Q_5={\bar{q_i}}^{\alpha}_R {q_j}^{\beta}_L \bar{q_i}^{\beta}_L {q_j}^{\alpha}_R,\hspace{1.8 cm}
\end{eqnarray}
$\tilde{Q}_{1,2,3}$ can be obtained from $Q_{1,2,3}$ by interchanging $L\leftrightarrow R$.

For computing $\eta_{2,4}$  we consider the new physics scale $M_H$ to be 500 GeV.  The evolution of the Wilson coefficients from $M_H$ down to the hadron scale $\mu$ is obtained from
\begin{equation}
C_r(\mu)=\sum_i \sum_s (b_i^{(r,s)}+\eta c_i^{(r, s)})\eta^{a_i}C_s(M_s).
\end{equation}
Here $\eta=\alpha_s(M_s)/\alpha_s(m_t)$. For our numerical study, we use the magic numbers $a_i$, $b_i^{(r, s)}$ and $c_i^{(r, s)}$ from Ref. \cite{magic1} for the $K$ meson system, from Ref. \cite{magic2} for the $B_{d, s}$ meson system and from Ref. \cite{magic3} for the $D$  meson system.
With $M_s=500 {\rm ~GeV}$, $m_t(m_t)=163.6 {\rm ~GeV}$  and $\alpha_s(m_Z) = 0.118$, we find $\eta=\alpha_s(0.5 {\rm ~TeV})/\alpha_s(m_t)=0.883$.

At the mass scale of the heavy Higgs bosons, only operators $Q_2$ and $Q_4$ are induced. After evolution to low energies for the $K^0$ system we find
\begin{eqnarray}
C_2(\mu)=C_2(M_s)\cdot  (2.552), ~~~~~C_4(\mu)=C_4(M_s)\cdot  (4.362),\nonumber\\
C_3(\mu)=C_2(M_s) \cdot (-7.43\times10^{-5}),~~~~~ C_5(\mu)=C_4(M_s)\cdot (0.157)~.
\end{eqnarray}
This leads to $\eta_2(\mu) = 2.552,~\eta_4(\mu) = 4.362$ at $\mu = 2$ GeV.  Note that although non-zero $C_3$ and $C_5$ are induced via operator mixing, their coefficients are relatively small.

Following the same procedure,  we compute the evolution of the Wilson coefficients for the $B^0_{d, s}$ system and obtain
\begin{eqnarray}
C_2(\mu)=C_2(M_s)\cdot (1.884), ~~~~~C_4(\mu)=C_4(M_s)\cdot  (2.824),\nonumber\\
C_3(\mu)=C_2(M_s) \cdot (-0.021),~~~~~ C_5(\mu)=C_4(M_s)\cdot (0.076),
\end{eqnarray}
leading to $\eta_2(\mu) = 1.884,~\eta_4(\mu) = 2.824$ at $\mu = M_B$.

Similarly, for the $D^0$ system we find
\begin{eqnarray}
C_2(\mu)=C_2(M_s)\cdot (2.174), ~~~~~C_4(\mu)=C_4(M_s)\cdot  (3.620),\nonumber\\
C_3(\mu)=C_2(M_s) \cdot (-0.011),~~~~~ C_5(\mu)=C_4(M_s)\cdot (0.128),
\end{eqnarray}
leading to $\eta_2(\mu) = 2.174,~\eta_4(\mu) = 3.620$ at $\mu = M_D$.
In all cases, we see that the induced operators $C_3$ and $C_5$ are negligible.


\vspace*{0.1in}
\noindent {\bf $K^0-\overline{K^0}$ mixing constraint}:
\vspace*{0.1in}

The neutral Higgs contributions will modify both the mass difference $\Delta M_K$ and the CP violation parameter $\epsilon_K$.  The mass splitting is obtained from the relation $\Delta m_{K}=2{\rm Re}(M^{K}_{12})$, while the CP violation parameter $\epsilon_K$ is given by $|\epsilon_K|\simeq\frac{{\rm Im} (M^{K}_{12})}{\sqrt{2}\Delta m_{K}}$. We demand that the new contributions to these quantities not exceed the experimental measurements:  $\Delta m_K \simeq (3.484\pm 0.006)\times10^{-15}$ GeV and  $|\epsilon_K|\simeq 2.232 \times10^{-3}$.  Measured values of the Kaon mass and decay constant are used:  $m_K=498 {~\rm MeV}$ and $f_K=160{~\rm MeV}$.  If we assume that $\tilde{Y}_{ds} = \tilde{Y}_{sd}$, and take these coupling to be real, we obtain the constraint
\begin{equation}\label{conK}
\frac{\tilde{Y}_{ds}^2}{10^{-10}} < \frac{1.12 \left(\frac{M_H}{500 \rm ~GeV}\right)^2}{0.18+\left[\left(\frac{M_H}{500 \rm ~GeV}\right)^2-0.062\right]\rm sin^2{(\alpha-\beta)}}
\end{equation}
If we now set $\sin(\alpha-\beta) = 0.4$ and $M_H = M_A = 500$ GeV, we get a limit of $|\tilde{Y}_{ds}| < 1.8 \times 10^{-5}$.  Writing $\tilde{Y}_{ds} = \sqrt{2} C_{ds} (m_d/v)$, as suggested by the modified Yukawa ansatz of Eq. (\ref{bj}), we find $C_{ds} < 1.16$.  If we use instead the CS ansatz and write $\tilde{Y}_{ds} = \sqrt{2} C_{ds} (\sqrt{m_dm_s}/v)$, we would get $|C_{ds}| < 0.26$.  These constraints are tabulated, along with other constraints, in Table \ref{tabcij}.

We have also derived the constraints on $C_{ds}$ from $\Delta m_{K}$ and  $|\epsilon_K|$ by assuming $Y_{ij} = Y_{ji}^{\star} = \frac{\sqrt{2}m_d }{v}C_{ij}e^{i\phi}$.  Here $\epsilon_K$ gives a much more stringent constraint on the phase $\phi$.  These results are 
summarized in Table \ref{tabcij3} where we present three benchmark points for the phase parameter. While both the CS ansatz and the modified ansatz require a relative small phase, the modified ansatz fares better than the CS ansatz.

\vspace*{0.1in}
\noindent {\bf $B_s^0-\overline{B_s^0}$ mixing constraint}:
\vspace*{0.1in}

For the $B_s^0-\overline{B_s^0}$ system we demand that the new contributions be less than the experimental value of \cite{charged1} $\Delta m_{B_s}=1.1688\times 10^{-11}{~\rm GeV}$. Using  $m_{B_s}=5.37{~\rm GeV}$ and $f_{B_s}=295{~\rm MeV}$, we obtain with $\tilde{Y}_{bs} = \tilde{Y}_{sb}$ the constraint
\begin{equation}\label{conBs}
\frac{\tilde{Y}_{bs}^2}{10^{-7}} < \frac{3.09 \left(\frac{M_H}{500 \rm ~GeV}\right)^2}{0.194+\left[\left(\frac{M_H}{500 \rm ~GeV}\right)^2-0.062\right]\rm sin^2{(\alpha-\beta)}}
\end{equation}
If we take $\tilde{Y}_{bs}$ to be real as well,  the limit is $|C_{bs}| \leq 3.17$.

\vspace*{0.1in}
\noindent {\bf $B_d^0-\overline{B_d^0}$ mixing constraint}:
\vspace*{0.1in}

As in the other cases, we demand the new contribution to the $B_d^0-\bar{B_d^0}$ mixing to be less than the experimental value of $\Delta m_{B_d}=3.12\times 10^{-13}{~\rm GeV}$.   
Using as input $m_{B_d}=5.281{~\rm GeV}, ~f_{B_d}=240{~\rm MeV}$ and with $\tilde{Y}_{db} = \tilde{Y}_{bd}$, the constraint is 
\begin{equation}\label{conBd}
\frac{\tilde{Y}_{bd}^2}{10^{-8}} < \frac{1.243 \left(\frac{M_H}{500 \rm ~GeV}\right)^2}{0.194+\left[\left(\frac{M_H}{500 \rm ~GeV}\right)^2-0.062\right]\rm sin^2{(\alpha-\beta)}}
\end{equation}
Under the assumption that the Yukawas couplings are also real,  we get $|C_{bd}| \leq 11.98$ in our modified ansatz.

\vspace*{0.1in}
\noindent {\bf $D^0-\overline{D^0}$ mixing constraint}:
\vspace*{0.1in}

For the $D^0-\overline{D^0}$ mixing, we use \cite{charged1} $\Delta m_{D}=6.25\times 10^{-15}{~\rm GeV}$. With $m_D=1.864 {\rm ~GeV}$, $f_D=200{\rm ~MeV}$, with $\tilde{Y}_{uc} = \tilde{Y}_{cu}$, we find
\begin{equation}\label{conD}
\frac{\tilde{Y}_{uc}^2}{10^{-10}} < \frac{4.04 \left(\frac{M_H}{500 \rm ~GeV}\right)^2}{0.193+\left[\left(\frac{M_H}{500 \rm ~GeV}\right)^2-0.062\right]\rm sin^2{(\alpha-\beta)}}
\end{equation}
When $\tilde{Y}_{uc}$ is also real, we get $|C_{uc}| < 4.9$.

Upper bound  on the coefficients $C_{ij}$ from $K^0-\overline{K^0}$, $B_s^0-\overline{B_s^0}$, $B_d^0-\overline{B_d^0}$ and $D^0-\overline{D^0}$ mixing constraints are summarized in Table \ref{tabcij} for the modified ansatz, and compare them with the constraints from the CS ansatz.  Since the coefficient $C_{sb}$ is near one, one could expect new sources of CP violation in $B_s-\overline{B}_s$, which is small in the SM.

\begin{table}[htbp]
\large
\centering
\setlength{\arrayrulewidth}{0.7mm}
\setlength{\tabcolsep}{17pt}
\renewcommand{\arraystretch}{1.5}
 
\adjustbox{max height=\dimexpr\textheight-1.0cm\relax,
           max width=\dimexpr\textwidth-0.4cm\relax}{
\begin{tabular}{||l|l|l||}
\hline
\multirow{2}{*}{\bf Upper bound  on $C_{ij}$ }  & \multirow{2}{*}{\bf Cheng-Sher Ansatz} & \multirow{2}{*}{\bf Our Ansatz}\\
                  &                   &                   \\ \hline
$K^0-\overline{K^0}$ mixing constraint  & 0.26             &      1.16          \\  \hline
$B_s^0-\overline{B_s^0}$ mixing constraint     &   0.436             &        3.171 \\ \hline
$B_d^0-\overline{B_d^0}$  mixing constraint   &     0.379          &      11.984       \\ \hline
$D^0-\overline{D^0}$ mixing constraint          &    0.222             &     4.893 \\ \hline
\end{tabular}
}
\caption{Upper bounds  on the coefficients $C_{ij}$ from $K^0-\overline{K^0}$, $B_s^0-\overline{B_s^0}$, $B_d^0-\overline{B_d^0}$ and $D^0-\overline{D^0}$ mixing constraints. Here we have set $Y_{ij}= Y_{ji}$, assumed the couplings to be real, and took $M_H = M_A = 500$ GeV. }
\label{tabcij}
\end{table}

\begin{table}[htbp]
\Huge
\renewcommand{\arraystretch}{1.6}
\begin{center}
\adjustbox{max height=\dimexpr\textheight-1.0cm\relax,
           max width=\dimexpr\textwidth-0.0cm\relax}{
\begin{tabular}{|| c | c | c |  c || }
\hline \hline
\centering 
\multirow{2}{*}{\shortstack{\bf Benchmark \\ \bf Points}} & \multirow{2}{*}{\shortstack{\bf Mass Matrix Ansatz}} &  \multirow{2}{*}{\bf Bound on $C_{ij}$  from $\Delta m_{K}$ } & \multirow{2}{*}{\bf Bound on $C_{ij}$  from $\epsilon_{K}$ }  \\ 
&   & & \\ \hline \hline
\multirow{2}{*}{Phase $\phi= 0.1$ } &  \bf Cheng-Sher Ansatz   & $\rm 0.269$ & $\rm 0.048$ \\
& \bf Our Ansatz &  $\rm 1.171$ & $\rm 0.210$\\
\hline
\multirow{2}{*}{Phase $\phi= 10^{-2}$ } &  \bf Cheng-Sher Ansatz  & $\rm 0.267$ & $\rm 0.152$ \\
& \bf Our Ansatz & $\rm 1.159$ & $\rm 0.661$\\
\hline
\multirow{2}{*}{Phase $\phi= 10^{-3}$ } &  \bf Cheng-Sher Ansatz  & $\rm 0.267$ & $\rm 0.481$ \\
& \bf Our Ansatz &  $\rm 1.159$ & $\rm 2.09$\\
\hline
 \hline 
\end{tabular}
}
\caption{Upper bound  on the coefficients $C_{ds}$ from $K^0-\overline{K^0}$ mixing and measurement of $CP$ violation parameter $|\epsilon_K|$.   Here we choose $Y_{ij} = Y_{ji}^{\star} = \frac{\sqrt{2}m_d }{v}C_{ij}e^{i\phi}$ in our modified ansatz, with the factor $m_d$ changed to $\sqrt{m_d m_s}$ for the CS ansatz. }
\label{tabcij3}
\end{center}
\end{table}

\vspace*{0.1in}
\noindent{\bf Constraints from $\mathbf{\mu \rightarrow e \gamma}$}
\vspace*{0.1in}

\begin{figure}[htb!]
$$
 \includegraphics[height=5cm,width=0.9\textwidth]{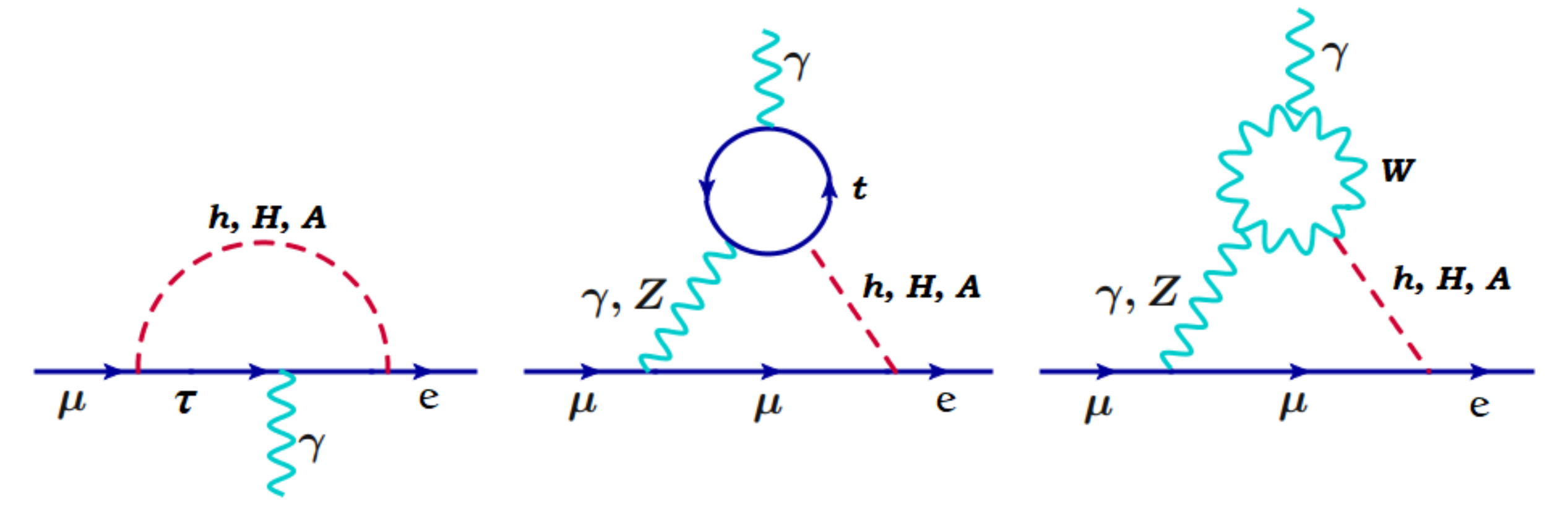}
 $$
 \caption{Representative one and two loop Feynman diagrams contributing to ($\mu \to e \gamma$) process in the 2HDM.
 }
\label{flavor2}
\end{figure}

Here we derive limits on the flavor violating leptonic Yukawa couplings from the loop-induced process $\mu \to e\gamma$.  Both Higgs doublets have couplings to charged leptons in our 2HDM framework.  As a result, there are 
one-loop and two-loop “Barr-Zee” contributions for  $\mu \to e \gamma$ decay. Representative one- and two-loop Feynman diagrams contributing to this decay in the 2HDM are shown in Fig. \ref{flavor2}. We follow the analytic results given in Ref. \cite{hou} to evaluate these diagrams. 
The leading one-loop contribution to $\mu \to e \gamma $  has the $\tau$ lepton and a neutral scalar inside the loop, with the photon radiated from the internal $\tau$ line. It was pointed out in Ref. \cite{weinberg} some time ago that  certain two-loop diagrams may in fact dominate over the one loop contributions owing to smaller chiral suppression.  The loop suppression is overcome by a chiral enhancement of the two-loop diagram, relative to the one-loop diagram.  This effect was noted by Barr and Zee \cite{barrzee} in the context of electric dipole moments.

\begin{figure}[htb!]
$$
 \includegraphics[height=7.5cm,width=0.55\textwidth]{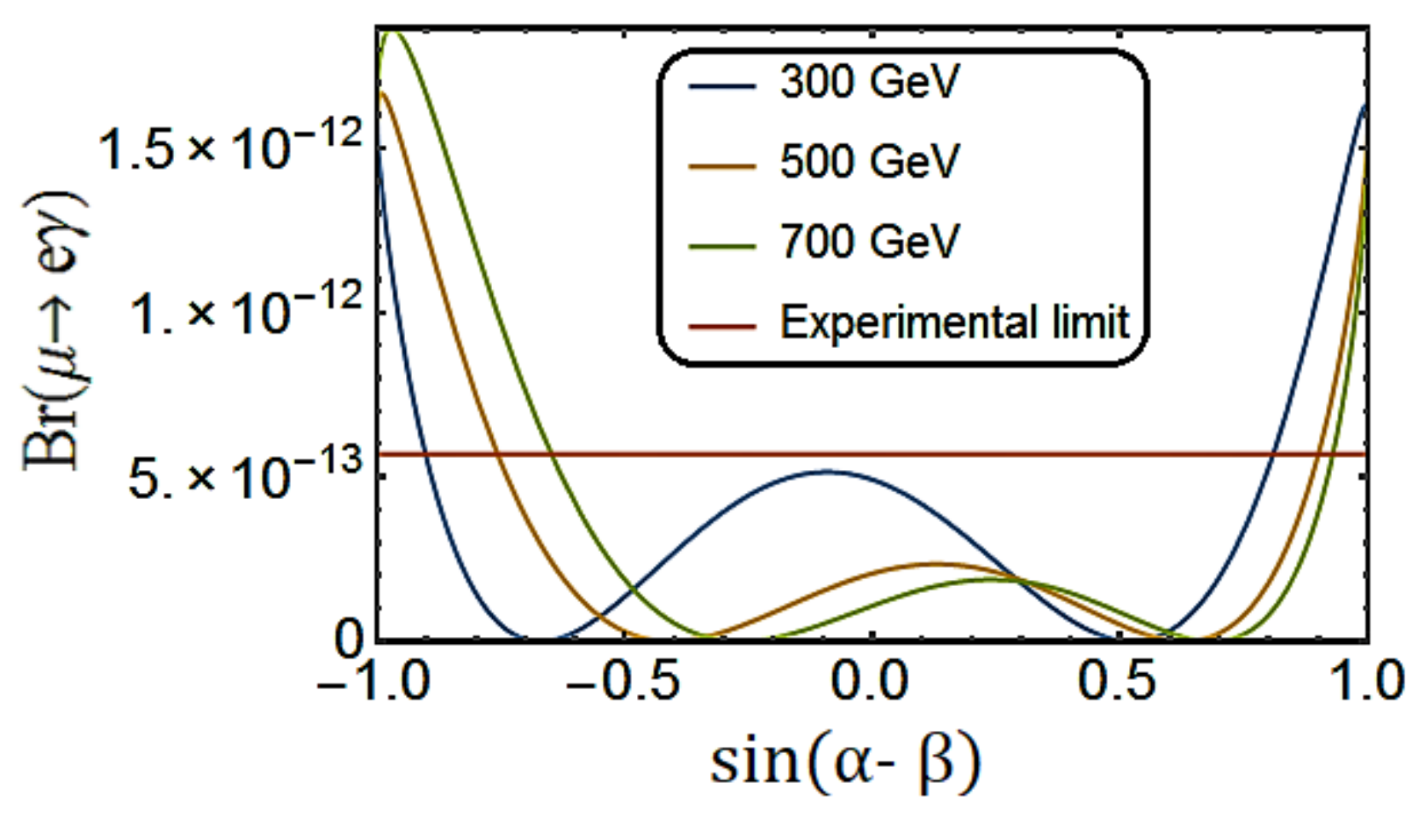}
  \includegraphics[height=7.2cm,width=0.47\textwidth]{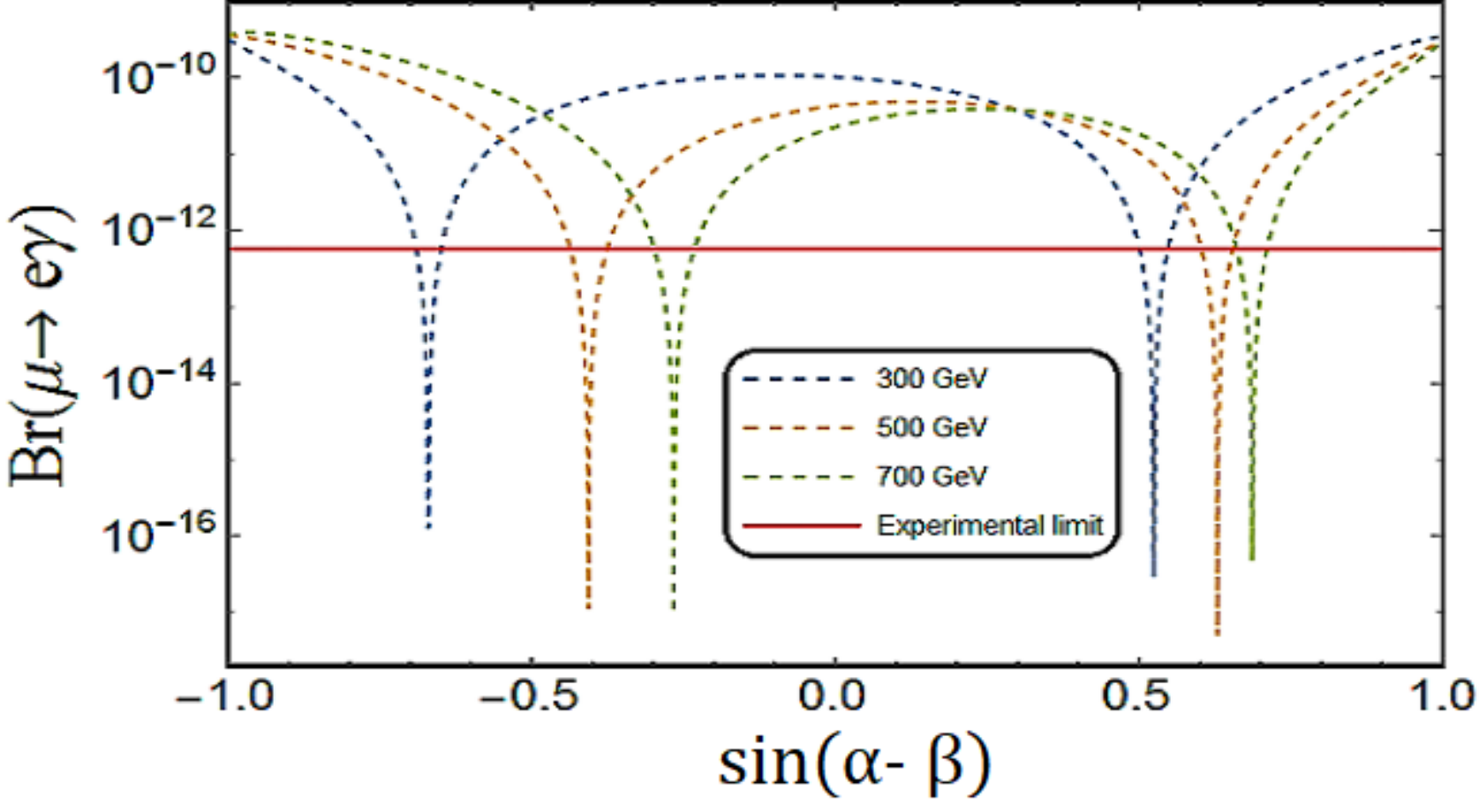}
 $$
 \caption{Branching ratio Br ($\mu \to e \gamma$) as a function  mixing $\sin(\alpha-\beta)$ in two scenarios: {Left:} our modified ansatz where amplitude goes as $m_e$, {Right:} Cheng-Sher ansatz where amplitude goes as $\frac{\sqrt{m_e m_{\mu}}}{v}$. Here we set  $\tilde{Y_t}=1$, and $C_{e\mu} = C_{\mu e} = 1$.
 }
\label{flavor3}
\end{figure}
In Fig. \ref{flavor3}, we compute the branching ratio Br ($\mu \to e \gamma$) as a function of the mixing angle $\sin(\alpha-\beta)$ in two scenarios; one following CS Yukawa coupling  ansatz, and the other following our proposed modified ansatz. For these plots we set
 $\tilde{Y}_t=1$. As can be seen from the figure, for  order one coefficient $C_{e\mu}$ in the CS ansatz, most of the parameter space is ruled out by experimental limit \cite{meg} $Br (\mu \to e \gamma)<4.2 \times 10^{-13} $. It is also clear from the figure that with the modified ansatz, $C_{e\mu}$ can be of order one, even when $\sin(\alpha-\beta)$ is as large as 0.6. For a specific choice of $M_H = 500$ GeV and $\sin(\alpha-\beta)=0.4$, we get $Br (\mu \to e \gamma)=2.5 \times 10^{-11} C_{e\mu}^2$ for the case of the CS ansatz, which requires $|C_{e\mu}| < 0.12$ in this case.  With our modified ansatz we get  $Br (\mu \to e \gamma)=1.21 \times 10^{-13} C_{e\mu}^2$ which leads to to a much weaker constraint $|C_{e\mu}|< 1.9$.  

Thus we see that the modified Yukawa ansatz fares better as regards the Higgs mediated FCNC compared to the CS ansatz.  We have already noted that both ansatze give reasonable values of the CKM matrix. Since in the modified ansatz $C_{e\mu}$ is close to one, we would expect the decay $\mu \to e\gamma$ to be potentially observable.

\section{  Collider Implications of the 2HDM}\label{sec:collider}

In this section we analyze the implications of the 2HDM at colliders.  We pay special attention to the allowed parameter space of the model from observed properties of the 125 GeV $h$ boson, and investigate possible deviations in $t\overline{t}h$, di-Higgs and $Zh$ production rates. As we shall see, in spite of the consistency of the $h$ boson with SM predictions, ample room remains for the above-mentioned signals 
deviating from the SM.

\subsection{Higgs observables at the LHC}

The propertis of the 125 GeV Higgs boson $h$ in various production modes and decays into various final states at LHC so far seem to agree with the SM predictions. But uncertainties still remain in some of these measurements. This encourages to explore potential deviations in certain observables, and their correlations.  One possibility that we have explored is to see if the new Yukawa couplings of the top quark in the 2HDM can lead to deviations in the  $t \bar{t} h$ production cross section at the LHC.  This has recently been observed by the CMS \cite{tthcms} and ATLAS \cite{tthatlas} collaborations.  
We numerically analyze the effects of anomalous top and bottom (and tau) Yukawa couplings on the $t \bar{t} h$ production as well as the signal strengths of Higgs boson decay modes for  $h \to \gamma\gamma, WW, ZZ, b\bar{b}, \tau \bar{\tau}, Z\gamma$.  Then we try to identify the  parameter space which is consistent with both the recent ATLAS and CMS results from the LHC Run-2 ($37$ $fb^{-1}$) data.  Then remaining within the allowed parameter region, we analyze possible conspicuous  signals such as enhanced di-Higgs boson production allowed by the 2HDM. 

The parameter space of our model relevant for LHC study is spanned by the three new  Yukawa couplings of $t$, $b$ and $\tau$ in the rotated Higgs basis,  mass of the heavy neutral Higgs boson $H$ and the mixing angle $\alpha-\beta$:\footnote{The masses of the pseudoscalar $A$ and the charged Higgs boson $H^{\pm}$  are nearly degenerate with the mass of  $H$, and thus not independent in our study.}
\beq
\left\lbrace \tilde{Y_t},\hspace*{0.2cm} \tilde{Y_b},\hspace*{0.2cm} \tilde{Y_\tau},\hspace*{0.2cm} M_H,\hspace*{0.2cm} \sin(\alpha-\beta)  \right\rbrace
\label{eq:parameters}
\eeq

There are several search channels for the 125 GeV $h$ \cite{higgs} at the LHC by ATLAS and CMS collaborations.  These results can give strong bounds on the free parameters of the 2HDM  affecting Higgs observables. While the properties of the $h$ boson are consistent with SM expectations, there is still enough room to look for new physics. To characterize the Higgs boson yields, the signal strength $\mu$ is defined as the ratio of the measured Higgs boson rate to its SM prediction. For a specific Higgs boson production channel and decay rate into specific final states, the signal strength is expressed as:

\begin{equation}
\mu^i_f =  \frac{\sigma^i \cdot BR_f}{(\sigma^i)_{SM} \cdot (BR_{f})_{SM}} = \mu^i\cdot\mu_f;
\label{eq:muif}
\end{equation}
where $\sigma^i$; $(i= ggF, VBF, Wh, Zh, t\bar{t}h)$ is the production cross section for $i\to h$  and $BR_f$; $(f = ZZ^{\star}, WW^{\star}, \gamma \gamma, \tau^+ \tau^-, b\bar{b}, \mu^+ \mu^-)$ is the branching ratio for different decay modes $h\to f$.
The current status on signal strengths constraints \cite{econf, higgs} for various decay modes are summarized in Table \ref{mustat}:  
\begin{table}[htbp]
\begin{center}
\renewcommand{\arraystretch}{1.0}
\adjustbox{max height=\dimexpr\textheight-12.7cm\relax,
           max width=\dimexpr\textwidth+1.0cm\relax}{
\begin{tabular}{||c|c|c|c|c|c|| }
\hline
\hline
\multicolumn{6}{|c|}{\bf \large Higgs Physics Constraints from LHC Run II Data}\\
\hline
\rule[-2ex]{0pt}{5.5ex} \textbf{ Collaboration} & \textbf{Luminosity}  ($\mathcal{L}$) [$fb^{-1}$]   & \textbf{Decay Channels}  & \textbf{Production Modes} & \textbf{Signal Strength Limit ($\mathbf{\mu}$)} &  \textbf{References} \\ 
\hline
\cline{2-6} \rule[-2ex]{0pt}{5.5ex}
&  & & $ggF$ & $1.15^{+0.21}_{-0.18}$ & \\ 
\cline{4-5} \rule[-2ex]{0pt}{5.5ex}
 & 36  & $\gamma\gamma$ & $VBF$ & $0.68^{+0.59}_{-0
.45}$ & \cite{gammarun21} \\ 
\cline{4-5} \rule[-2ex]{0pt}{5.5ex}
&  & & $Wh$ & $3.71^{+1.49}_{-1.35}$ & \\ 
\cline{4-5} \rule[-2ex]{0pt}{5.5ex}
&  & & $Zh$ & $0.0^{+1.13}_{-0.00}$ & \\ 
\cline{2-6} \rule[-2ex]{0pt}{5.5ex}
 &  & & $ggF$ & $1.22^{+0.24}_{-0.21}$ & \\ 
\cline{4-5} \rule[-2ex]{0pt}{5.5ex}
 & 36  & $ZZ^{\star}$ & $VBF$ & $-0.09^{+1.02}_{-0
.76}$ & \cite{zrun21} \\ 
\cline{4-5} \rule[-2ex]{0pt}{5.5ex}
\textbf{CMS}  &  & & $Wh$ & $0.0^{+2.32}_{-0.00}$ & \\ 
\cline{4-5} \rule[-2ex]{0pt}{5.5ex}
  &  & & $Zh$ & $0.0^{+4.26}_{-0.00}$ &
\\
\cline{2-6} \rule[-2ex]{0pt}{5.5ex}
&  & & $ggF$ & $1.35^{+0.20}_{-0.19}$ & \\ 
\cline{4-5} \rule[-2ex]{0pt}{5.5ex}
 & 36  & $WW^{\star}$ & $VBF$ & $0.28^{+0.64}_{-0
.60}$ & \cite{w3} \\ 
\cline{4-5} \rule[-2ex]{0pt}{5.5ex}
&  & & $Wh$ & $3.91^{+2.26}_{-2.01}$ & \\ 
\cline{4-5} \rule[-2ex]{0pt}{5.5ex}
&  & & $Zh$ & $0.96^{+1.81}_{-1.46}$ & \\
\cline{2-6} \rule[-2ex]{0pt}{5.5ex}
&  & & $ggF$ & $1.05^{+0.53}_{-0.47}$ & \\ 
\cline{4-5} \rule[-2ex]{0pt}{5.5ex}
 & 36  & $\tau^{+} \tau^{-}$ & $VBF$ & $1.12^{+0.45}_{-0
.43}$ & \cite{tau2} \\ 
\cline{4-5} \rule[-2ex]{0pt}{5.5ex}
&  & & $ggF+VBF+Vh$ & $1.06^{+0.25}_{-0.24}$ & \\ 
\cline{2-6} \rule[-2ex]{0pt}{5.5ex}
& 36 & $b\overline{b}$ & $Vh$ & $1.06^{+0.31}_{-0.29}$ & \cite{b2} \\ 
\cline{2-6} \rule[-2ex]{0pt}{5.5ex}
& 36  & $\mu^{+}\mu^{-}$ & $ggF+VBF+Vh$ & $0.7^{+1.0}_{-1.0}$ & \cite{mucms}\\
\hline \hline
 \rule[-2ex]{0pt}{5.5ex}
&  & & $ggF$ & $0.80^{+0.19}_{-0.18}$ & \\ 
\cline{4-5} \rule[-2ex]{0pt}{5.5ex}
 & 36  & $\gamma\gamma$ & $VBF$ & $2.10^{+0.60}_{-0
.60}$ & \cite{gammarun22} \\ 
\cline{4-5} \rule[-2ex]{0pt}{5.5ex}
&  & & $Vh$ & $0.70^{+0.90}_{-0.80}$ & \\ 
\cline{2-6} \rule[-2ex]{0pt}{5.5ex}
 &  & & $ggF$ & $1.11^{+0.23}_{-0.27}$ & \\ 
\cline{4-5} \rule[-2ex]{0pt}{5.5ex}
 & 36  & $ZZ^{\star}$ & $VBF$ & $4.0^{+2.10}_{-0
.18}$ & \cite{zrun22} \\ 
\cline{4-5} \rule[-2ex]{0pt}{5.5ex}
\textbf{ATLAS}  &  & & $Vh$ & $0.0^{+1.90}_{-1.90}$ & \\ 
\cline{2-6} \rule[-2ex]{0pt}{5.5ex}
&  & & $ggF$ & $1.02^{+0.29}_{-0.26}$ & \\ 
\cline{4-5} \rule[-2ex]{0pt}{5.5ex}
 & 6  & $WW^{\star}$ & $VBF$ & $1.70^{+1.1}_{-0
.90}$ & \cite{w1,w2} \\ 
\cline{4-5} \rule[-2ex]{0pt}{5.5ex}
&  & & $Vh$ & $3.2^{+0.44}_{-4.2}$ & \\ 
\cline{2-6} \rule[-2ex]{0pt}{5.5ex}
&  & & $ggF$ & $2.0^{+0.80}_{-0.80}$ & \\
\cline{4-5} \rule[-2ex]{0pt}{5.5ex}
 & 36  & $\tau^{+} \tau^{-}$ & $VBF +Vh$ & $1.24^{+0.58}_{-0
.54}$ & \cite{tau2} \\ 
\cline{4-5} \rule[-2ex]{0pt}{5.5ex}
&  & & $ggF+VBF+Vh$ & $1.43^{+0.43}_{-0.37}$ & \\ 
\cline{2-6} \rule[-2ex]{0pt}{5.5ex}
& 36 & $b\overline{b}$ & $Vh$ & $0.9^{+0.28}_{-0.26}$ & \cite{b1} \\ 
\cline{2-6} \rule[-2ex]{0pt}{5.5ex}
& 36  & $Z\gamma$ & $ggF+VBF+Vh$ & $0.0^{+3.4}_{-3.4}$ & \cite{zgamma1}\\ 
\cline{2-6} \rule[-2ex]{0pt}{5.5ex}
& 36  & $\mu^{+}\mu^{-}$ & $ggF+VBF+Vh$ & $-0.10^{+1.50}_{1.50}$ & \cite{muatlas}\\
\hline \hline
\end{tabular}
}
\caption{ Signal strength constraints from recently reported 13 TeV  LHC data along with references. }
\label{mustat} 
\end{center}
\end{table}

Now the partial decay widths for various SM Higgs decay modes within the 2HDM are calculated as: 
\beqa
\Gamma_{h \to \gamma\gamma} &=& \kappa_{\gamma\gamma}^2 \Gamma_{h \to \gamma\gamma}^{{\rm SM}}, \\
\Gamma_{h \to b\bar{b}} &=& \kappa_b^2 \Gamma_{h \to bb}^{{\rm SM}}, \\
\Gamma_{h \to WW^*} &=& \kappa^2_{W}\Gamma_{h \to WW^*}^{{\rm SM}}, \\
\Gamma_{h \to ZZ^*} &=& \kappa^2_{Z}\Gamma_{h \to ZZ^*}^{{\rm SM}}, \\
\Gamma_{h \to gg} &=& \kappa^2_{g}\Gamma_{h \to gg}^{{\rm SM}}, \\
\Gamma_{h \to \tau^+\tau^-} &=& \kappa_\tau^2\Gamma_{h \to \tau\tau}^{{\rm SM}}, \\
\Gamma_{h \to c\bar{c}} &=& \Gamma_{h \to cc}^{{\rm SM}}, \\
\Gamma_{h \to Z\gamma} &=& \kappa_{Z\gamma}^2 \Gamma_{h \to Z\gamma}^{{\rm SM}}, 
\eeqa
where one can find the SM partial decay widths in Ref. \cite{Gunion}.

 In order to study the constraints from the current LHC data, the scaling factors which show deviations in the Higgs coupling from the SM in the 2HDM are defined as:
\beqa
\kappa_{W,Z} &=& \cos{(\alpha-\beta)}, \\
\kappa_t &=& \left[ \cos{(\alpha-\beta)} +\frac{\tilde{Y_t}v}{\sqrt{2}m_t} \sin(\alpha-\beta) \right], \label{kappat}\\
\kappa_b &=& \left[ \cos{(\alpha-\beta)} +\frac{\tilde{Y_b}v}{\sqrt{2}m_b} \sin(\alpha-\beta) \right], \\
\kappa_\tau &=& \left[ \cos{(\alpha-\beta)} +\frac{\tilde{Y_\tau}v}{\sqrt{2}m_\tau} \sin(\alpha-\beta) \right], \\
\kappa_{\gamma\gamma} &=&  \left| \frac{\frac{4}{3} \kappa_t F_{1/2}(m_h) + F_1(m_h) \cos{(\alpha-\beta)} + \frac{v \lambda_{hH^+H^-}F_{0}(m_h)}{2m^2_{H^{+}}}}{\frac{4}{3} F_{1/2}(m_h) + F_1(m_h)}  \right|, \\
\kappa_{Z\gamma} &=&  \left| \frac{ \frac{2}{\cos \theta_W} \left( 1- \frac{8}{3} \sin^2 \theta_W \right) \kappa_t F_{1/2}(m_h) + F_1(m_h)\cos{(\alpha-\beta)} + \frac{v \lambda_{hH^+H^-}\lambda_{ZH^+H^-}F_{0}(m_h)}{2m^2_{H^{+}}}}{ \frac{2}{\cos \theta_W} \left( 1- \frac{8}{3} \sin^2 \theta_W \right) F_{1/2}(m_h) + F_1(m_h)}  \right|, \\
\kappa_g &=& \left[ \frac{1.034 \kappa_t + \epsilon_b \kappa_b}{1.034+\epsilon_b} \right], \\
\eeqa
where the loop function are given by:
 \begin{eqnarray}
F_1(x) &=& -x^2\left[2x^{-2}+3x^{-1}+3(2x^{-1}-1)f(x^{-1})\right]\ ,\\
F_{1/2}(x) &=& 2  \, x^2 \left[x^{-1}+ (x^{-1}-1)f(x^{-1})\right] \ , \\
 f(x) &=& \arcsin^2 \sqrt{x}  \, \\
 \epsilon_b &=& -0.032+0.035 i\,,
\end{eqnarray}
 with $x_i \equiv 4m_i^2/m_h^2$ ($i=t, W$).

In the 2HDM, the charged Higgs boson will also contribute to $h \to \gamma \gamma$ and $h \to Z \gamma$ decay via loops, in addition to the top quark and W boson loops. These  contributions to $h \to \gamma \gamma$ and $h \to Z \gamma$ are negligible, as long as the mass of $H^\pm$ is kept above 300 GeV to be consistent with the current experimental constraints \cite{charged1, ch1, ch2, ch3}.

\begin{figure}[htb!]
$$
 \includegraphics[height=7.5cm,width=0.49\textwidth]{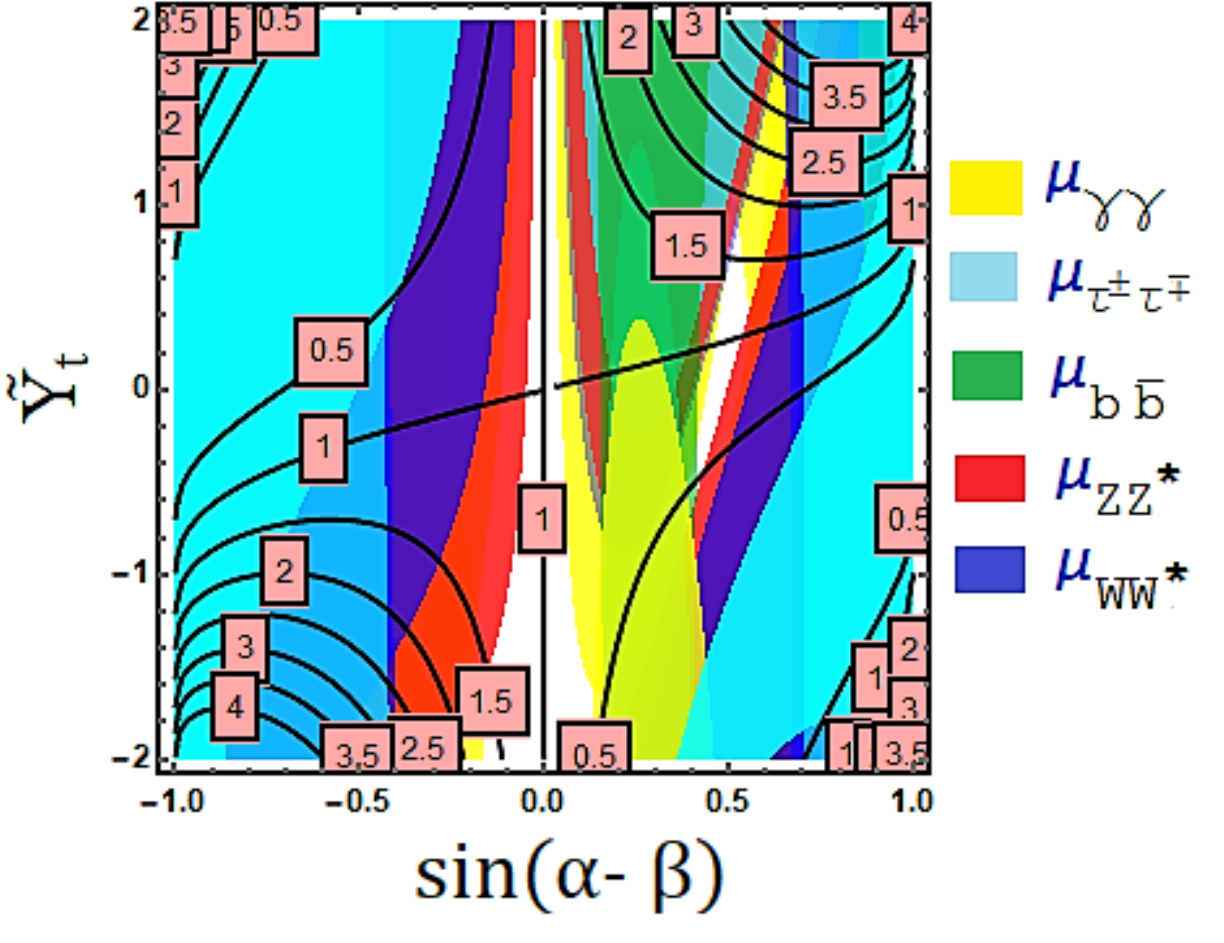}
  \includegraphics[height=7.5cm,width=0.49\textwidth]{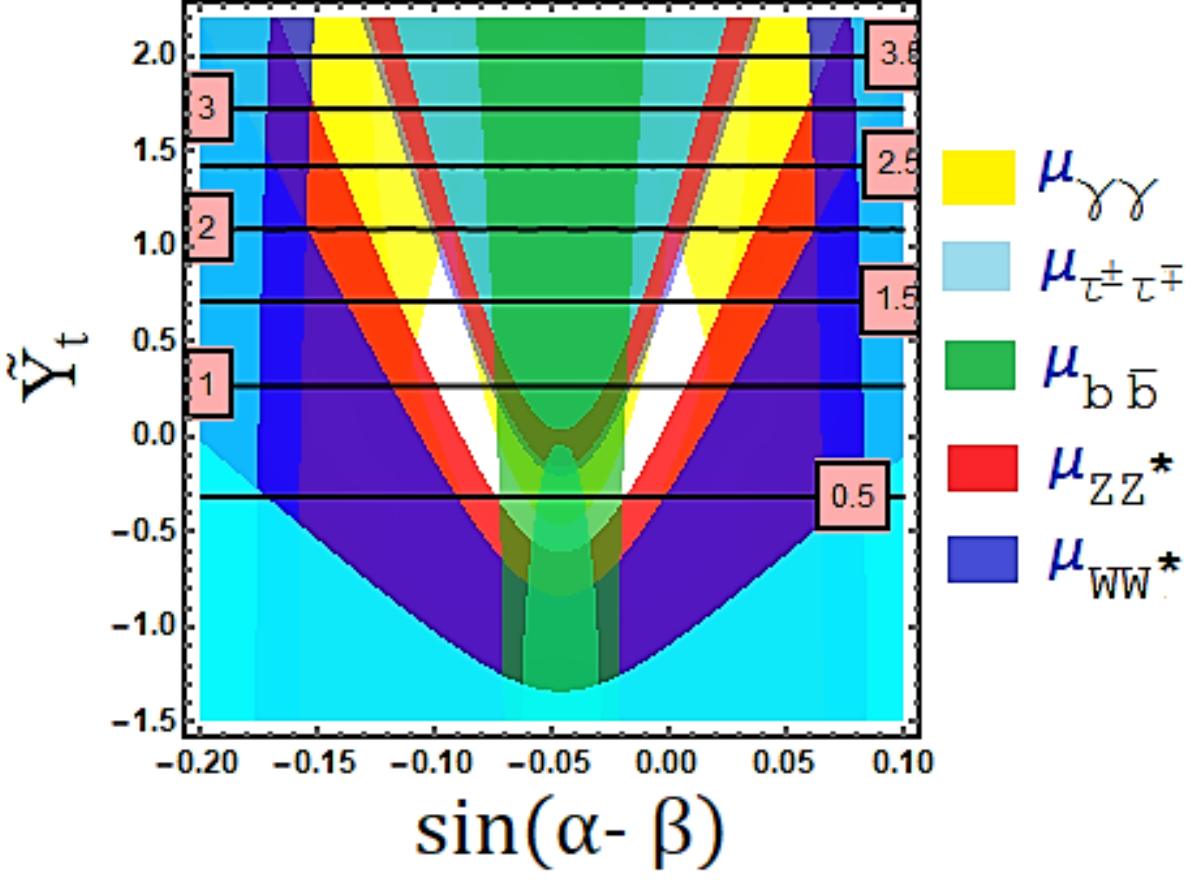}
 $$
 $$
\includegraphics[height=7.2cm]{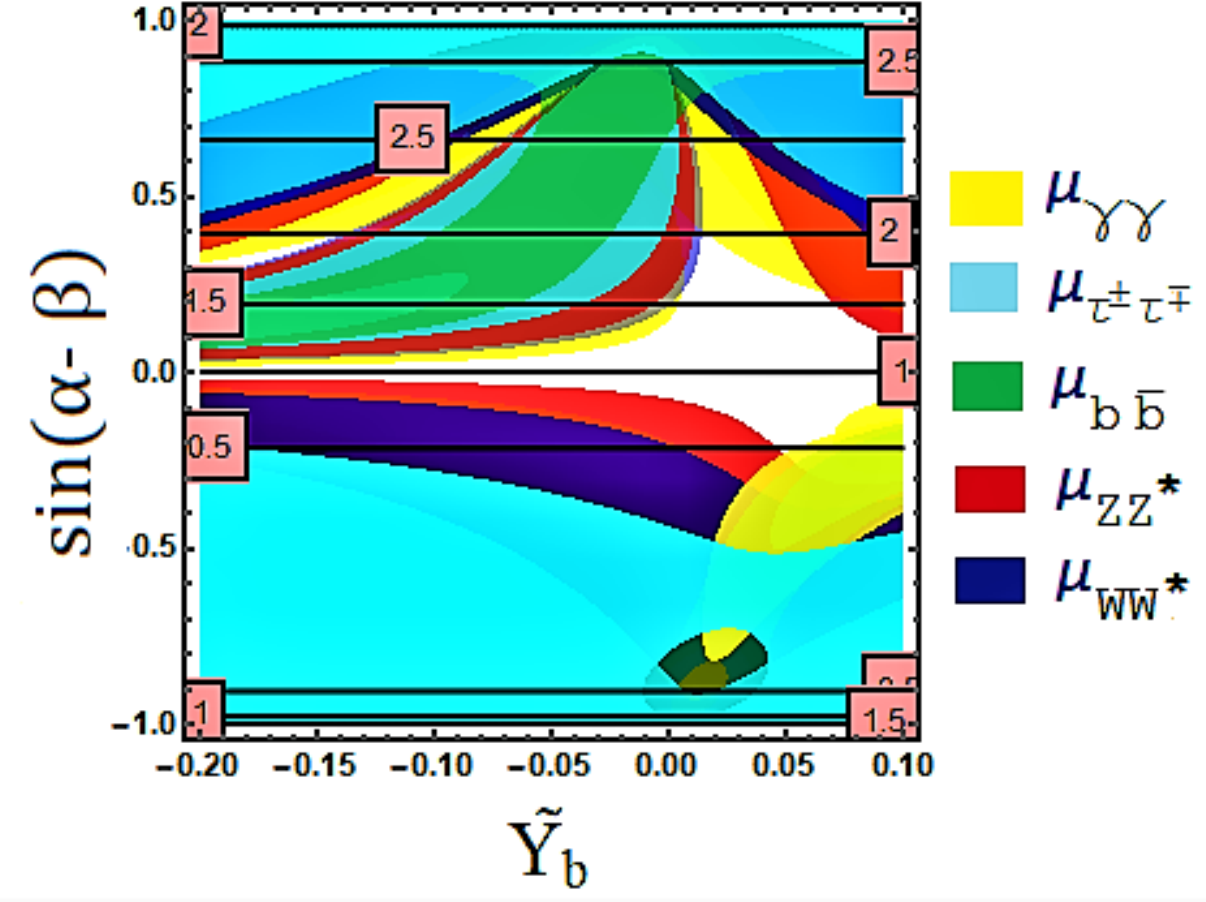}
 $$
 \caption{{Top Left:} Contour plot of $\mu^{t\bar{t}h}$ in $\lbrace\tilde{Y_t}, \sin(\alpha-\beta)\rbrace$ plane. Here $\tilde{Y_b}= -0.09$ is kept fixed. 
{Top Right:} Contour plot of $\mu^{t\bar{t}h}$ in $\lbrace\tilde{Y_t}, \tilde{Y_b}\rbrace$ plane. Here we choose $\sin(\alpha-\beta) = 0.5$. {Bottom:} Contour plot of $\mu^{t\bar{t}h}$ in $\lbrace\sin(\alpha-\beta), \tilde{Y_b}\rbrace$ plane, with $\tilde{Y_t}=1.25$ fixed. The yellow, cyan, green, red and purple shaded regions are excluded from the signal strength limits for various decay modes ($\gamma\gamma, \tau \tau, b\bar{b}, ZZ^\star, WW^\star$) respectively. The white shaded region simultaneously satisfies all the experimental constraints.  
 }
\label{1}
\end{figure}

The Run-1 data reported by ATLAS and CMS collaborations have been combined \footnote{Since Run-II 37 $fb^{-1}$ data of ATLAS and CMS collaborations have not combined and reported yet, we have used the following formalism to combine the ATLAS and CMS reported results for the signal strength for a specific production and decay mode: If $x_{1}, . . . , x_{N}$ are unbiased measurements of the same unknown
quantity $x$ with different variances $\sigma^2_1 . . . \sigma^2_N$,  then the weighted average is:
$X = \frac{\sum_{i=1}^{N} w_i x_i}{\sum_{i=1}^{N} w_i}$, where $w_i = \frac{1}{\sigma_i^2}$ is unbiased ($<X>= x$) with variance
$\sigma_X^2 = \frac{1}{\sum_{i=1}^{N} w_i}$.} and analyzed using the signal strength formalism and the results are presented in Ref. \cite{higgs}. Recently, ATLAS and CMS collaborations have reported the results \cite{econf} on Higgs searches based on 36 $fb^{-1}$ data at 13 TeV LHC. The individual analysis by each experiment examines a specific Higgs boson decay mode corresponding to the various production processes which are $h \to \gamma\gamma$ \cite{gammarun21,gammarun22,gamma1,gamma2}, $h \to ZZ^\star$ \cite{zrun21,zrun22,z1,z2}, $h \to WW^\star$ \cite{w1,w2,w3}, $h \to \tau \tau$ \cite{tau1,tau2}, $h \to b\bar{b}$ \cite{b1,b2} and $h \to Z\gamma$ \cite{zgamma1,zgamma2}.  In Fig \ref{1}, we have used the most updated constraints on signal strengths reported by ATLAS and CMS collaboration for all individual production and decay modes  as shown in Table \ref{mustat} at 95$\%$ confidence level. As a consistency check, we have used also the combined signal strength value for a specific decay mode considering all the production modes and which is relatively stable with the results of Fig. \ref{1}.

\subsection{Deviations in \boldmath{$h$} Yukawa couplings and LHC constraints}

\begin{figure}[htb!]
$$
 \includegraphics[height=6.0cm,width=0.54\textwidth]{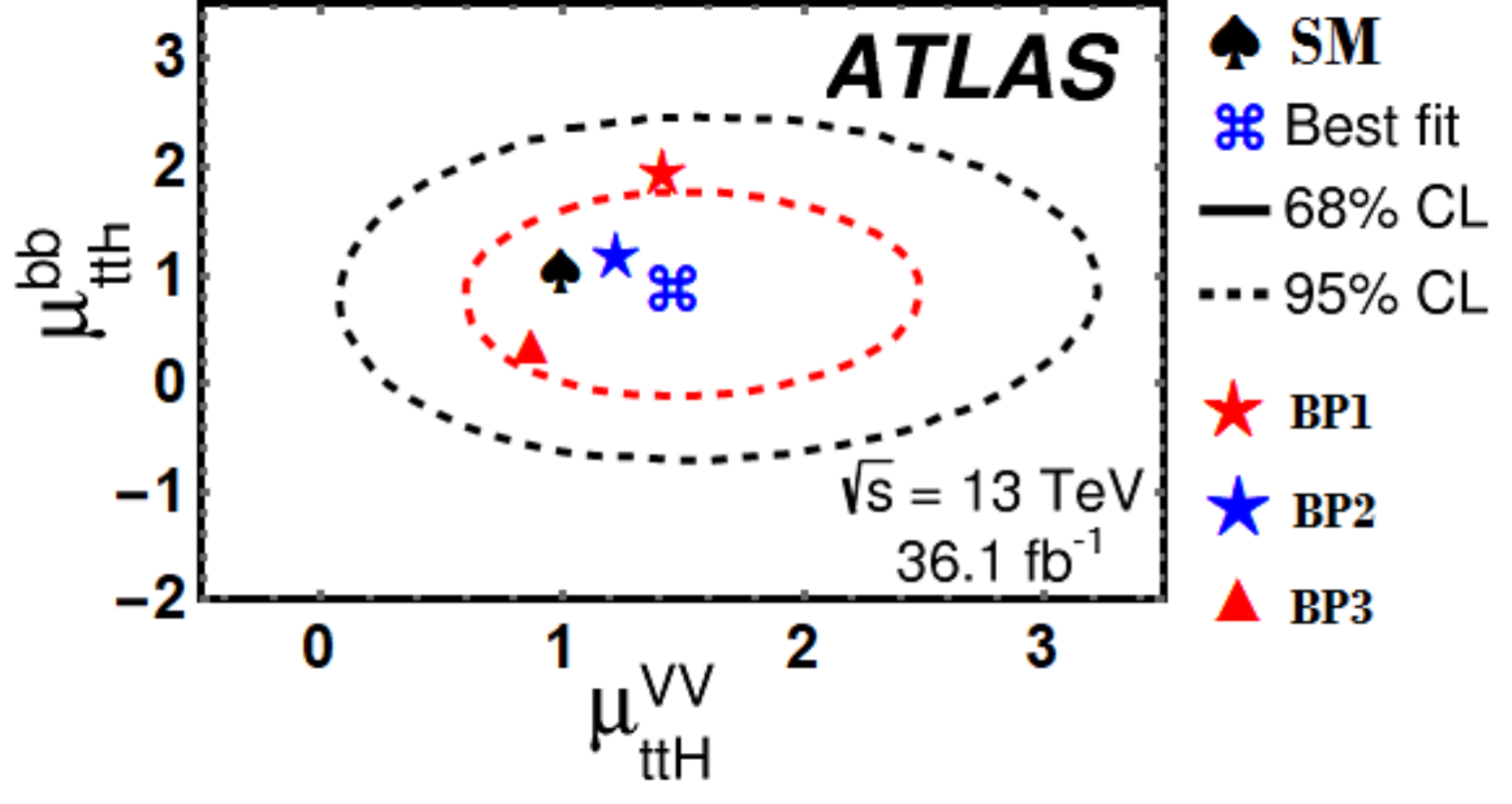}
  \includegraphics[height=6.0cm,width=0.54\textwidth]{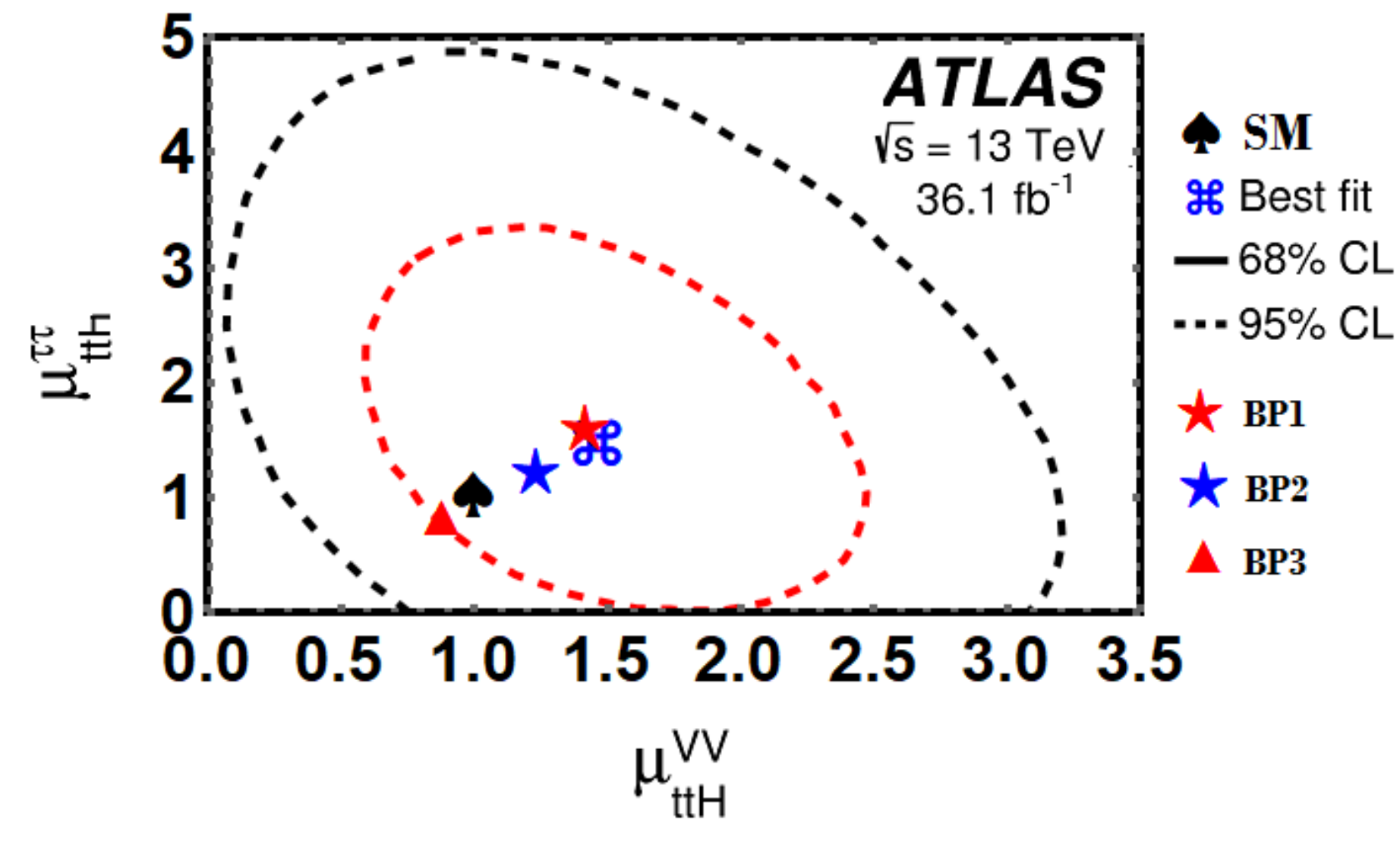}
 $$
 \caption{: The two-dimensional best-fit of the signal strength modifiers for the processes $t\bar{t}h, h \to b\bar{b}$ versus $t\bar{t}h, h \to VV^{\star},(V=W,Z)$ (left) and $t\bar{t}h, h \to   \tau^{+} \tau^{-}$ versus $t\bar{t}h, h \to VV^{\star}, (V=W,Z)$ (right). Three benchmark points (BP) are also shown in this contour plot.  
 }
\label{new2}
\end{figure}


Now, we evaluate the signal strength $\mu^{t\bar{t}h}$ $(= \kappa_t^2)$ for the production of SM Higgs associated with the top quark pair. Since the tau Yukawa term, $\tilde{Y_\tau}$, has no significant contribution to the total decay width of SM Higgs compared to $\tilde{Y_b}$, we set the value of $\tilde{Y_\tau}$ small and equal to $10^{-3}$ for the rest of our analysis. This choice does not affect the phenomenology that we focus on. The upper left segment of Fig. \ref{1} shows the contour plot of $\mu^{t\bar{t}h}$ in $\lbrace\tilde{Y_t}, \sin(\alpha-\beta)\rbrace$ plane for a fixed value of $\tilde{Y_b}=-0.09$, while the right segment shows the contour plot of $\mu^{t\bar{t}H}$ in $\lbrace\tilde{Y_t}, \tilde{Y_b}\rbrace$ plane for a fixed value of $\sin(\alpha-\beta) = 0.5$. The bottom segment of Fig. \ref{1} shows the contour plot of $\mu^{t\bar{t}h}$ in $\lbrace\tilde{Y_b}, \sin(\alpha-\beta)\rbrace$ plane for a fixed value of $\tilde{Y_t}=1.25$. Fig. \ref{1} clearly indicates that within the 2HDM, $t\bar{t}h$ can be produced upto 1.9 times the SM  predicted cross-section at the LHC, satisfying all the current experimental constraints from the 125 GeV Higgs boson searches within our model as we allow a variation of $\tilde{Y_t}$ between $-2$ and $2$. As we shall see in the next section, $\mu_{t\bar{t}h}$ will be further constrained from experimental limits on heavy Higgs boson searches. It is also to be noted that $t\bar{t}h$ production rate can be as low as 0.5 times weaker than the SM predicted values within the 2HDM. We see from Fig. \ref{1} that $\mu_{t\bar{t}h}$ value gets suppressed if either $\sin(\alpha-\beta)$ or $\tilde{Y_t}$ is negative  as shown in Eq. \ref{kappat}.  Due to the different interference patterns between Yukawas ($\tilde{Y_t}, \tilde{Y_b}, \tilde{Y_{\tau}} $) and the mixing $\sin(\alpha-\beta)$ in different decay modes, these plots are not symmetric about the central axes. It should be noted that, within our model there are additional modes of $t\bar{t}h$ production via SM Higgs $h$ production in association with the pseudoscalar $A$ or heavy Higgs $H$, followed by the decay of $A$ and/or $H$ to  $t\bar{t}$. Since the $hA$ production via quark fusion is dictated by the coupling $ZhA$, which is suppressed by $\sin(\alpha-\beta)$, and also due to the significant loss in quark luminosity compared to gluon luminosity in the production, we have found its contribution in $t\bar{t}h$ production to be less than $1\%$. On the other hand, $hH$ or $hA$ production via gluon gluon fusion will occur through triangle and box diagrams with top quarks. Due to the destructive interference between these two diagrams, the production rate will be small. In addition,
 there will be another suppression in the subsequent branching ratios for $H \to t\bar{t}$, or $A \to t\bar{t}$. Although its contribution to the total $t\bar{t}h$ production is found to be less than $2\%$, we take these effects into account. The white shaded region in Fig. \ref{1} simultaneously satisfies all the experimental constraints.
 
  We have also scanned the parameter space for negative $\tilde{Y_t}$ and  found that the most of the parameter space is ruled out by the current experimental constraints  provided that $t\bar{t}h$ production is decreased with compared to the SM. We also calculate the signal strength for $ Z \gamma$ channel which is well consistent with the available experimental data \cite{zgamma1,zgamma2}. The signal strength in $ Z \gamma$ channel can vary from 0.7 to 1.4 satisfying all the constraints. Considering the best possible scenario in the available allowed parameter space for the Higgs boson production associated with a top quark pair, followed by the Higgs boson decays to $WW^\star$ and $ZZ^\star$, the limit will go upto 1.6 times of the SM, where we allow $\mid \tilde{Y_t} \mid$ values upto 2, since it will be suppressed by an extra $\cos^2 {(\alpha-\beta)}$ term due to the different $hWW$ and $hZZ$ coupling and which can simultaneously explain the recently reported ATLAS and CMS results  \cite{tthcms, tthatlas} on $t\bar{t}h$ production. The two-dimensional best-fit of the signal strength modifiers for the processes $t\bar{t}h, h \to b\bar{b}$ versus $t\bar{t}h, h \to VV^{\star},(V=W,Z)$ (left) and $t\bar{t}h, h \to   \tau^{+} \tau^{-}$ versus $t\bar{t}h, h \to VV^{\star}, (V=W,Z)$ (right) is shown in Fig. \ref{new2}. Three benchmark points (BP), mentioned later in next sub-section, are also shown in this contour plot.

\begin{table}[htbp]
\centering
\label{my-label}
\begin{tabular}{||l|l|l||}
\hline
\multirow{2}{*}{\textbf{Channel}} & \multicolumn{2}{l|}{\textbf{Observed Limit on Signal Strength $(\mu)$}} \\ \cline{2-3} 
                  &     \textbf{ATLAS}      & \textbf{CMS}           \\ 
                  \hline
                 Run-1 Combination & \multicolumn{2}{l|}{\quad \quad \quad \quad  $\mu_{t\bar{t}h}$ =  $2.3\substack{+0.7 \\ -0.6}$  } \\ 
                 \hline
                 $b\bar{b}$ &   $\mu_{t\bar{t}h}$ =  $0.84\substack{+0.64 \\ -0.61}$          &     $\mu_{t\bar{t}h}$ =  $0.91\substack{+0.45 \\ -0.43}$        \\ 
                 \hline
                 Multilepton &   $\mu_{t\bar{t}h}$ =  $1.6\substack{+0.5 \\ -0.4}$         &      $\mu_{t\bar{t}h}$ =  $1.60\substack{+0.66 \\ -0.59}$    \\ 
                 \hline
                 ZZ &   $\mu_{t\bar{t}h} <$  7.7       &    $\mu_{t\bar{t}h}$ =  $0.00\substack{+1.51 \\ -0.00}$      \\ 
                 \hline
                  $\gamma \gamma$&    $\mu_{t\bar{t}h}$ =  $0.5\substack{+0.6 \\ -0.6}$        &    $\mu_{t\bar{t}h}$ =  $2.14\substack{+0.87 \\ -0.74}$     \\
                  \hline
                  Combined analysis&    $\mu_{t\bar{t}h}$ =  $1.32\substack{+0.28 \\ -0.26}$        &    $\mu_{t\bar{t}h}$ =  $1.26\substack{+0.31 \\ -0.26}$     \\
                  \hline
\end{tabular}
\caption{Current summary of the observed signal strength $\mu$ measurements and $t\bar{t}h$ production significance from individual analyses and the combination as reported by ATLAS and CMS collaboration \cite{tthcms,tthatlas}. }
\end{table}


\subsection{Constraints from measurements of flavor violating Higgs boson couplings}
\noindent { \textit{Constraint from the exotic decay of top quark $t\rightarrow h c$ :}}
\vspace{0.5cm}

In the 2HDM, an exotic top quark decay $t\rightarrow h c$ will be generated for non-zero $\tilde{Y}_u^{tc}$ as well as $\tilde{Y}_u^{ct}$.
The decay branching ratio for $t\rightarrow h c$ is:
\begin{align}
  {\rm BR}(t\rightarrow h c)&=\frac{\sin^2{(\alpha - \beta)} (|\tilde{Y}_u^{tc}|^2+|\tilde{Y}_u^{ct}|^2)}{64\pi}\frac{m_t}{\Gamma_t}
  \left(1-\frac{m_h^2}{m_t^2}\right)^2, \\
  &\simeq 3\times 10^{-3}~\left(\frac{\tilde{Y}_u^{tc} \sin{(\alpha-\beta)}}{0.15}\right)^2.
  \label{tDecay}
\end{align}
Here we adopt $\Gamma_t=1.41~{\rm GeV}$ for the total decay rate of the top quark. The current experimental  bound at the 95\% C.L.~\cite{Khachatryan:2016atv} is reported as:
\begin{align}
  {\rm BR}(t\rightarrow h c)\le 4\times 10^{-3}.
\end{align}
Our model is well consistent with this constraint as it predicts very suppressed branching ratio $ {\rm BR}(t\rightarrow h c)= 7.33 \times 10^{-6} \sin^2(\alpha - \beta )$ for order one coefficient $C_{tc}$.

\vspace{0.5cm}
\noindent{\textit{Constraints from lepton flavor violating Higgs boson decays:}}
\vspace{0.5cm}

Searches for the lepton flavor violating Higgs boson decays $h\to e\tau,~h \to \mu\tau$ constrain the lepton flavor violating Yukawa couplings $\tilde{Y}_e^{e\tau}$ and $\tilde{Y}_e^{\mu\tau}$. CMS collaboration recently reported their updated results 
with an integrated luminosity of 35.9 fb$^{-1}$ at $\sqrt{s}=13$ TeV:
\begin{align}
  {\rm BR}(h\rightarrow \mu\tau)&\le0.25~\%,\\
  {\rm BR}(h\rightarrow e \tau)&\le0.61~\%,
\end{align}
at 95\% C.L.~\cite{LFVmutau}. Our model predicts the branching ratio ${\rm BR}(h\rightarrow \mu\tau)$ to be 
\begin{align}
  {\rm BR}(h\rightarrow \mu\tau)&=\frac{\sin^2{(\alpha - \beta)}\left(|\tilde{Y}_e^{\mu\tau}|^2+|\tilde{Y}^{\tau\mu}_e|^2\right)m_h}
  {16\pi \Gamma_h},\\
  &=0.24\% \left(\frac{\tilde{Y}_e^{\mu\tau}\sin{(\alpha-\beta)}}{2\times 10^{-3}}\right)^2,
\end{align}
where  the total decay rate of the Higgs boson is given by $\Gamma_h=4.1$ MeV. Similarly the 2HDM predicts
\begin{align}
  {\rm BR}(h\rightarrow e\tau)&=0.62\% \left(\frac{\tilde{Y}_e^{e\tau}\sin{(\alpha-\beta)}}{3.2\times 10^{-3}}\right)^2.
\end{align}
Hence lepton flavor violating Higgs boson decays do not put any significant bound on the parameters of our 2HDM framework.


\subsection{$\mathcal{CP}$-- even  Higgs phenomenology}
\begin{figure}[htb!]
$$
 \includegraphics[height=7cm,width=0.45\textwidth]{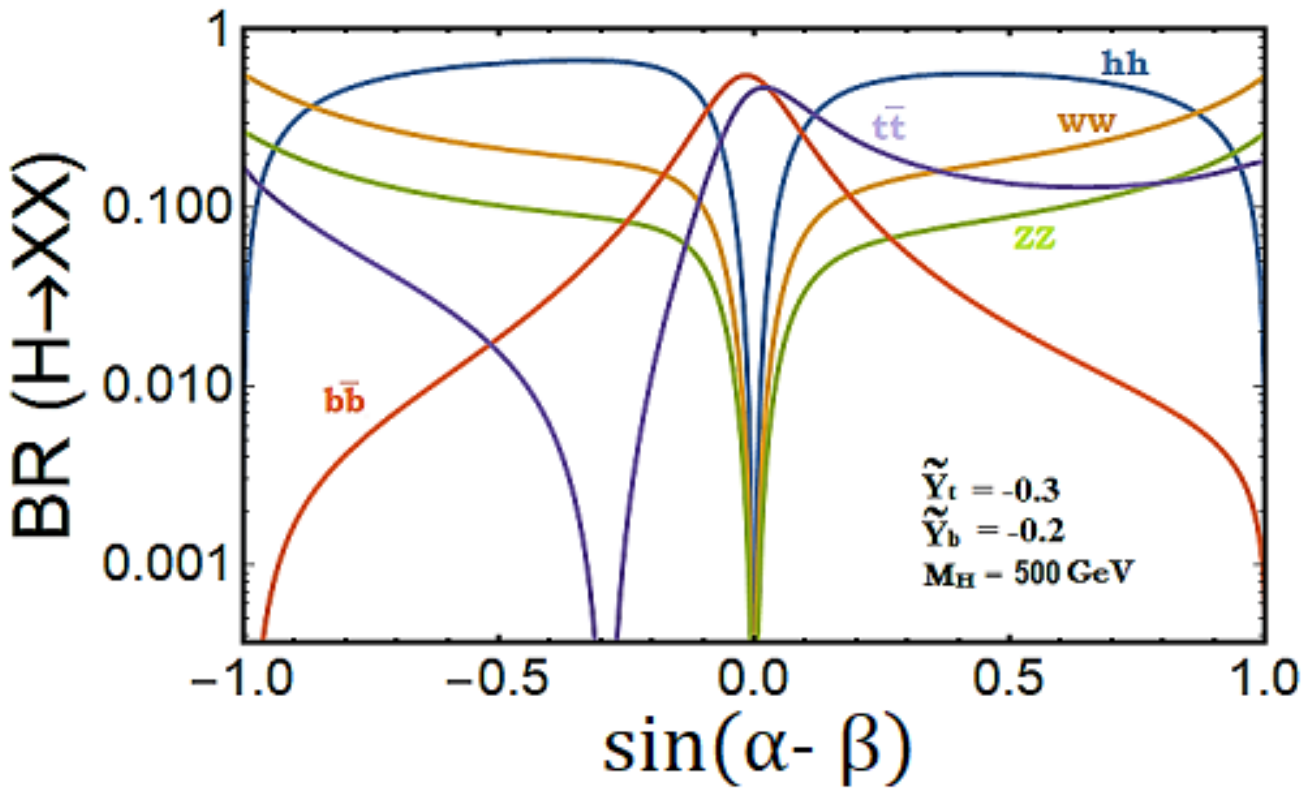} \hspace{0.3 in}
  \includegraphics[height=7cm,width=0.45\textwidth]{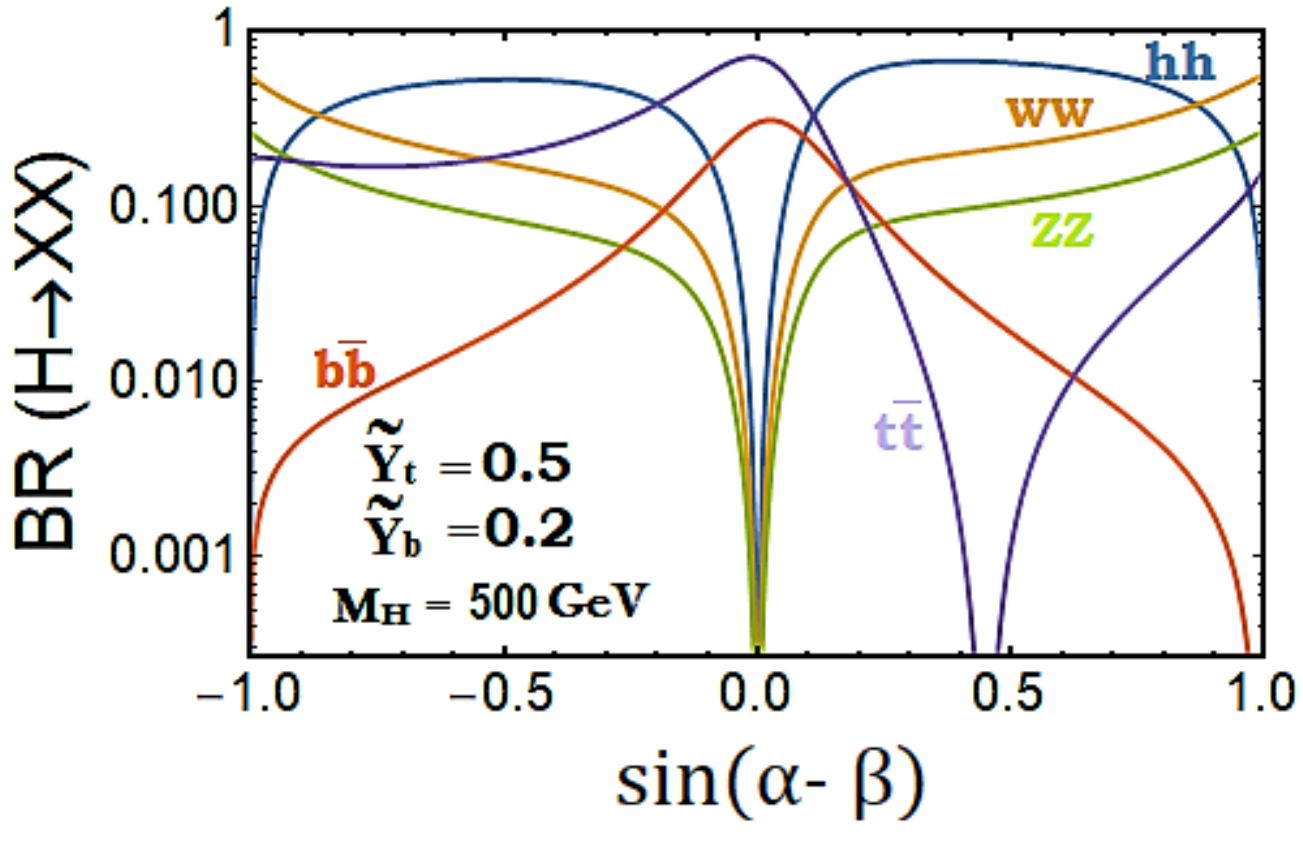}
 $$
 $$
\includegraphics[height=7cm,width=0.45\textwidth]{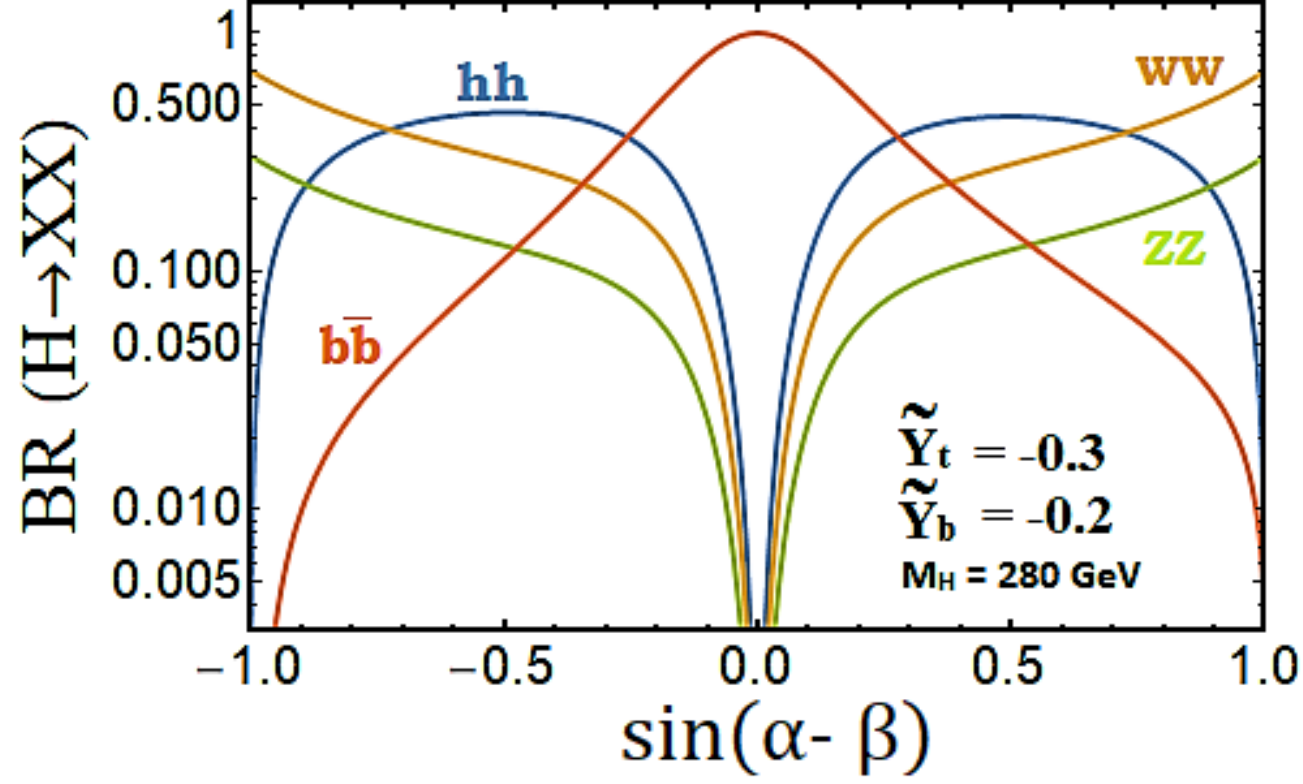}\hspace{0.3 in}
\includegraphics[height=7cm,width=0.45\textwidth]{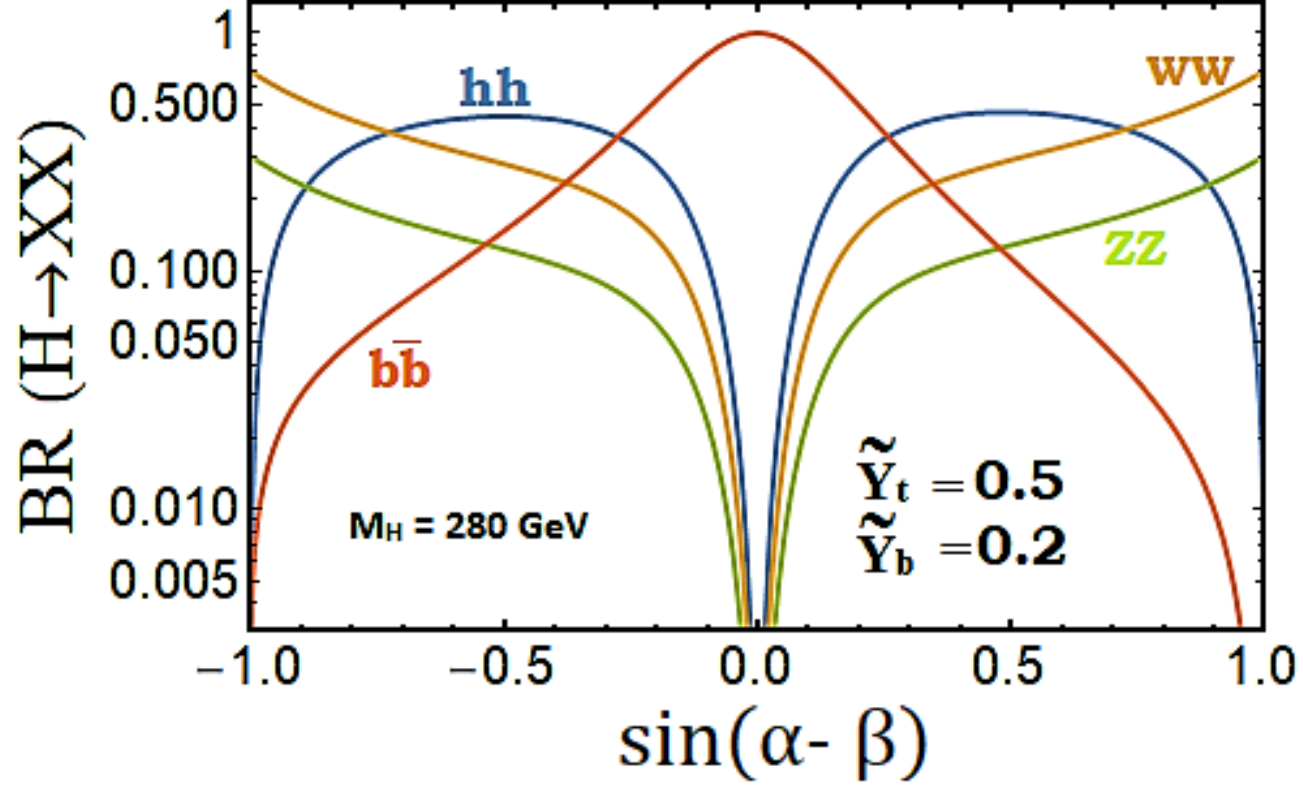}
 $$
 \caption{Branching ratios of $H$  as a function of mixing parameter $\sin(\alpha-\beta)$.
 }
\label{BrH}
\end{figure}

\begin{figure}[htb!]
\centering
 \includegraphics[height=6.6cm]{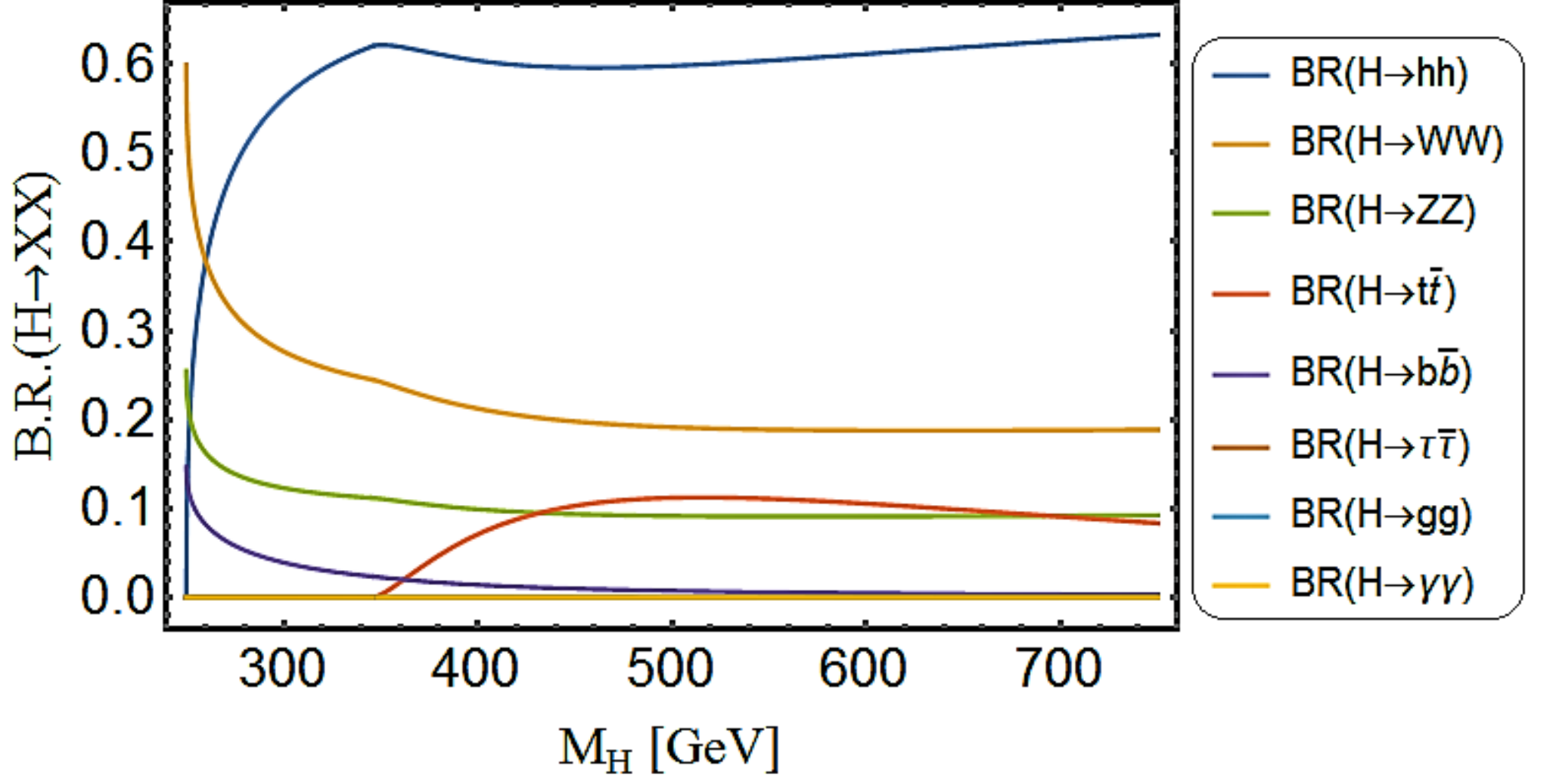}
 \caption{Branching ratios of $H$ as a function of its mass M$_{H}$.}
\label{2}
\end{figure}

Here we turn to the production and decay of the heavy neutral scalar, $H$, in the context of the LHC experiments. The most relevant interaction, in the context of collider phenomenology of $H$, is the $\Lamf$ term in the scalar potential involving $Hhh$ coupling which gives rise to tree-level decay of $H$ into Higgs pairs, $ H \to hh$. There are also contributions to $W^\pm W^\mp$ and $ZZ$ decay modes of $H$ arising from its mixing with the SM Higgs boson.
$H$ has a tree level Yukawa couplings with the quarks and leptons.  As a consequence, $H$ is also allowed to decay into a pair of top quarks, bottom quarks, tau leptons and even into a pair of gluons. The $HW^\pm W^\mp, Ht\bar{t}, HH^+H^-$ couplings allow $H$ to decay into $\gamma \gamma$, $Z\gamma$, $ZZ$, $W^\pm W^\mp$, $gg$, $hh$ and $Zh$ pairs at the one loop level. However, all the loop induced decays are too suppressed compared to tree level $hh, W^\pm W^\mp, ZZ, t\bar{t}, b\bar{b}, \tau^+ \tau^-$. $H$  has loop induced coupling to a pair of gluons due to its coupling to $t\bar{t}H$ at tree level. As a result $H$ will be dominantly produced via gluon gluon fusion at the LHC. Branching ratios to different decay modes of $H$ as a function of the  mixing term $\sin(\alpha-\beta) $  is shown in the Fig. \ref{BrH} for two different masses of $H$ ($M_H=$ 500 and 280 GeV). As we can see, below top quark mass threshold ($M_H <2 M_t$), heavy Higgs $H$ mostly decays to $hh, W^+W^-, ZZ, b\bar{b}$. On the other hand, when $M_H > 2 M_t$, one of the dominant decay modes become $H \to t\bar{t}.$  However, we see that di-Higgs mode is the most dominant decay in all scenarios. In Fig. \ref{2}, we have shown the branching ratios to different decay modes of $H$ as a function of the mass $M_{H}$. Here we fix the value of three Yukawas and mixing  as: $\tilde{Y_t} = 1.25, \tilde{Y_b} = -0.09$, $\tilde{Y_\tau} = 10^{-3}$, and $\sin(\alpha-\beta) = 0.5$) to be consistent with the Fig. \ref{1} constraints from properties of $h$. Throughout our analysis, the charged Higgs boson ($H^\pm$) mass is kept almost degenerate with $H$ and well above 300 GeV to be consistent with the current experimental constraints \cite{charged1}. We also keep the mass splitting between charged and neutral member of the $H_2$ to be below 100 GeV \cite{report}.

\subsubsection{\textbf{Di-Higgs boson production}}


\begin{figure}[htb!]
$$
 \includegraphics[height=7.2cm]{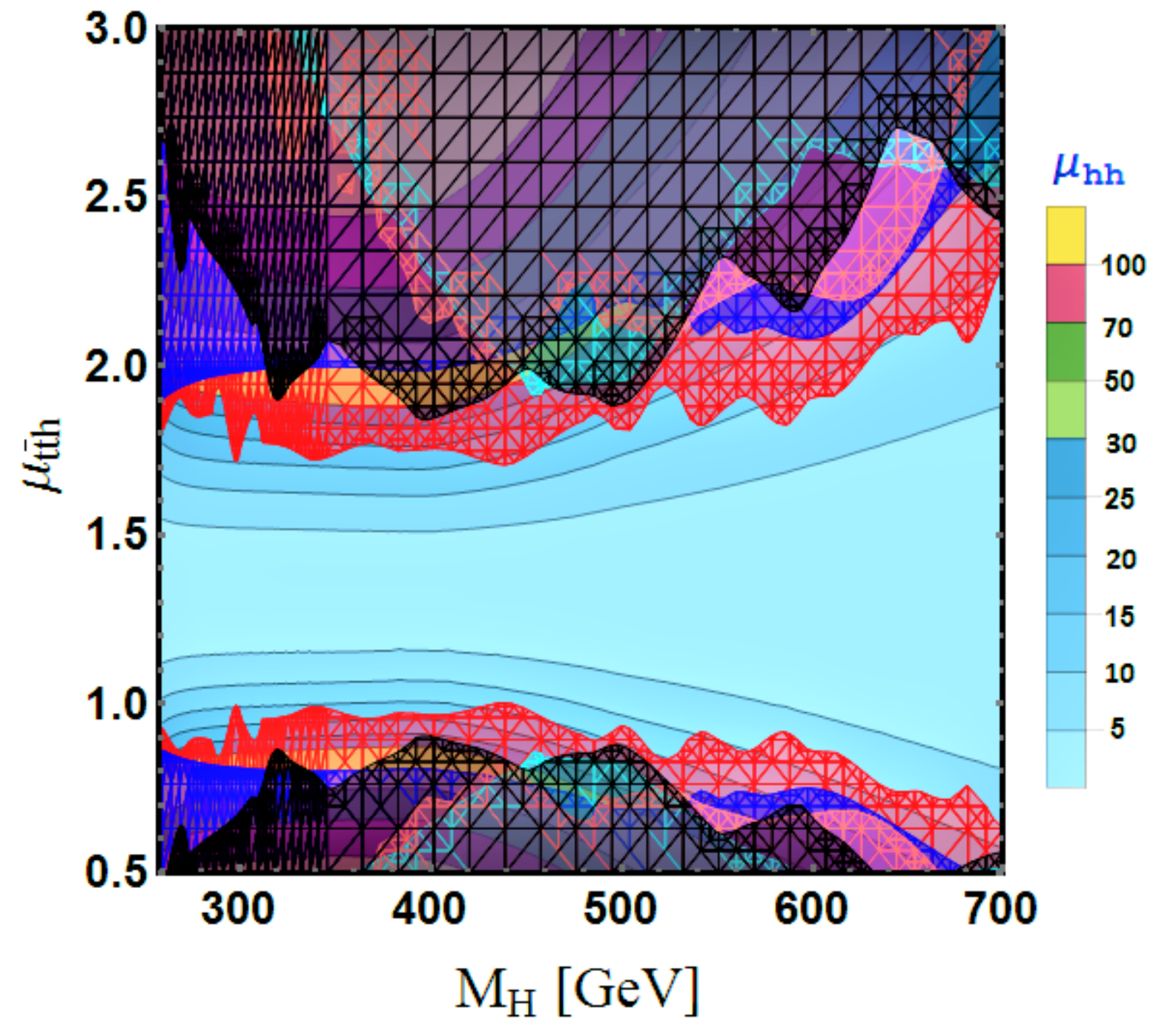}
 \includegraphics[height=7.2cm]{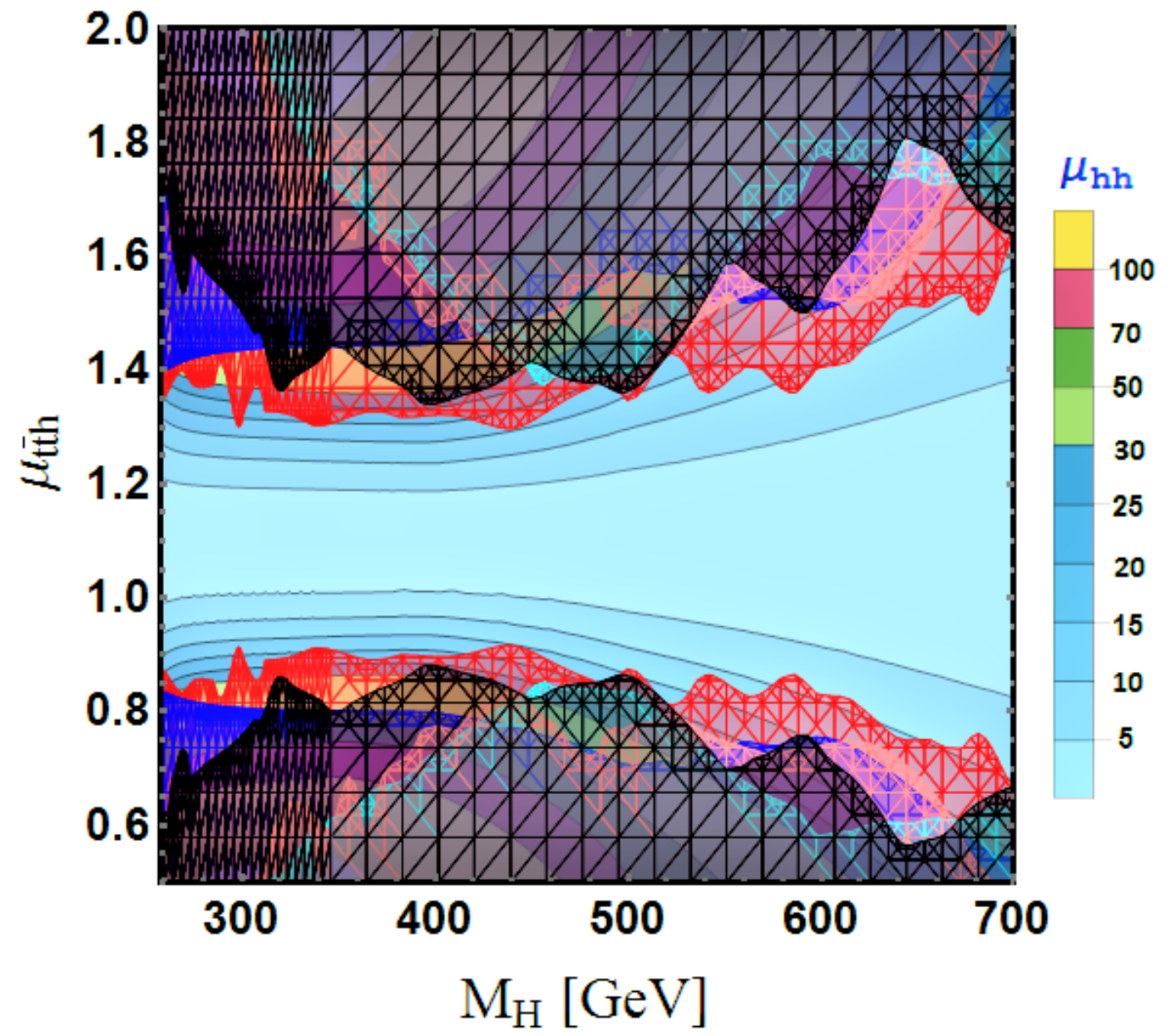}
 $$
 $$
 \includegraphics[height=7.2cm]{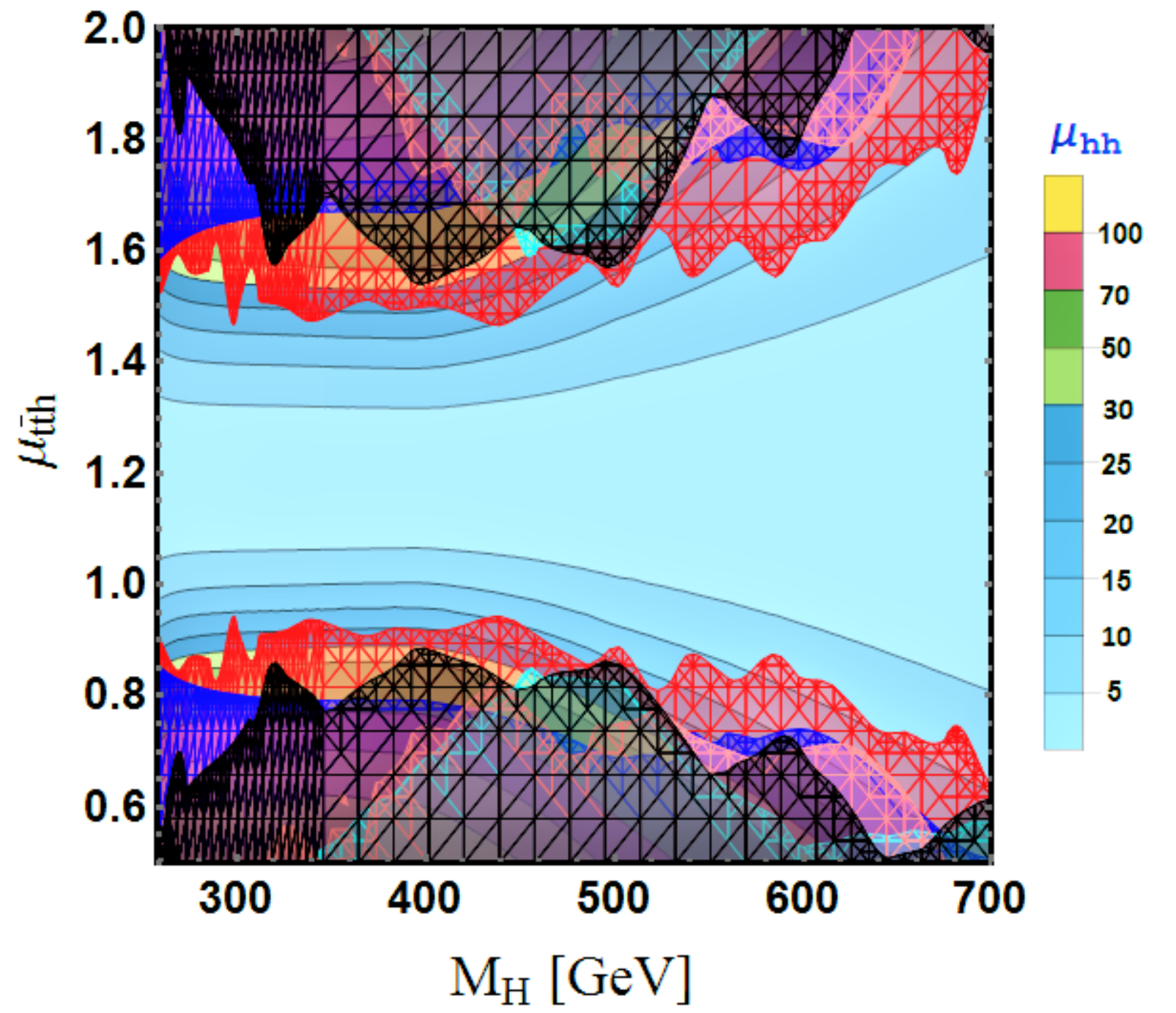}
 \includegraphics[height=7.2cm]{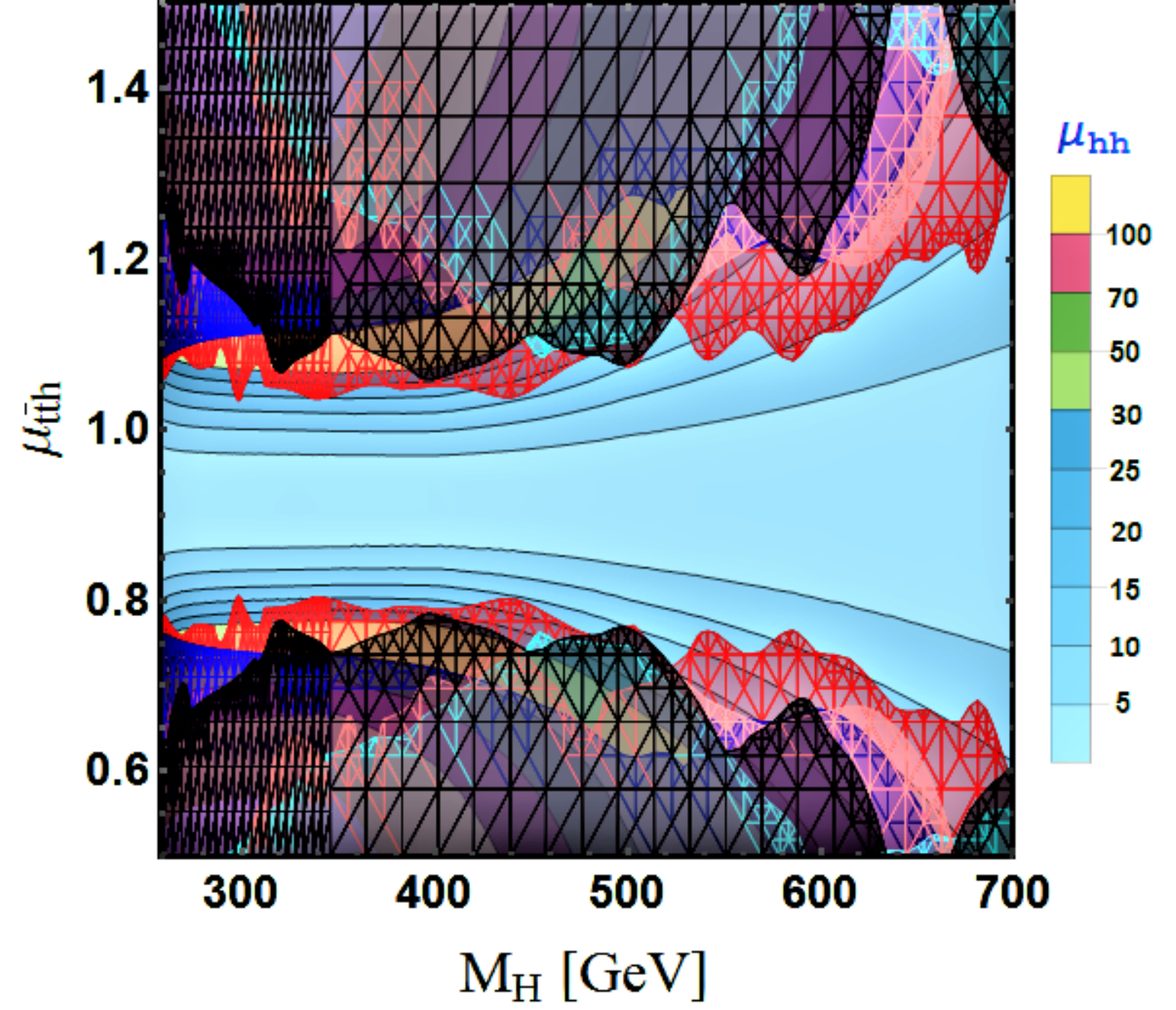}
 $$
 \caption{ Contour plot of $\mu_{hh}$ in $\mu_{t\bar{t}h}$-M$_H$ plane. The scaling of $\mu_{hh}$ is shown on right side of the each figure.  Black, pink and cyan colored meshed zones are excluded parameter space from current di-Higgs limit looking at different final states $b\bar{b}\gamma\gamma, b\bar{b}b\bar{b}$ and  $b\bar{b}\tau^{+} \tau^{-}$ respectively; red and blue meshed zone is the excluded parameter space from the resonant $ZZ$ and $W^+W^-$ production constraints. We have used a typical set of parameters ($\sin(\alpha-\beta) = 0.5, \tilde{Y_b} = -0.09, \tilde{Y_\tau} = 10^{-3}$) for top left; ($\sin(\alpha-\beta) = 0.3, \tilde{Y_b} = -0.09, \tilde{Y_\tau} = 10^{-3}$) for top right; ($\sin(\alpha-\beta) = 0.4, \tilde{Y_b} = 0.02, \tilde{Y_\tau} = 10^{-3}$) for bottom left and ($\sin(\alpha-\beta) = -0.2, \tilde{Y_b} = 0.04, \tilde{Y_\tau} = 10^{-3}$) for bottom right. }
\label{3}
\end{figure}


The di-Higgs production has drawn a lot of attentions \cite{lewis,lithh0,lithh1,lithh3,lithh4,lithh5,lithh6,lithh7,lithh8,lithh9} since it is the golden channel to directly probe the triple Higgs-boson self-interaction within the SM, and therefore, tests the EW symmetry breaking mechanism. In the SM, the 125 GeV Higgs boson is pair produced through a triangle and a box diagram. The di-Higgs boson production rate in the SM is very small mainly due the smallness of the strength of individual diagrams, and also due to the negative interference between the triangle and box diagrams. At the $13{~\rm TeV}$ LHC, the $hh$ production cross section is about $33.5~{\rm fb}$, which is almost 1295 times weaker than the single Higgs production and it cannot be measured at current luminosity  owing to the small branching fraction of $h$ decaying into $ZZ^*$ and $W W^*$ and the large SM background. In the SM, di-Higgs production is thoroughly studied in Ref. \cite{higgswg}. 

Within the 2HDM framework, extra contribution to di-Higgs production arises from the decay of $H$ after being resonantly produced mainly via gluon gluon fusion process at the LHC. Also, change in the $t\bar{t}h$ coupling compared to the SM could give a significant deviation on di-Higgs production cross-section. These effects could significantly enhance the di-Higgs production rate and make it testable at the LHC. Therefore, it is important to analyze how largethe cross section can be, consistent with SM Higgs boson properties. The most promising signal for di-Higgs search is the $b\bar{b}\gamma\gamma$, since it benefits from the large branching ratio of  $h\rightarrow b\bar{b}$ decay ($\sim$ 58$\%$) and also due to the clean diphoton signal (due to high $m_{\gamma\gamma}$ resolution) on top of the smooth continuum diphoton SM background. On the other hand, due to the higher branching ratios of the SM Higgs boson decays to $b\bar{b}$ and $\tau^{+} \tau^-$, the $b\bar{b}b \bar{b}$ and $b\bar{b}\tau^{+} \tau^-$ channels have larger signals, but they suffer from large QCD background.

The signal strength relative to the SM expectation $\mu_{hh}$ is defined as following: 
\begin{equation}
\mu_{hh}=\frac{\sigma(pp\to hh)_{2HDM}}{\sigma(pp\to hh)_{SM}}=\frac{\left[\sigma^{Res}(pp\to hh)+\sigma^{Non-Res}(pp\to hh)\right]_{2HDM}}{\sigma(pp\to hh)_{SM}},
\end{equation}
where, 
\beqa
\sigma^{Res}(pp\to hh) &=& \sigma(pp\to H) \times Br (H \to hh)
\\ \sigma(pp\to H) &=& \sigma (pp\to h(M_H)) \times \left(-\sin(\alpha-\beta) + \frac{v\tilde{y_t}}{\sqrt{2}m_t}\cos{(\alpha-\beta)}  \right)^2 
\eeqa
In  the 2HDM, di-Higgs ($hh$) production will occur both resonantly and non-resonantly. Non-resonant di-Higgs ($hh$) production will be largely affected by the deviation of $t\bar{t}h$ and $hhh$ couplings, whereas the resonant production of $hh$, absent in SM, will occur significantly due to the large $Hhh$ coupling which exists at the tree level. In Fig.~\ref{3}, we show the correlation between $t\bar{t}h$  enhancement and $hh$ production enhancement when compared to the SM. Here we implement the current experimental limits \cite{econf, mor, CMShh, CMShh2, ATLAShh} on di-Higgs production  which  is indicated by the black, pink and cyan colored meshed zone looking at different final states $b\bar{b}\gamma\gamma, b\bar{b}b\bar{b}$ and  $b\bar{b}\tau^{+} \tau^{-}$ respectively. Although, we calculate all the current experimental limits from $hh, t\bar{t}, W^+W^-$ and $ZZ$ resonant production, the most stringent limit is occurs from resonant di-Higgs production limit \cite{econf, mor, CMShh, ATLAShh} and from resonant $ZZ$ and $W^+W^-$ production limits \cite{zz}. We choose four different set of parameters \{$\sin(\alpha-\beta) = 0.5, \tilde{Y_b} = -0.09, \tilde{Y_\tau} = 10^{-3} \rbrace$ for top left, $\lbrace \sin(\alpha-\beta) = 0.3, \tilde{Y_b} = -0.09, \tilde{Y_\tau} = 10^{-3} \rbrace$ for top right, \{$\sin(\alpha-\beta) = 0.4, \tilde{Y_b} = 0.02, \tilde{Y_\tau} = 10^{-3}$\} for bottom left and \{$\sin(\alpha-\beta) = -0.2, \tilde{Y_b} = 0.04, \tilde{Y_\tau} = 10^{-3}$\} for bottom right].  Fig.~\ref{3} clearly indicates that we can satisfy enhanced $\mu_{t\bar{t}h}$ value of upto 1.9 and the $\mu_{hh}$ value as big as 25 depending upon the heavier Higgs mass. This is significant enough to observe the $hh$ pair production in the upcoming run or at the high luminosity LHC. These  plots signify that if $t\bar{t}h$ signal strength remains higher than the SM, the 2HDM is a great platform to explain it with a smoking gun signal of di-Higgs production at the LHC. In a recent study\cite{tg}, it is shown that the SM like di-Higgs production with a cross section of 33.45 fb can be observed with $3.6 \sigma$ significance \cite{tg}, if LHC luminosity is upgraded to 3 $ab^{-1}$. IN our 2HDM scenario, the enhanced di-Higgs production rate is so large that it can potentially be observed with the 100 fb$^{-1}$ LHC luminosity, which is close to the data set currently analyzed. The newly proposed  future hadron-hadron circular collider (FCC-hh) or  super proton-proton collider (SppC), designed to operate at 100~TeV centre of mass energy, can easily probe most of the parameter space in 2HDM through the $hh$ pair production~\cite{ATL-PHYS-PUB-2014-019,Gomez-Ceballos:2013zzn,CEPC-SPPCStudyGroup:2015csa}.


\begin{table}[htb!]
\centering
\setlength{\arrayrulewidth}{0.7mm}
\setlength{\tabcolsep}{17pt}
\renewcommand{\arraystretch}{1.8}
\adjustbox{max height=\dimexpr\textheight-2.0cm\relax,
           max width=\dimexpr\textwidth-0.0cm\relax}{
\begin{tabular}{ ||p{1.8cm}|p{0.5cm}|p{0.5cm}|p{0.4cm}|p{1.7cm}|p{1.5cm}|p{2.7cm}|p{0.6cm}|p{0.6cm}||  }

\hline
\textbf{Benchmark Points} & $\tilde{\textbf{Y}}_\textbf{t}$ & $\tilde{\textbf{Y}}_\textbf{b}$ & $\tilde{\textbf{Y}}_{\boldsymbol{\tau}}$ & $\boldsymbol{\sin(\alpha-\beta)}$ & $\boldsymbol{M_H   [GeV]}$ & \textbf{Scaling Factors} & $\boldsymbol{\mu_{t\bar{t}h}}$ & $\boldsymbol{\mu_{hh}}$\\
\hline
\textbf{BP1} & $+1.01$ & $-0.10$ & $10^{-3}$ &  $+0.50$ & 500  & $\kappa_{W} = 0.866$ \newline $\kappa_{Z} = 0.866$ \newline $\kappa_{t} = 1.374 $ \newline $\kappa_{b} = -1.001 $ \newline $\kappa_{\tau} = 0.915$ \newline $\kappa_{\gamma\gamma} = 0.723$ \newline $\kappa_{Z \gamma} = 0.778$  & 1.89 & 15\\
\hline
\textbf{BP2} & $-1.0$ & $+0.01$ & $10^{-3}$ & $-0.10$ & 600 & $\kappa_{W} = 0.995$ \newline $\kappa_{Z} = 0.995$ \newline $\kappa_{t} = 1.096 $ \newline $\kappa_{b} = 0.958 $ \newline $\kappa_{\tau} = 0.985$ \newline $\kappa_{\gamma\gamma} = 0.966$ \newline $\kappa_{Z \gamma} = 0.976$  & 1.2 & 10\\
\hline
\textbf{BP3} & $1.25$ & $+0.05$ & $10^{-3}$ & $-0.20$ & 680 & $\kappa_{W} = 0.980$ \newline $\kappa_{Z} = 0.980$ \newline $\kappa_{t} = 0.728 $ \newline $\kappa_{b} = 0.61 $ \newline $\kappa_{\tau} = 0.960$ \newline $\kappa_{\gamma\gamma} = 1.05$ \newline $\kappa_{Z \gamma} = 1.08$  & 0.53 & 11\\
\hline
\end{tabular}
}
\caption{Sample points on parameter space and corresponding $\mu_{hh}$ and $\mu_{t\bar{t}h}$.}

\label{table1}

\end{table}

 A few benchmark points within the 2HDM and the corresponding $\mu_{hh}$ and $\mu_{t\bar{t}h}$ are summarized in Table \ref{table1}. Let us focus on one of the benchmark points, say BP1 in detail. Since $\kappa_{t}$ is enhanced by a factor $\sim 37 \%$, resonant SM Higgs boson production will also be enhanced in gluon gluon fusion. But, due to tiny enhancement in the $hb\bar{b}$ coupling and a decrease in the effective couplings $h\gamma\gamma, hWW, hZZ$ and $h \tau \bar{\tau}$, the branching ratios for the decay modes $h \to \gamma\gamma, WW, ZZ, \tau \bar{\tau}$ will be suppressed to $\sim 52 \%, 75 \%, 75\%, 84\%$ respectively.  Overall, production times branching ratio will get adjusted within the signal strength constraints at 95$\%$ confidence level. 
 
 Since the effective $hZZ$ or $hWW$ couplings are suppressed in these scenario via mixing, it may lead to tension in $VBF$, $Wh$ and $Zh$ production and subsequent decay of $h$ to $WW$ or $ZZ$, as the signal strength will be suppressed naively by a factor of $\cos^4{(\alpha-\beta)}$ due to mixing.  Although, there is a huge uncertainty in these measurements accrording to the updated results of 13 TeV 36 $fb^{-1}$ data, the central value prefers suppressed signal strength as low as 0.05 \cite{zrun21}.  Such a suppression, if needed, can be  achieved in our scenario compared to the SM.  The suppression factor is 0.56 in our case for BP1. Hence, our allowed parameter space can satisfy all the experimental constraints \cite{econf} at $2 \sigma$ level and simultaneously can lead to enhanced di-Higgs and $t\bar{t}h$ production rates as well.


\begin{figure}[htb!]
$$
 \includegraphics[height=10.0cm]{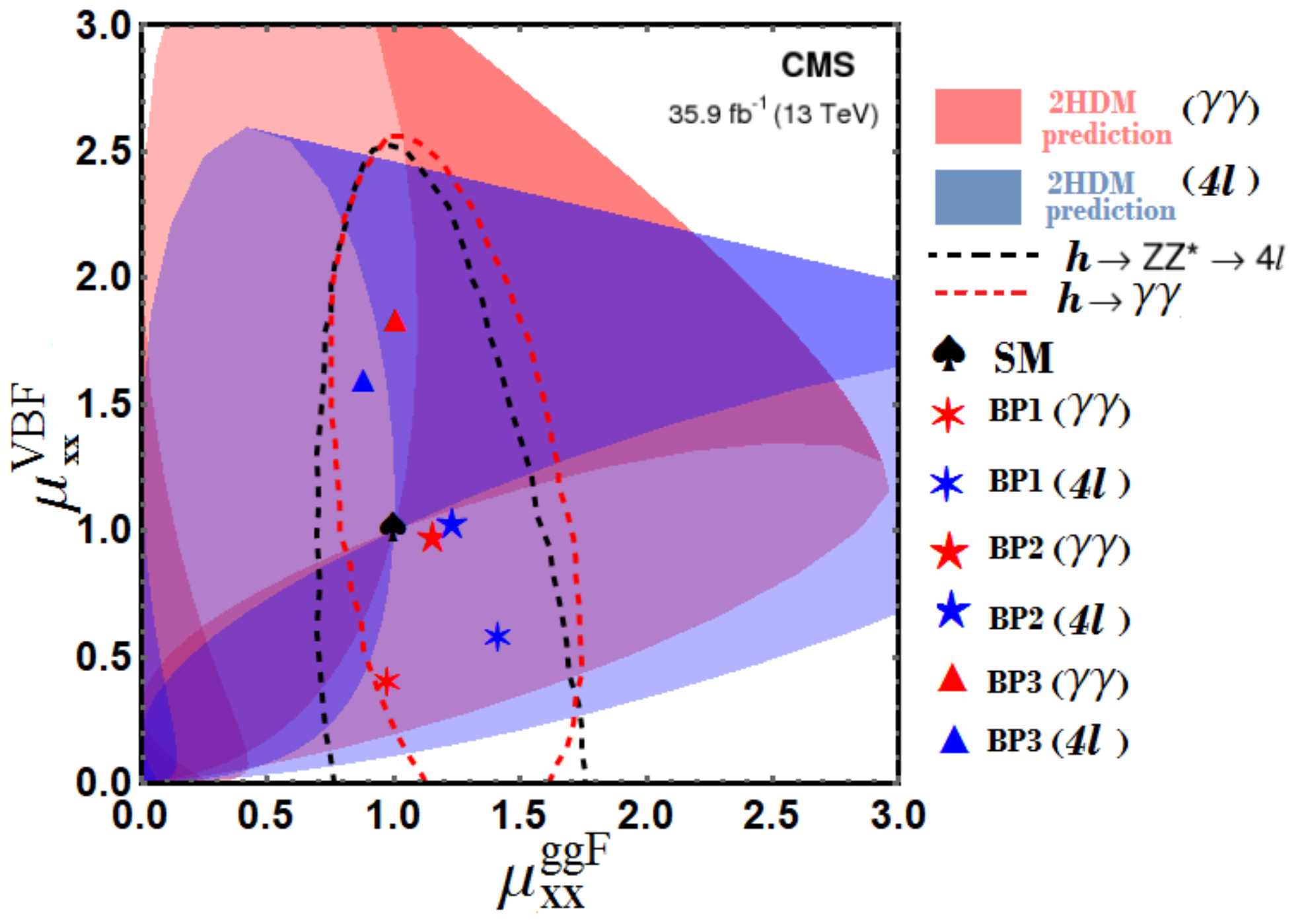}
 $$
 \caption{: The two-dimensional best-fit of the signal strengths for ggF and VBF production modes compared to the SM expectations (black spade) looking at $h \to \gamma \gamma$ \cite{gammarun21} and $h \to ZZ^{\star}\to 4l$ \cite{zrun21} channels.  The dashed red and black line represent the 2$\sigma$ standard deviation confidence region for  $h \to \gamma \gamma$ \cite{gammarun21} and $h \to ZZ^{\star}\to 4l$ \cite{zrun21} channels respectively. Red and blue shaded regions are 2HDM predicted region. Three benchmark points (BP) are also shown in this contour plot.  }
\label{new}
\end{figure}


The two-dimensional best-fit for the signal strengths for gluon gluon fusion (ggF) and vector boson fusion (VBF) production modes compared to the SM expectations (black spade) looking at $h \to \gamma \gamma$ \cite{gammarun21} and $h \to ZZ^{\star}\to 4l$ \cite{zrun21} channels are shown in Fig. \ref{new} using 36 fb$^{-1}$ data of 13 TeV LHC. Uncertainty in these two channels are least compared to other channels and hence impose the most stringent limits. ggF is the most dominant and VBF is the second dominant production mode for single $h$ production. In the SM, the ratio of resonant single Higgs boson ($h$) production in VBF production to the ggF production is $\sim$ 0.085. In the 2HDM this ratio can  deviate largely compared to the SM prediction which is reflected in the Fig. \ref{new}. For three of the benchmark points listed, this ratio becomes 0.034,  0.07 and 0.154 respectively. As the branching ratios also deviate from the SM simultaneously, these points lie within the 2$\sigma$  confidence region for  the two-dimensional best-fit of the signal strengths for ggF and VBF production modes.

 In two Higgs doublet model with additional discrete $Z_2$ symmetry, there will be resonant di-Higgs production which has been extensively studied in \cite{hhz2}. The resonant di-Higgs production rate 
 is much larger in our framework compared to the type-II 2HDM. 
 It is easy to  understand this difference. 
 In the type-II 2HDM, resonant production cross section of $H$ is suppressed since $H$ has no direct coupling to top quark.  Such a coupling is induced proportional to the Higgs mixing angle $\sin(\alpha-\beta)$, which is strongly constrained from the properties of $h$.  In our case, $H$ has direct coupling to the top quark.

The viability of a scenarios where the sign of the $b$-quark coupling to $h$ is opposite to that of the Standard Model (SM), while other couplings are close to their SM value, has been   studied in Ref. \cite{wrongsignyuk} in the context of type-I and type-II 2HDM. 
Our analysis here includes such effects, as we allow both signs for $\tilde{Y}_b$.   Using scans over the full parameter space, subject to basic theoretical and experimental constraints as described previously, we found that a sign change in the down-quark Yukawa couplings can be accommodated in the context of the current LHC data set at 95$\%$ C.L. as shown in the benchmark points of Table \ref{table1}.  We have shown that such a scenario is consistent with all LHC observations.

\subsection{Pseudoscalar Higgs phenomenology}
   \begin{figure}[htb!]
$$
 \includegraphics[height=8cm,width=1.0\textwidth]{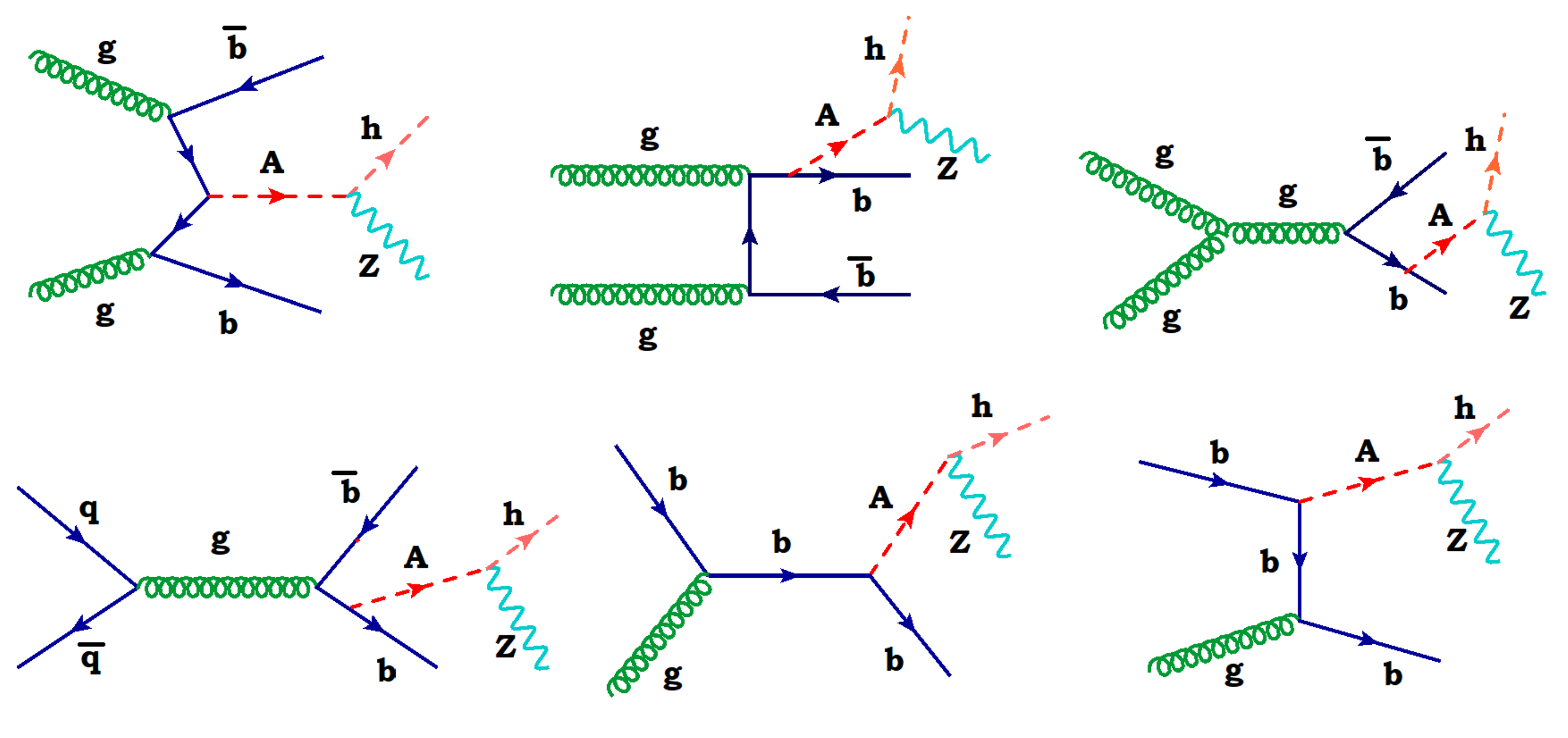}
 $$
 \caption{ Representative leading order Feynman diagrams for pseudoscalar $A$ production in association with b quarks and simulataneous decay to $Z$ and $h$ boson. The diagrams that can be obtained by crossing the initial state gluons, or radiating the Higgs off an antibottom quark are not shown.
 }
\label{A1}
\end{figure}
 
\begin{figure}[htbp]
$$
 \includegraphics[height=7cm,width=0.45\textwidth]{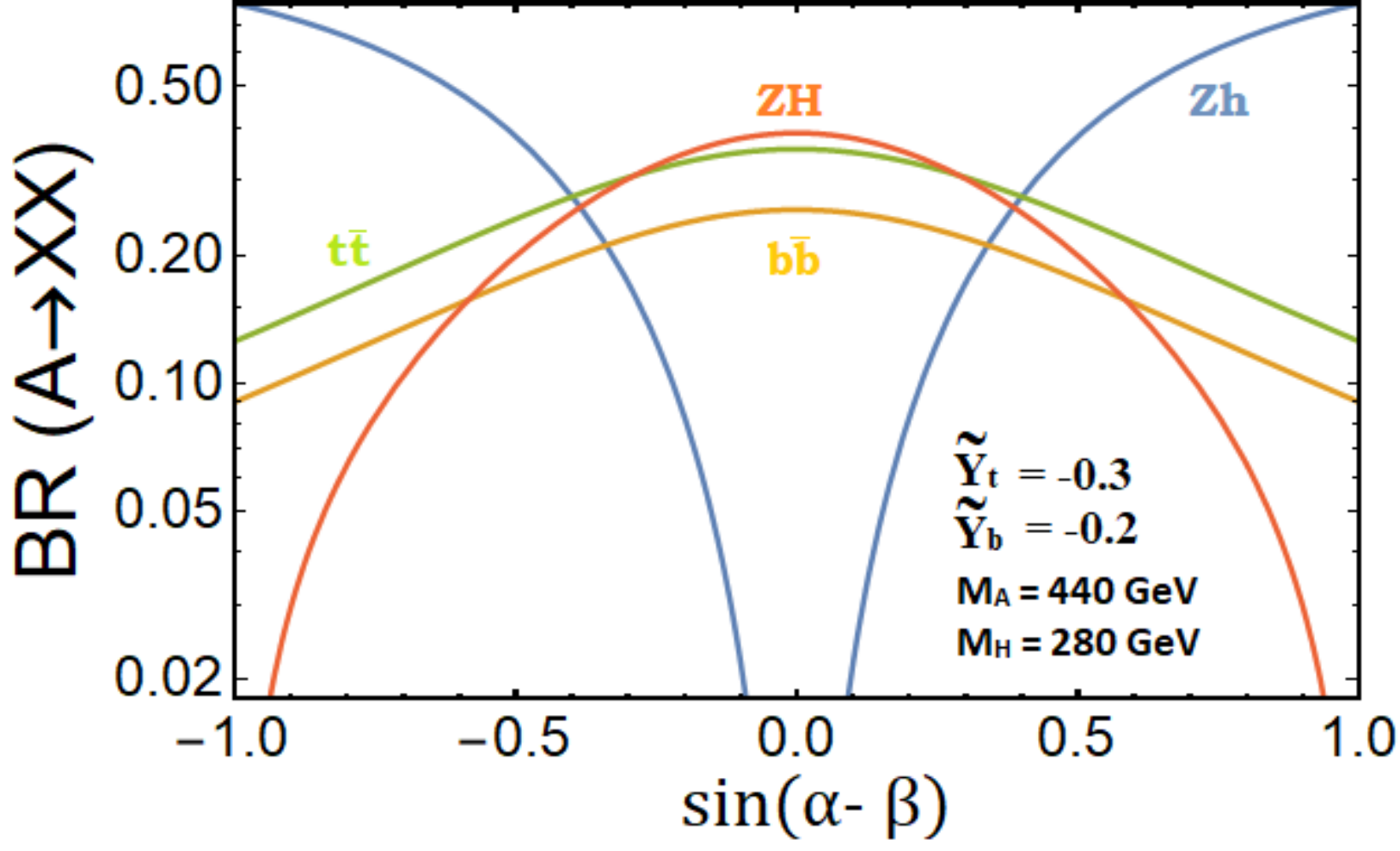} \hspace{0.3in}
  \includegraphics[height=7cm,width=0.45\textwidth]{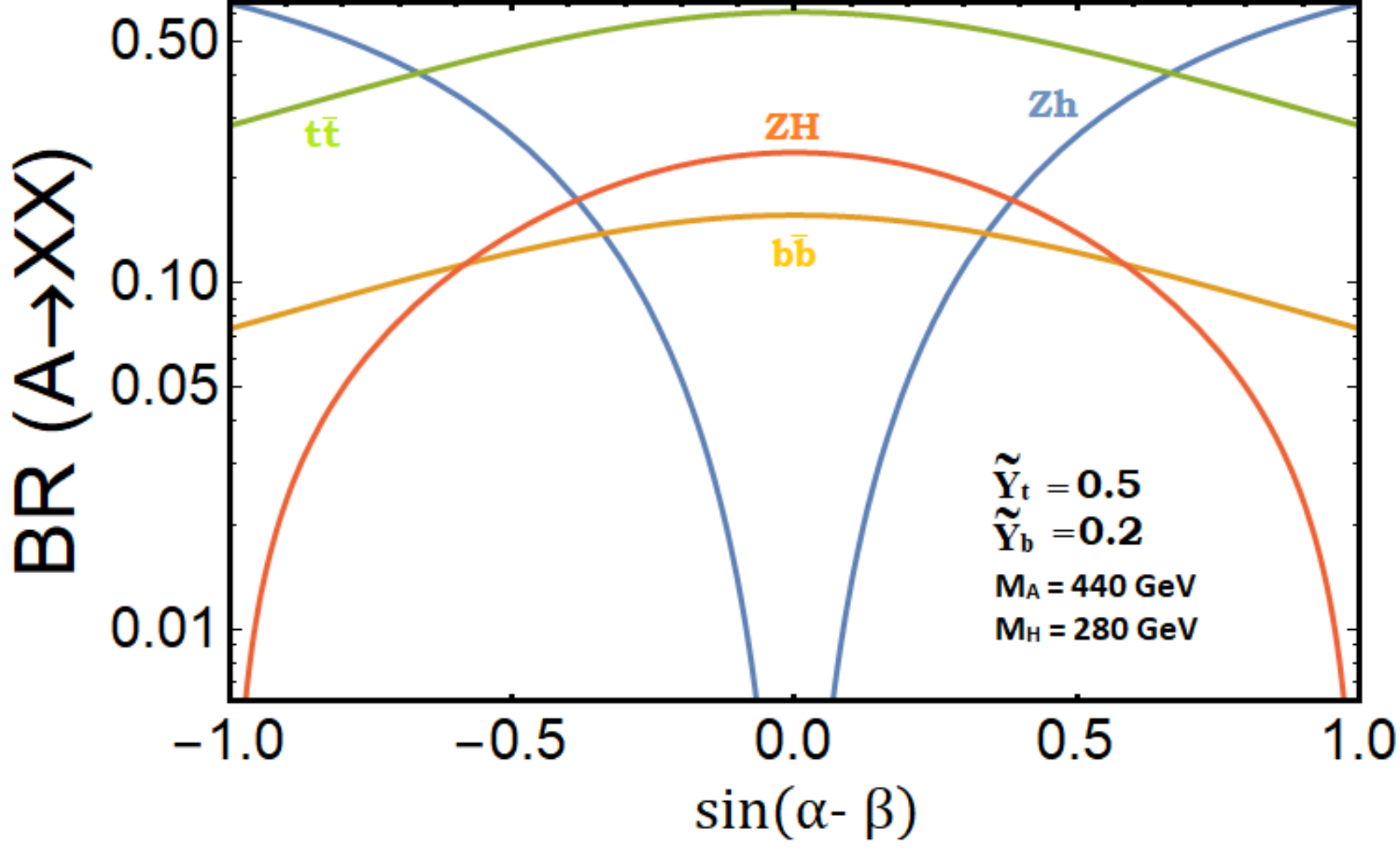}
 $$
 $$
\includegraphics[height=7cm,width=0.45\textwidth]{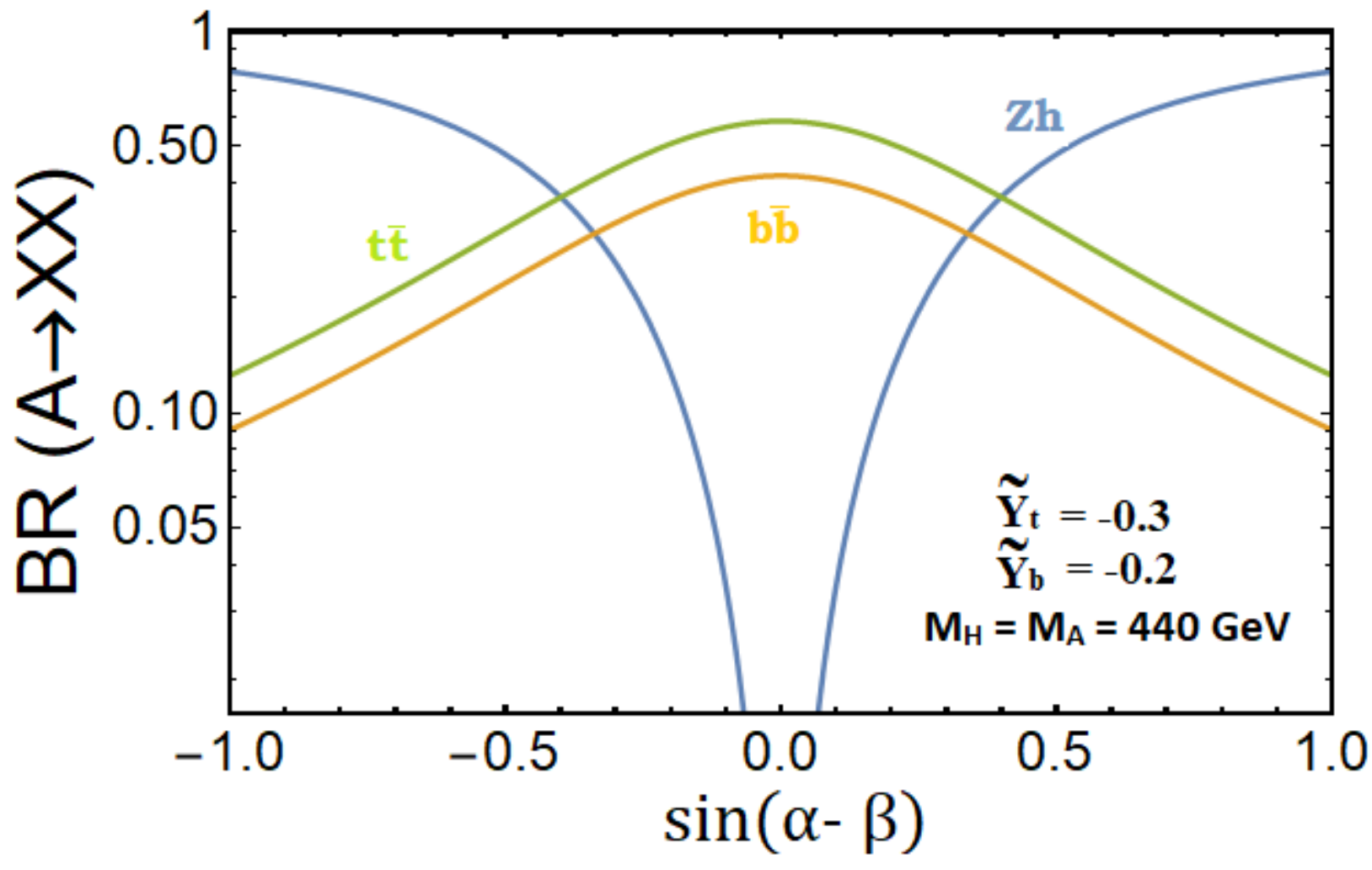} \hspace{0.3in}
\includegraphics[height=7cm,width=0.45\textwidth]{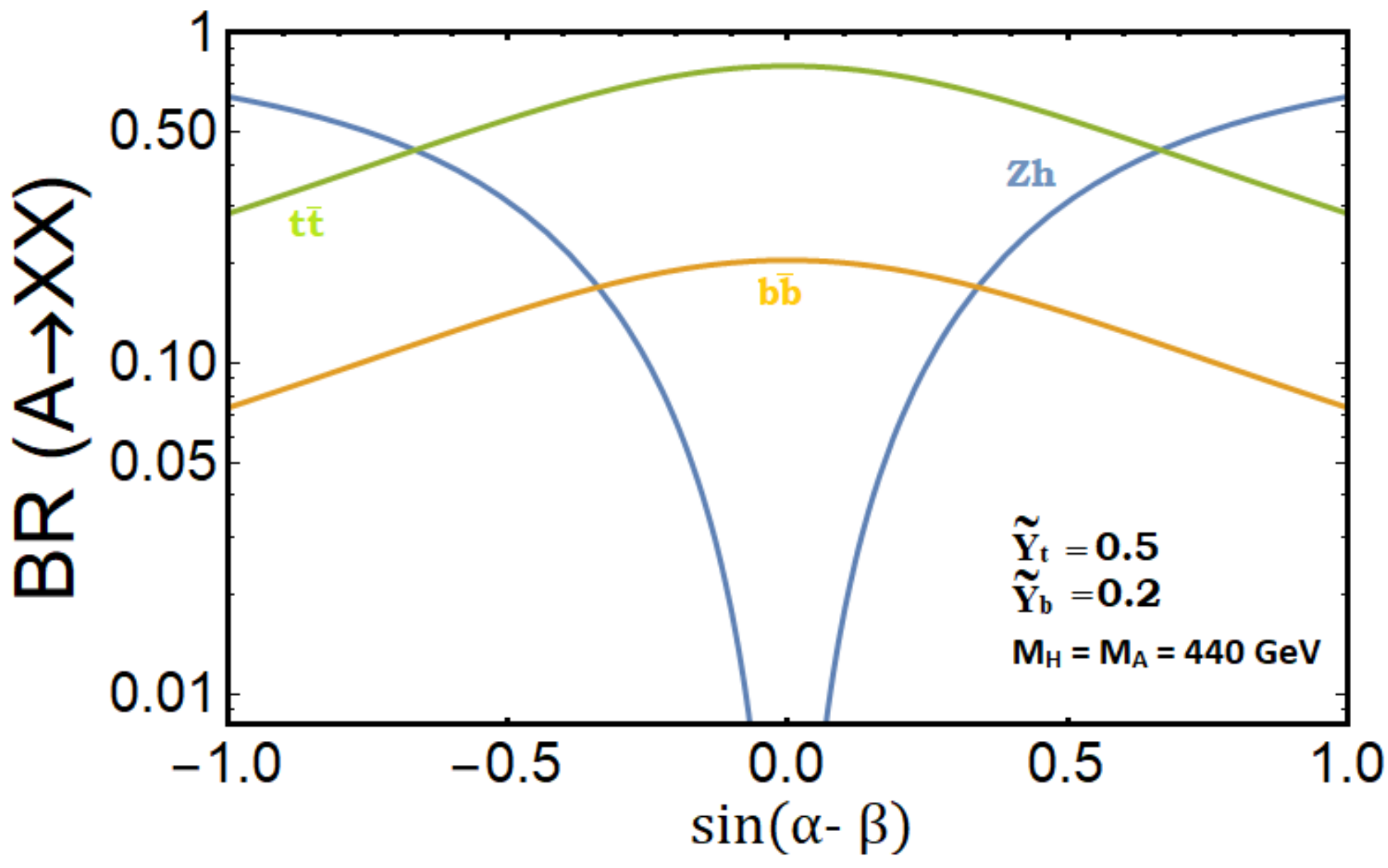}
 $$
 \caption{Branching ratios $A$ decaying into various modes as a function of $\sin(\alpha-\beta)$ for $M_A = 500$ GeV.
 }
\label{A2}
\end{figure}

 
 Now we turn to the production and decay of the heavy pseudoscalar-$A$ in the context of the LHC experiments. The most promising signal is the tree-level decay of $A$ into di-boson pairs ($Zh, ZH$). Although, the decay mode $A \to Zh$ arises due to the mixing ($\sin(\alpha-\beta)$) between heavy Higgs H and SM Higgs $h$, this leads to a smoking gun signal of the 2HDM in the channel $pp \to A \to Zh \to l^+ l^- b \bar{b}$. The pseudoscalar-$A$  has loop induced coupling to a pair of gluons due to its  $t\bar{t}A$ coupling at tree level. It will be then dominantly produced via gluon gluon fusion at the LHC. On the other hand, for sufficiently large bottom Yukawa coupling, $\tilde{Y}_b \sim 0.1$, another promising mode is the production of $A$ in association with two bottom quarks. Representative leading order Feynman diagrams for pseudoscalar $A$ production in association with $b$ quarks and subsequent decay to $Zh$  is shown in Fig. \ref{A1}. After being produced at the LHC, it will dominantly decay to $Zh, ZH, t\bar{t}, b\bar{b}$. The unique signals at the LHC will be resonant production of pseudoscalar $A$ and its subsequent decay to SM Higgs in association with $Z$ boson ($pp \to A \to Zh$) and di-Higgs production in association with $Z$ boson $pp \to A \to ZH \to Zhh$.  If the mass splitting between $H$ and $A$ is kept small, for large mixing, $A \to Zh$ will be the most promising mode and we will focus on this scenario from here on. In Fig. \ref{A2}, branching ratios of $A$ to different decay modes are shown.
 
 
\begin{figure}[htb!]
$$
 \includegraphics[height=7.2cm]{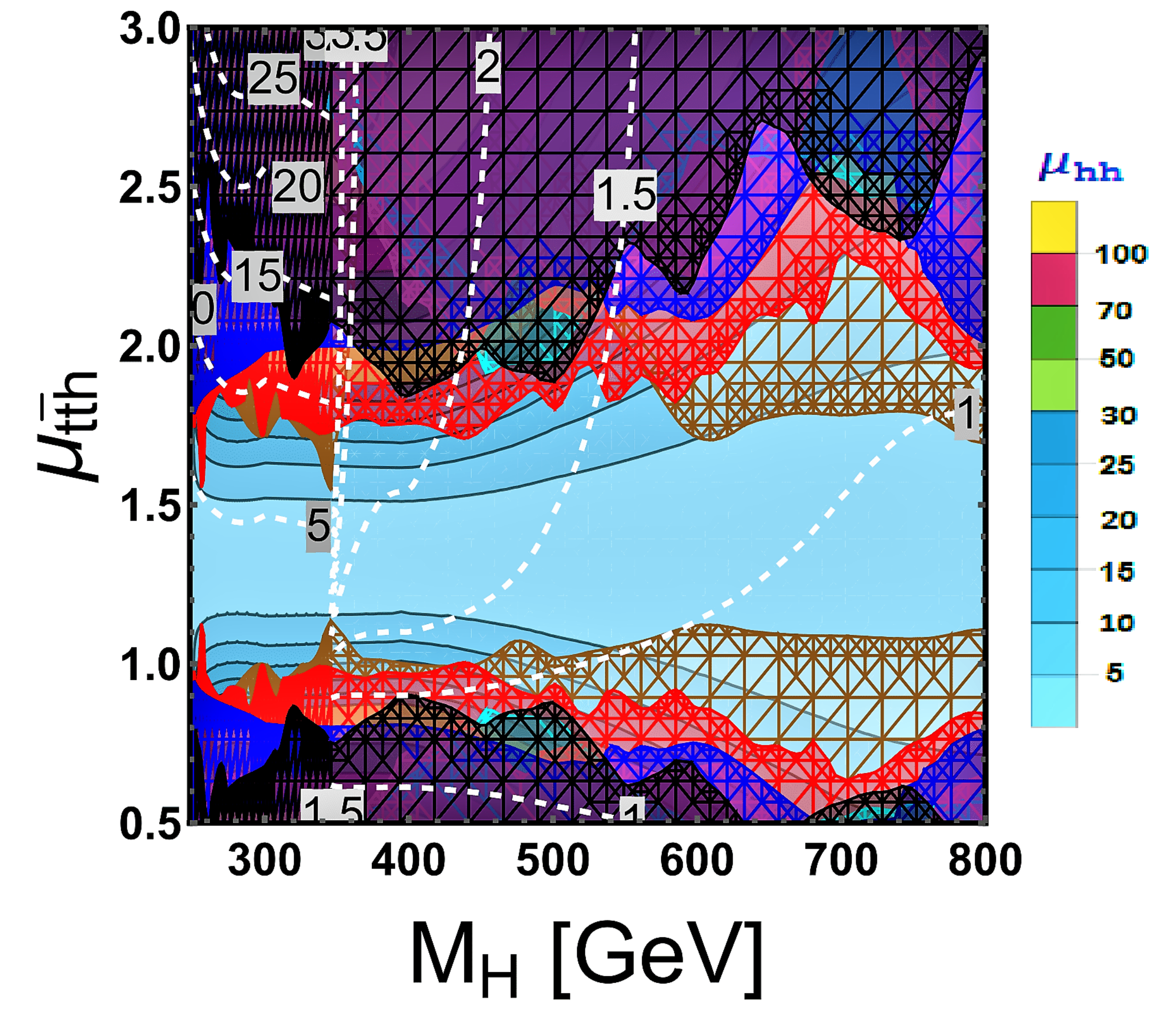}
 \includegraphics[height=7.2cm]{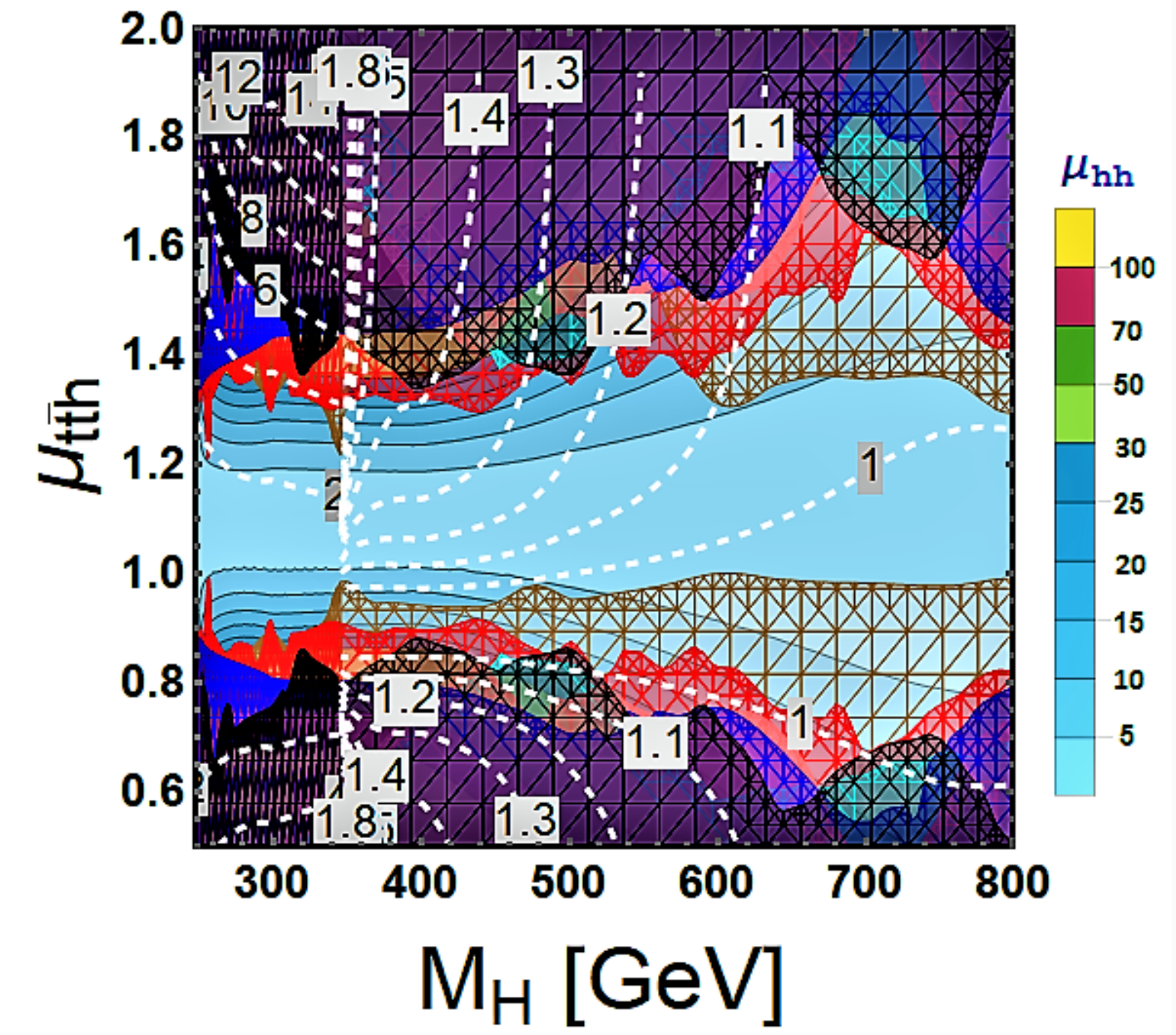}
 $$
 $$
 \includegraphics[height=7.2cm]{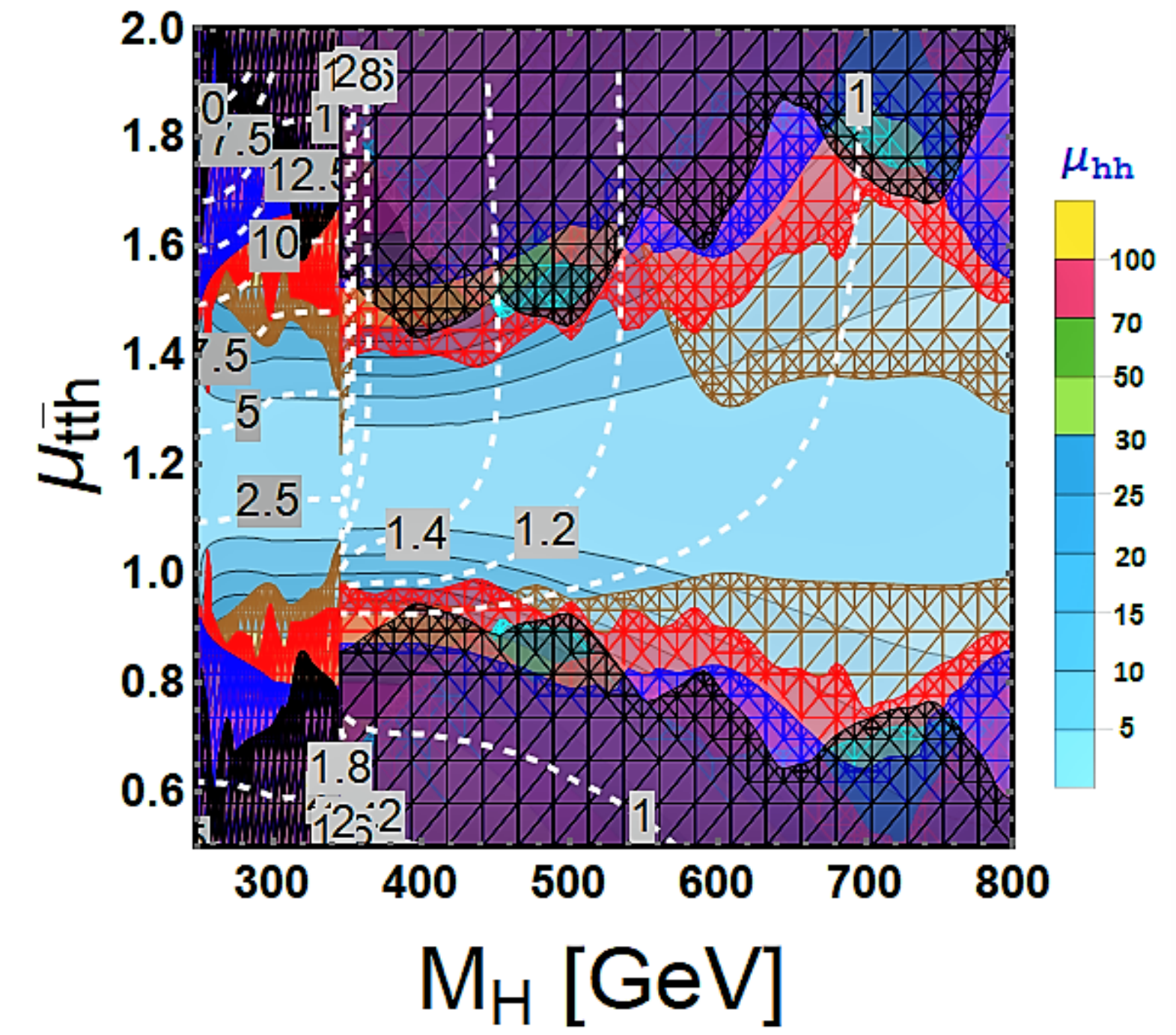}
 \includegraphics[height=7.2cm]{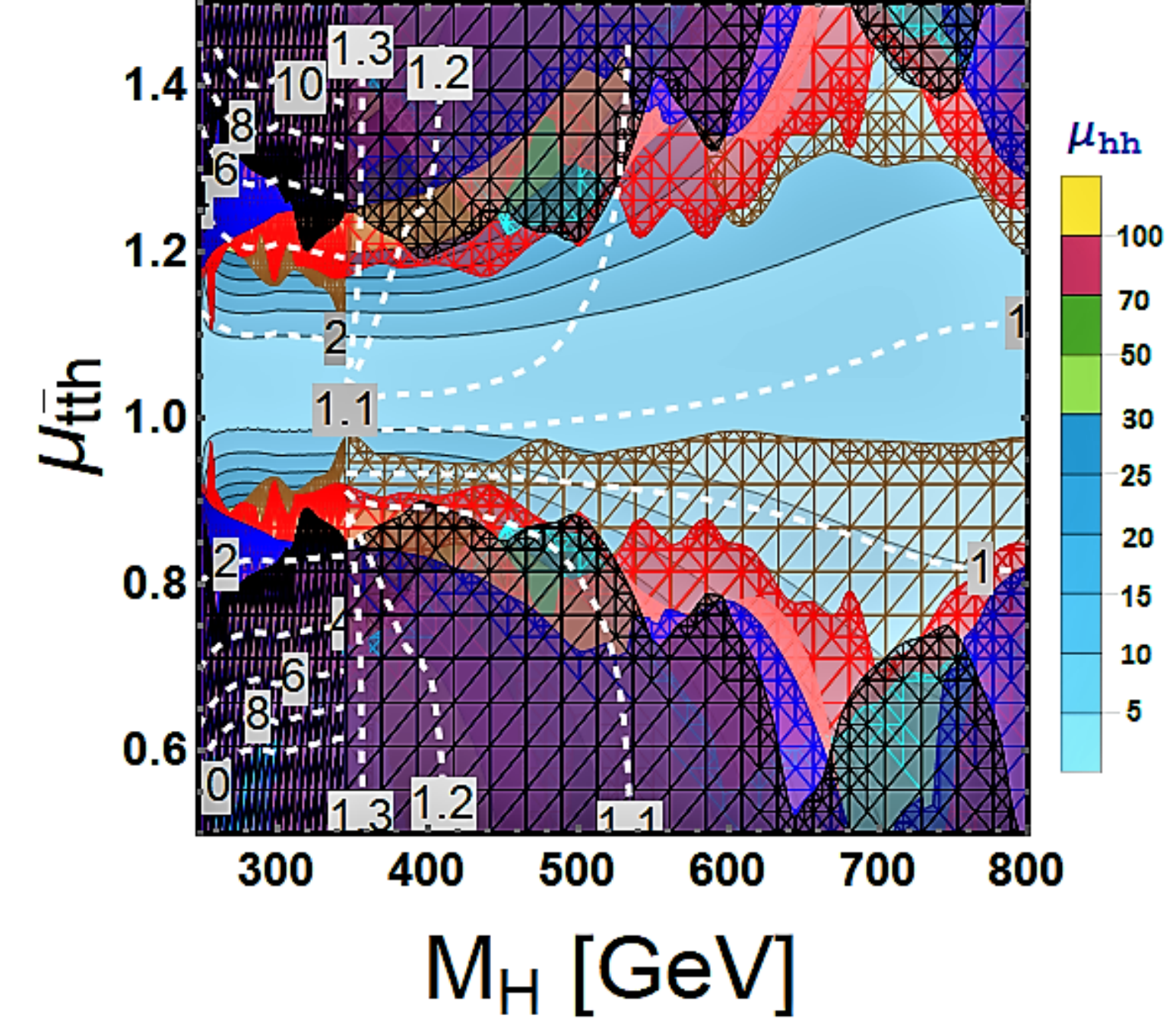}
 $$
 \caption{ Contour plot of $\mu_{Zh}$ and $\mu_{hh}$ in $\mu_{t\bar{t}h}$-M$_H$ plane. The scaling of $\mu_{hh}$ is shown on right side of the each figure, whereas the boxed numbers for the dashed contours indicate different values of $\mu_{Zh}$. Black, pink and cyan colored meshed zones are excluded parameter space from current di-Higgs limit looking at different final states $b\bar{b}\gamma\gamma, b\bar{b}b\bar{b}$ and  $b\bar{b}\tau^{+} \tau^{-}$ respectively; red, blue and brown meshed zone is the excluded parameter space from the resonant $ZZ$, $W^+W^-$ and $Zh$ production constraints. We have used a typical set of parameters ($\sin(\alpha-\beta) = 0.5, \tilde{Y_b} = -0.09, \tilde{Y_\tau} = 10^{-3}$) for top left; ($\sin(\alpha-\beta) = 0.3, \tilde{Y_b} = -0.09, \tilde{Y_\tau} = 10^{-3}$) for top right; ($\sin(\alpha-\beta) = 0.4, \tilde{Y_b} = 0.02, \tilde{Y_\tau} = 10^{-3}$) for bottom left and ($\sin(\alpha-\beta) = -0.2, \tilde{Y_b} = 0.04, \tilde{Y_\tau} = 10^{-3}$) for bottom right. }
\label{A3}
\end{figure}

  The Higgs boson production in association with a $Z$ boson has drawn a lot of attentions as it is the channel to probe the $ZZh$ coupling in the SM, and therefore, tests the electroweak symmetry breaking mechanism. In the SM, the $hZ$ production rate at the 13 TeV LHC is 869 fb.  In the 2HDM, $Zh$ production cross-section can significantly deviate from the SM value. Within the 2HDM, extra contributions to $Zh$ production arises from the decay of $A$ after being resonantly produced via gluon gluon fusion process. Hence the signal strength $\mu_{Zh}$ relative to the SM expectation can deviate and 
  can be as large as the current experimental limit \cite{440}, large mixing $\sin{(\alpha - \beta)}$ is allowed within our framework. Contourplost of $\mu_{Zh}$ and $\mu_{hh}$ in $\mu_{t\bar{t}h}$-M$_H$ plane are shown in Fig. \ref{A3}. For simplicity, we consider the case of degenerate $H$ and $A$. The white dashed line 
  indicates the different values of $\mu_{Zh}$. As we can see, there is strong correlation between $\mu_{hh},\mu_{t\bar{t}h}$ and $\mu_{Zh}$. Black, pink and cyan colored meshed zones are excluded parameter space from current di-Higgs limit looking at different final states $b\bar{b}\gamma\gamma, b\bar{b}b\bar{b}$ and  
  $b\bar{b}\tau^{+} \tau^{-}$ respectively; red,blue and brown meshed zones are the excluded parameter space from the resonant $ZZ$, $W^+W^-$ and $Zh$ production constraints. Here we also impose the constraint from $pp \to A \to Zh$ searches 
  \cite{440} at the LHC and the  brown meshed zone is the exclusion region from that. We have used a typical set of parameters ($\sin(\alpha-\beta) = 0.5, \tilde{Y_b} = -0.09, \tilde{Y_\tau} = 10^{-3}$) for top left; ($\sin(\alpha-\beta) = 0.3, \tilde{Y_b} = -0.09, \tilde{Y_\tau} = 10^{-3}$) for top right; ($\sin(\alpha-\beta) = 0.4, \tilde{Y_b} = 0.02, \tilde{Y_\tau} = 10^{-3}$) for bottom left and ($\sin(\alpha-\beta) = -0.2, \tilde{Y_b} = 0.04, \tilde{Y_\tau} = 10^{-3}$) for 
  bottom right. For better illustration, we concentrate on one of the figures (top left). Here we see that if $M_H= M_A =$ 500 GeV, we can get simultaneous enhancement in $t\bar{t}h, hh$ and $Zh$ channels ($\mu_{t\abar{t}h}=1.7, \mu_{hh}=10$ and $\mu_{Zh}=1.5$). If any of these above mentioned modes is discovered at the LHC, our framework would indicate a correlated enhancement on the other signals, which can be tested at the LHC.


  \begin{figure}[htb!]
 $$
\includegraphics[height=7cm,width=0.6\textwidth]{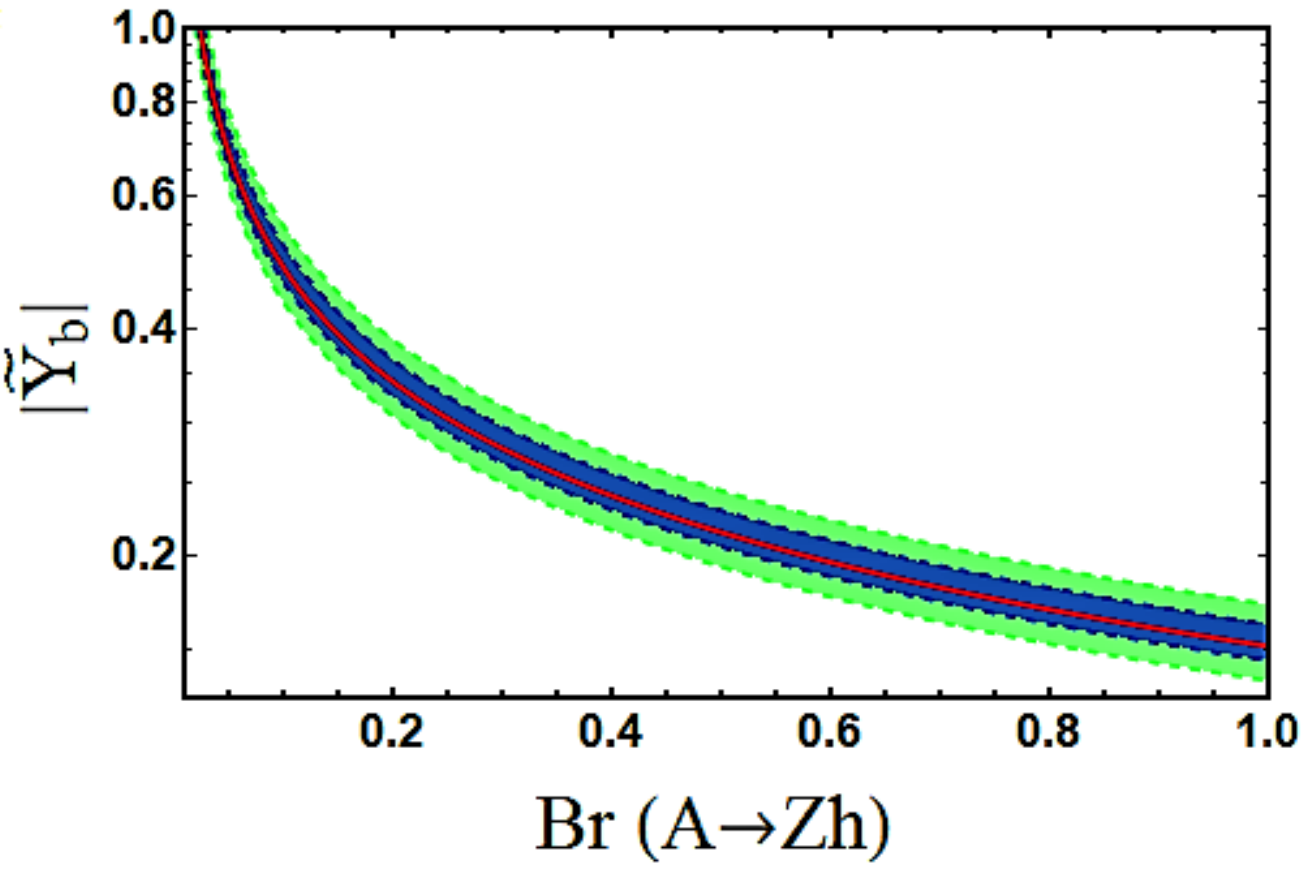}
 $$
 \caption{The parameter space in bottom Yukawa ($\tilde{Y}_b$) and Br ($A\to Zh$) plane consistent with 440 GeV excess. The blue and green regions correspond to the $1\sigma$ and $2\sigma$ bands on the observed limit. 
 }
\label{A12}
\end{figure}
\begin{figure}[htb!]
$$
 \includegraphics[height=8cm,width=0.5\textwidth]{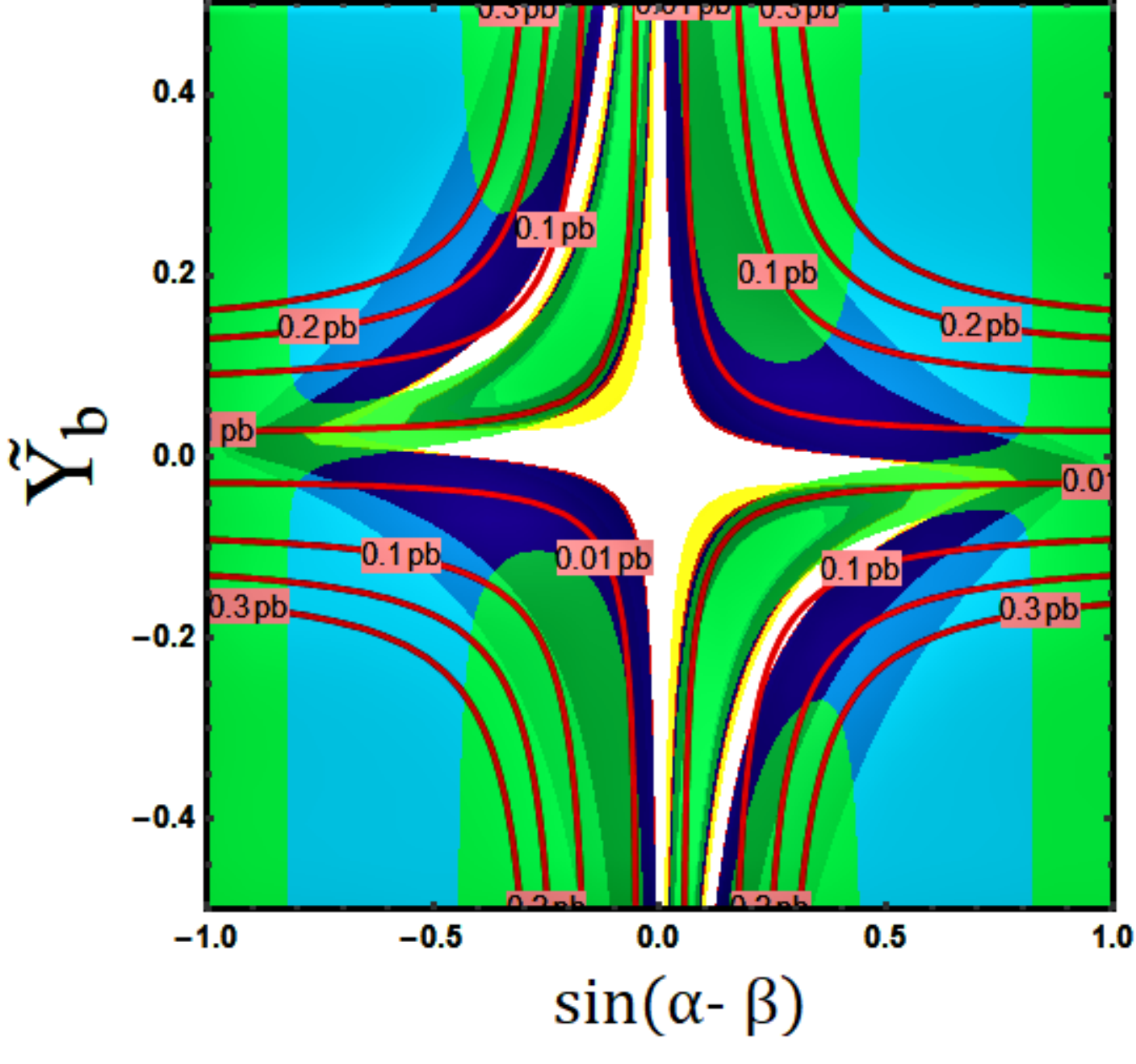}
  \includegraphics[height=8cm,width=0.5\textwidth]{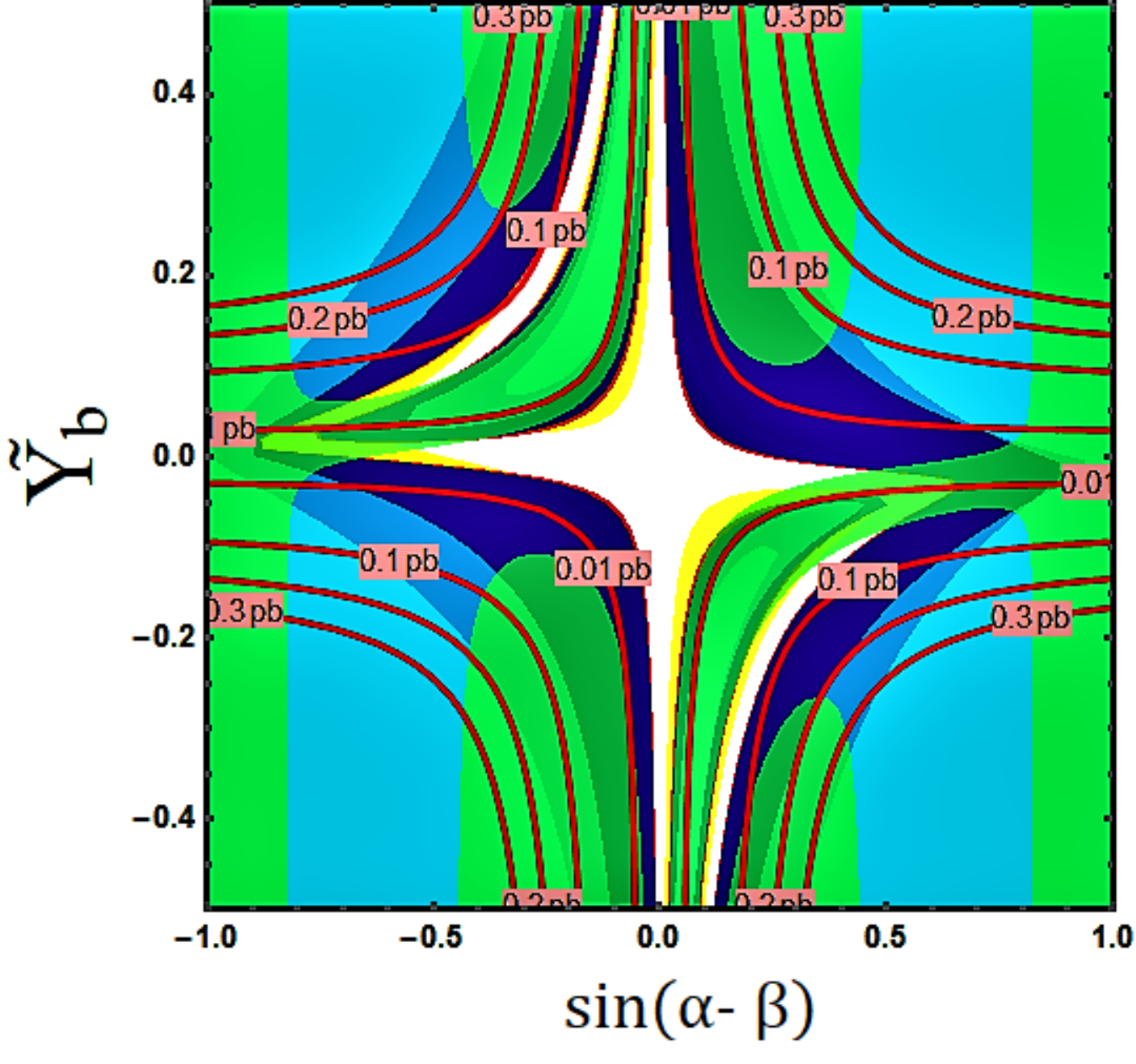}
 $$
 $$
\includegraphics[height=8cm,width=0.5\textwidth]{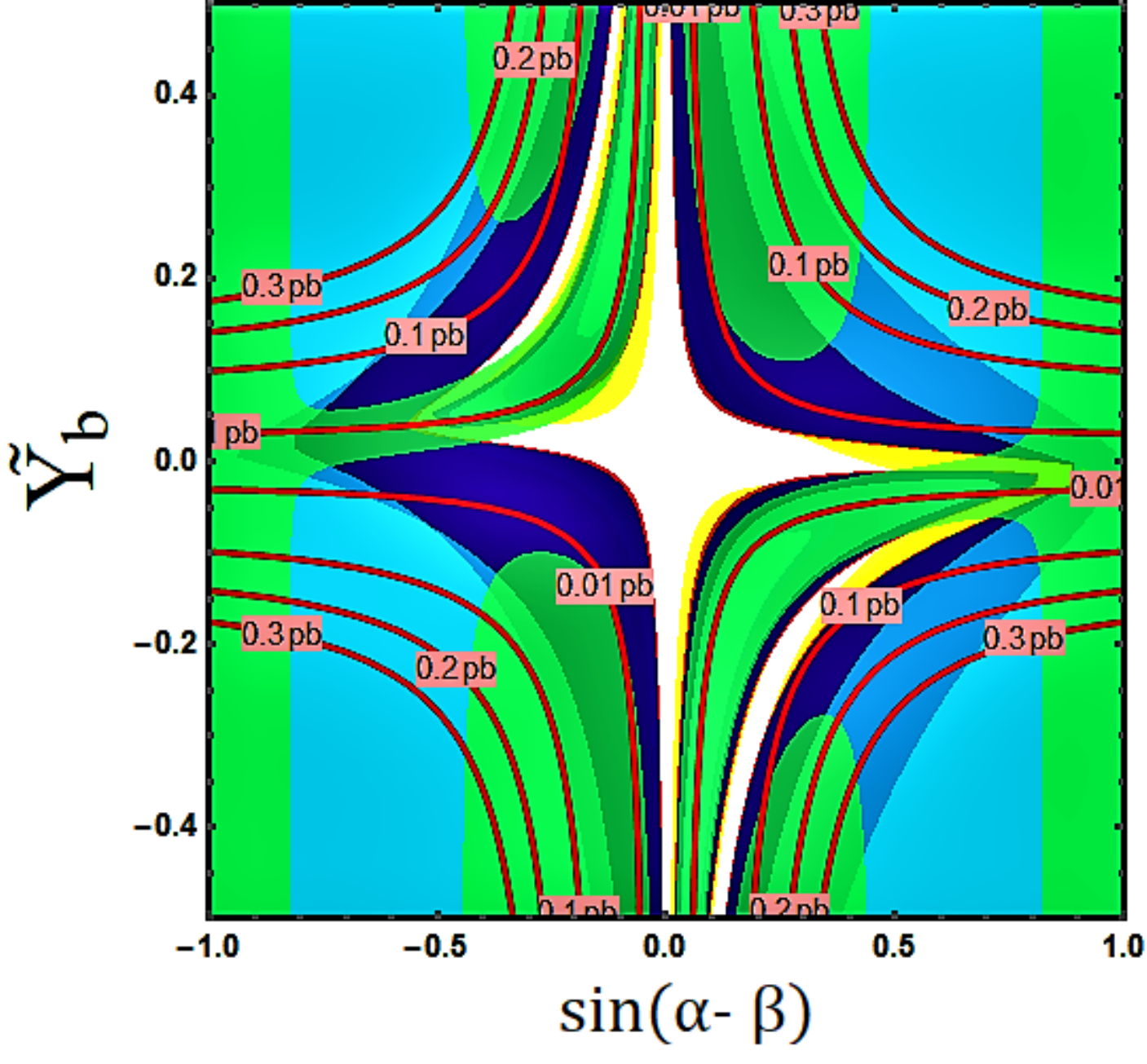}
 $$
 \caption{Cross-section in bottom Yukawa coupling $\tilde{Y_b}$ and mixing angle $\sin(\alpha-\beta)$ plane for pseudoscalar $A$ production in association with $b$ quarks and subsequent decay to $Zh$. The colored region is excluded from SM Higgs properties. We set top Yukawa $\tilde{Y}_t$ = 0 (top left), $-0.3$ (top right) and $0.5$ (bottom).
 }
\label{Azh}
\end{figure}

Recently, ATLAS collaboration reported a small excess of 440 GeV resonance while searching \cite{440} for a heavy, CP-odd Higgs boson decaying into a Z boson and a CP-even Higgs boson h with a mass of 125 GeV  at a centre-of-mass energy of 13 TeV
corresponding to an integrated luminosity of 36 fb$^{-1}$. The local significance of this excess is estimated to be 3.6$\sigma$ and the global significance is 2.4$\sigma$. The statistical significance is larger in case the pseudoscalar $A$ is produced through bottom-quark annihilation rather than gluon fusion, but both production processes indicate a deviation. In Fig. \ref{A1}, we show the representative leading order Feynman diagrams for pseudoscalar $A$ production in association with $b$ quarks and subsequent decay to $Zh$. The diagrams that can be obtained by crossing the initial state gluons, or radiating the Higgs off an antibottom quark are not shown. To explain this excess two ingredients are needed: (a) one bottom quark Yukawa coupling ($b\bar{b}A$) and (b) $AZh$ coupling which can occur through mixing. In Fig. \ref{A12}, we show the parameter space in bottom Yukawa ($\tilde{Y}_b$) and Br ($A\to Zh$) plane consistent with a 440 GeV excess. The blue and green regions correspond to the $1\sigma$ and $2\sigma$ bands on the observed limit. If any model can reproduce this parameter space consistent with the other experimental data, one can explain the excess. The 
statistical significance of this excess is too low to indicate anything meaningful. Since in our 2HDM framework we can have large mixing and hence larger $A \to Zh $ production rate, we can have significant production rate of the pseudoscalar $A$ in association with bottom quarks and subsequent decay to $Zh$. We compute the production cross section for the process $pp \to b\bar{b}A \to b\bar{b}Zh$ and subsequent decay of $h$ to bottom quarks. The cross-section in bottom Yukawa ($\tilde{Y}_b$) and mixing($\sin(\alpha-\beta)$) plane for is shown in Fig. \ref{Azh} for different values of top Yukawa $\tilde{Y}_t$ = 0(top left), -0.3 (top right) and 0.5(bottom). The colored region is excluded from SM Higgs properties. As we can see from Fig. \ref{Azh}, we get maximum enhancement in the production rate either for the wrong sign of bottom Yukawa or a negative sign of mixing $\sin{(\alpha - \beta)}$ term. While we are setting same sign large bottom Yukawa($\tilde{Y}_b$) to get the larger production rate, it modifies SM $hb\bar{b}$ coupling  due to the presence of non-zero mixing. On the other hand, if we start with either wrong sign  bottom Yukawa($\tilde{Y}_b$) or negative sign mixing $\sin{(\alpha - \beta)}$, the deviation factor $\kappa_b$ will be close to one with different sign without effecting SM Higgs properties. The thin white bands  with cross-section 0.1-0.3 pb above the
SM background in the second and the fourth quadrants can simultaneously explain the observed excess.

 \subsection{Other collider implications for heavy Higgs searches}
 
 For completeness, we discuss  the  prospects  for  charged  Higgs  boson and other pseudoscalar searches  at  the  LHC within the 2HDM.
LHC  experiments  have  already  set  very  strong  bounds  on  the  singly charged Higgs mass in the low mass region $M_{H^+} < m_t$ for the $pp \to H^{+} \bar{t} b$ process, assuming  the  decay $H^+ \to \tau^+ \nu$ \cite{ch1,ch2,ch3,charged1}.  To be consistent with the current experimental constraints, we consider the singly charged Higgs mass well above 300 GeV. However, the smallness of $T$ parameter does not allow for a large mass splitting between $H^+$, $H$ and $A$. A variety of production mechanism is involved in singly charged Higgs production. It can be pair produced at the LHC via gluon gluon fusion through the Higgs portal, via quark fusion through $s$ channel $Z$ or $\gamma$ exchange and also via photon initiated processes  \cite{babujana1}. $H^+$ can also be produced in association with fermions via gluon gluon fusion. There is another production mode of $H^+$ in association with the SM Higgs $h$ via s-channel $W$ boson exchange. The significant decay modes of $H^+$ are $H^+ \to \tau^+ \nu$, $H^+ \to t \bar{b}$, $H^+ \to W^+ h$ in the higher mass region. We choose mass splitting such that  $H^+ \to W^+ H$, $H^+ \to W^+ A$ are not kinematically allowed. In this scenario, it mostly decays to $t\bar{b}$ since $H^+ \to W^+ h$ is suppressed partially by the mixing angle $\sin(\alpha-\beta)$. The detailed phenomenology of charged Higgs is beyond the scope of this study. We will give some naive estimates of cross-sections and required luminosity to discover it at the LHC for the three sample points shown in Table \ref{table1}. For simplicity, we choose $M_{H^+}=M_{A}=505, 705$ and $605$ GeV for the the three sample points (BP1,BP2 and BP3) respectively. Pair production cross-section of $H^\pm$ corresponds to 1.35 fb, 0.5 fb and 0.48 fb respectively. This signal strength is too small, while the $H^+ \to t \bar{b}$ channel suffers from high QCD background. The tau channel benefits from being rather clean  compared to the other channels. $hH^+$ production will give rise to $t\bar{b}h$ signals. The production rate of $hH^+$ gets a suppression due to mixing. For the three sample points, the estimated $hH^+$ production cross-sections are 0.628 fb, 0.211 fb and 0.049 fb respectively.  We consider also  the signal $pp \to H^{+} \bar{t} b \to \bar{t} b \tau^+ \nu$. For this process, the total cross-section is 0.043 fb, 0.013 fb and 0.016 fb respectively for the three BP. After naive estimation of the background, we get that we need atleast $\sim ab^{-1}$ luminosity at the upcoming run of the LHC for the discovery of $H^\pm$. 

Similar to $H$ the pseudoscalar $A$ is also produced resonantly via gluon-gluon fusion through the triangle loop with the top quark since $A$ directly couples to $t$. After being produced resonantly, it mainly decays to $t\bar{t}$ and $b\bar{b}$. These channels are very challenging from the point of background elimination. Due to mixing, it also decays to Zh. We restrict ourselves such that $A$ is not kinematically allowed to decay to $W^+H^-$ by keeping the mass splitting between $A$ and $H$ very small ($\sim 5 GeV$). The detailed LHC phenomenology of $A$ and $H$ is beyond of the scope of this study and will be presented in our future work. 

$b\bar{b}h$ coupling is also modified in the 2HDM framework. $b\bar{b}h$ production rate will change compared to the SM. On the other hand, as $b\bar{b}h$ coupling plays the most significant role in SM Higgs branching ratio to different decay modes, large deviation from SM $b\bar{b}h$ coupling is not possible as it  could highly constrain the parameter space. For three of the benchmark points used in Table \ref{table1}, the signal strength for $b\bar{b}h$ production is $\mu_{b\bar{b}h}=1.002, 1.11$ and $0.819$ respectively. Similar to $t\bar{t}h$ and  $b\bar{b}h$ production, there will be also $t\bar{t}H$  and $b\bar{b}H$ production in this model. For three of the benchmark points, the $t\bar{t}H$  production rate at the 13 TeV LHC is 2.26 fb, 2.05 fb and 4.13 fb respectively. $b\bar{b}H$ production will give a unique signature of six $b$-quarks via the process $pp \to b\bar{b}H \to b\bar{b}hh \to b\bar{b}b\bar{b}b\bar{b}$. The production cross section for $b\bar{b}H$ at the 13 TeV LHC turns out to be 0.897 fb, 0.140 fb and 0.0478 fb respectively for three of the sample points. This is too small to probe at the 13 TeV LHC for the current luminosity. It requires very high luminosity ($\sim ab^{-1}$) to get the discovery reach limit via these channels.

 After being resonantly produced via gluon fusion at the LHC, the heavy Higgs boson $H$ can decay via lepton flavor violating processes such as $H \to \mu \tau$ \cite{Hmt1, Hmt2}, which is a very clean signal at the LHC. But, due to the small Yukawa coupling ($\sim m_{\mu} / v$), the branching fraction for $H \to \mu \tau$ process is highly suppressed. For a benchmark point ($M_H = 500 \rm ~GeV, ~\tilde{Y}_t = -0.3,  ~\tilde{Y}_b = -0.2,  ~\tilde{Y}_{\tau} = 10^{-3}, \sin{(\alpha - \beta)}=0.3$), the branching ratio for $H \to \mu \tau$ process is computed to be $10^{-7}$. Since the branching ratio is highly suppressed, even if for $\mathcal{O}(\sim pb)$ resonant production rate of heavy Higgs $H$, it requires very high luminosity ($\sim ab^{-1}$) to observe a single event. This is too small to probe at the 13 TeV LHC for the current luminosity.


\begin{figure}[htb!]
$$
 \includegraphics[height=6.2cm]{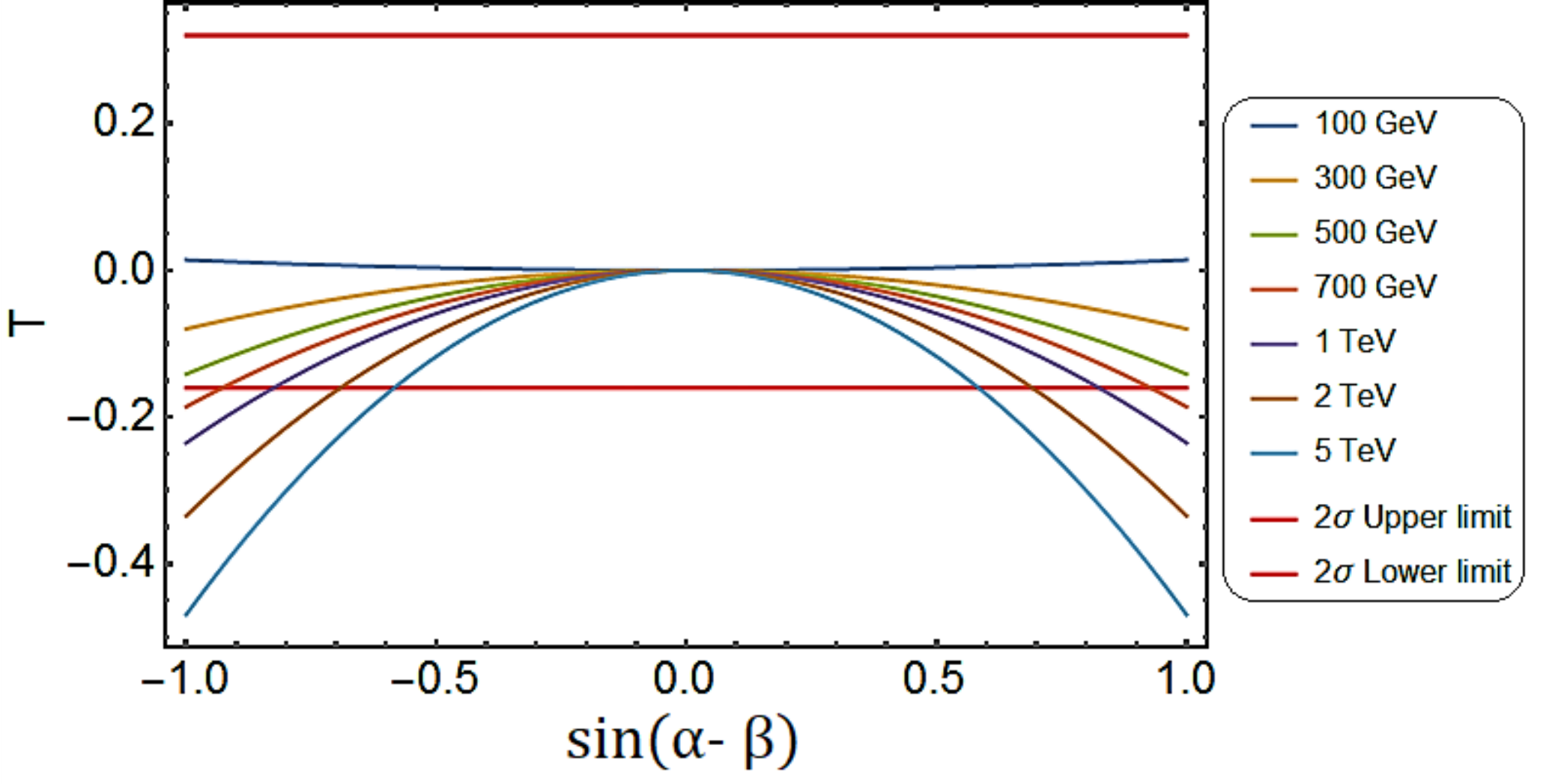}
 $$
 $$
 \includegraphics[height=6.2cm]{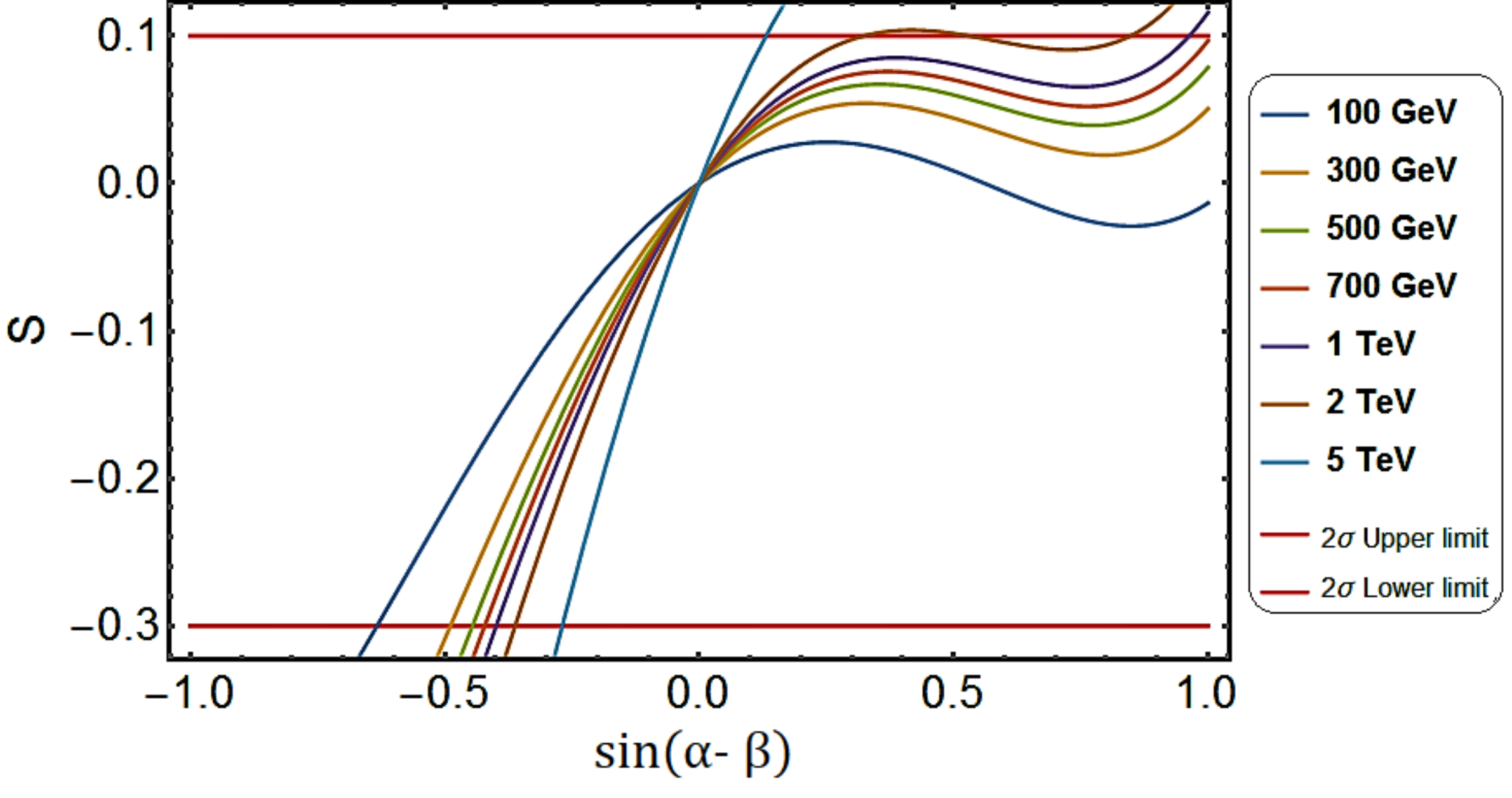}
 $$
 \caption{ Top: $T$ parameter as a function of $\sin(\alpha-\beta)$ for different masses of the (degenerate) heavy Higgs bosons. Bottom: $S$ parameter as a function of $\sin(\alpha-\beta)$ for different masses of the heavy Higgs. }
\label{4}
\end{figure}


\color{black}

\section{Electroweak precision constraints, boundedness and unitarity}\label{ewpt}
The oblique parameters $S$, $T$  and $U$ provide important constrains from electroweak precisiton data on many models beyond the SM.   These parameters have been calculated at the one loop in the two Higgs doublet model \cite{stu1, stu2, stu3}. We focus on the scenario where w$H, A$ and $H^+$ are nearly degenerate in mass. The condition for the degeneracy is  $\Lamd=\Lame = (m_h/v)^2 - \Lama $. With this assumption, we scan for the predicted values of $S$ and $T$ as a
function of the mixing angle $\sin(\alpha-\beta)$. Our results are shown in Fig. \ref{4}.  We see that the constraints from $S$ and $T$ parameters \cite{charged1} allow large parameter space of interest in the 2HDM which can lead to observable signals at the LHC.

To ensure the existence of a stable vacuum, the 2HDM scalar potential must be bounded from below, i.e. it must take positive values for any direction for which the the value of any field tends to infinity.
This places some restrictions on the allowed values of the quartic scalar couplings.   We require $\Lamf < 3.5$  in order to avoid non-perturvative regimes.  Note that $\Lamf$ term is related to $M_H$ and $\sin(\alpha-\beta)$ via Eq. (\ref{exactbma}).
 As a result, for larger mixing ($\sin(\alpha-\beta) \sim 0.5$), we can go upto a mass of $\sim$ 700 GeV for the mass of the heavy Higgs $H$, whereas for
smaller mixing ($\sin(\alpha-\beta) \sim 0.1$), $M_H$ can be as large as $\sim$ 1.5 TeV.  

We have also checked all the boundedness conditions \cite{boundedness}.  To ensure that the scalar potential is bounded from below, we evaluate the eigenvalues and eigenvectors of the following matrix: 
\[
\begin{bmatrix}
    \frac{1}{4}(\Lama+\Lamb+2\Lamc) & -\frac{1}{2}(\Lamf+\Lamg) & 0 & -\frac{1}{4}(\Lama-\Lamb) \\
    \frac{1}{2}(\Lamf+\Lamg) & -\frac{1}{2}(\Lamd+\Lame) & 0 & -\frac{1}{2}(\Lamf-\Lamg) \\
   0 & 0 & -\frac{1}{2}(\Lamd-\Lame) & 0 \\
    \frac{1}{4}(\Lama-\Lamb) & -\frac{1}{2}(\Lamf-\Lamg) & 0 & -\frac{1}{4}(\Lama+\Lamb+2\Lamc) 
\end{bmatrix}
\]
where we choose all the quartic couplings to be real.
Here we present one set of values of the quartic couplings: $\Lama=1.4 , \Lamb=0.01 , \Lamc=1 , \Lamd=0.1 , \Lame=0.001 , \Lamf=3 $ and $\Lamg=-1.2 $. For this set, we found all the eigenvalues of the matrix to be $\lbrace 2.0527, -1.75315, \newline 0.649943, -0.0495 \rbrace$ . This satisfies the boundedness conditons. For this specific choice, we obtain the three mass eigenvalues for the neutral scalar masses to be $\lbrace$ 125 GeV, 751 GeV, 706 GeV  $\rbrace$ from Eq. (\ref{massmhh}) and the mixing angle $\sin(\alpha-\beta =0.458$. The unitarity bounds in the most general two Higgs doublet model without any discrete $Z_2$ symmetry has been studied here Ref. \cite{unitarity}. Our choice of parameters are consistent with these unitarity constraints.

\section{Conclusion }\label{con}

In this paper we have undertaken a detailed analysis of the collider implications of the two Higgs doublet model.  In our framework, both doublets are treated on equal footing, which implies that they have  Yukawa couplings with fermions with comparable strengths.  Such a scenario would have Higgs mediated flavor changing neutral currents, which can potentially be excessive.  It is customary to assume a discrete symmetry to forbid such FCNC.  Here we have proposed a simple ansatz for the Yukawa couplings of each of the Higgs doublets that is consistent with the observed CKM mixing angles and also compatible with FCNC constraints.  Our ansatz, where the Yukawa couplings are taken to have a form $Y_{ij}^{(a)} =\sqrt{2} C_{ij}^{(a)} {\rm min}\{m_i,\,m_j\}/v$, with $C_{ij}$ being order one coefficients and $m_i$ being the mass of fermion $i$, is an improvement over the Cheng-Sher (CS) ansatz \cite{Cheng} which assumes  $Y_{ij}^{(a)} = \sqrt{2} C_{ij}^{(a)} \sqrt{m_i m_j}/v$. 

We have studied the flavor phenomenology of the new ansatz and shown that all Higgs mediated FCNC are within acceptable limits.  We have also compared the FCNC constraints with those from the CS ansatz, and found that the modified ansatz fares much better.  

We have studied the collider implications of the 2HDM framework with both Higgs doublets coupling to fermions.  The LHC phenomenology is sensitive mostly to the couplings of the top and bottom quarks and the tau lepton.  Taking these couplings to be comparable for the two doublets, we have shown that resonant di-higgs boson production rate  can be enhanced by a factor of 25 compared to the SM.  Such a large deviation is possible, as the properties of the 125 GeV Higgs boson $h$ can be within experimentally allowed range with large Yukawa couplings of the $t$ and $b$ quarks to both doublets.  

We have studied the correlations between an enhanced di-higgs cross section with $t \overline{t} h$ and $Zh$ production rates.  These rates can also be modified by as much as 100\% compared to the SM, with such modifications inter-correlated.  Possible signals of a resonant structure in the $Zh$ production can be explained consistently in terms of a pseudoscalar $A$ within this framework.  Thus, the 2HDM offers a variety of tests at the LHC.

\section*{Acknowledgement}
 This work is supported in part by the US Department of Energy Grant No. de-sc0016013. The authors would like to thank the theory group at Fermilab for  hospitality during a summer visit under a Fermilab Distinguished Scholar program. We thank A. Belyaev, B. Dobrescu, Z. Liu and M. Sher for helpful discussions.

\begin{appendix}
\section{ \large Scalar potential}
\label{appendix1}
The most general gauge invariant 2HDM scalar potential is given by  :
\beqa  \label{pot}
\mathcal{V}&=& m_{11}^2\Phi_1^\dagger\Phi_1+m_{22}^2\Phi_2^\dagger\Phi_2
-[m_{12}^2\Phi_1^\dagger\Phi_2+{\rm h.c.}]\nonumber\\[6pt]
&&\quad +\half\lambda_1(\Phi_1^\dagger\Phi_1)^2
+\half\lambda_2(\Phi_2^\dagger\Phi_2)^2
+\lambda_3(\Phi_1^\dagger\Phi_1)(\Phi_2^\dagger\Phi_2)
+\lambda_4(\Phi_1^\dagger\Phi_2)(\Phi_2^\dagger\Phi_1)
\nonumber\\[6pt]
&&\quad +\left\{\half\lambda_5(\Phi_1^\dagger\Phi_2)^2
+\big[\lambda_6(\Phi_1^\dagger\Phi_1)
+\lambda_7(\Phi_2^\dagger\Phi_2)\big]
\Phi_1^\dagger\Phi_2+{\rm h.c.}\right\}\,,
\eeqa
where $m_{11}^2$, $m_{22}^2$, and $\lam_1,\cdots,\lam_4$ are real parameters. In general, $m_{12}^2$, $\lambda_5$, $\lambda_6$ and $\lambda_7$ are complex. 

\vspace*{0.1in}
\noindent{\bf  Correspondence of parameters}
\vspace*{0.1in}

The transformation relations of the scalar potential parameters between the original (Eq. (\ref{pot})) and rotated basis (Eq. (\ref{higgsbasis})) are given by:
\beqa
M_{11}^2&=&m_{11}^2\cb^2+m_{22}^2\sb^2-\Re(m_{12}^2 e^{i\xi})\stwob\,,
\label{maa}\\
M_{22}^2&=&m_{11}^2\sb^2+m_{22}^2\cb^2+\Re(m_{12}^2 e^{i\xi})\stwob\,,
\label{mbb}\\
M_{12}^2 e^{i\xi}&=&
\half(m_{11}^2-m_{22}^2)\stwob+\Re(m_{12}^2 e^{i\xi})\ctwob
+i\,\Im(m_{12}^2 e^{i\xi})\,.\label{mab}
\eeqa
and
\beqa
\!\!\!\!\!\!\!\!\!\!\!\!\!\!\Lama&=&
\lam_1\cb^4+\lam_2\sb^4+\half\lamtil\stwob^2
+2\stwob\left[\cb^2\Re(\lam_6 e^{i\xi})
+\sb^2\Re(\lam_7e^{i\xi})\right]\,,
\label{Lam1def}  \\
\!\!\!\!\!\!\!\!\!\!\!\!\!\!\Lamb &=&
\lam_1\sb^4+\lam_2\cb^4+\half\lamtil\stwob^2
-2\stwob\left[\sb^2\Re(\lam_6 e^{i\xi})
+\cb^2\Re(\lam_7e^{i\xi})\right]\,,
\label{Lam2def}      \\
\!\!\!\!\!\!\!\!\!\!\!\!\!\!\Lamc &=&
\quarter\stwob^2\left[\lam_1+\lam_2-2\lamtil\right]
+\lam_3-\stwob\ctwob\Re[(\lam_6-\lam_7)e^{i\xi}]\,,
\label{Lam3def}      \\
\!\!\!\!\!\!\!\!\!\!\!\!\!\!\Lamd &=&
\quarter\stwob^2\left[\lam_1+\lam_2-2\lamtil\right]
+\lam_4-\stwob\ctwob\Re[(\lam_6-\lam_7)e^{i\xi}]\,,
\label{Lam4def}      \\
\!\!\!\!\!\!\!\!\!\!\!\!\Lame e^{2i\xi}&=&
\quarter\stwob^2\left[\lam_1+\lam_2-2\lamtil\right]+\Re(\lam_5 e^{2i\xi})
+i\ctwob\Im(\lam_5 e^{2i\xi})
\nonumber \\
&&\qquad
-\stwob\ctwob\Re[(\lam_6-\lam_7)e^{i\xi}]
-i\stwob\Im[(\lam_6-\lam_7)e^{i\xi})]\,,
\label{Lam5def}      \\
\!\!\!\!\!\!\!\!\!\!\!\!\Lamf e^{i\xi}&=&
-\half\stwob\left[\lam_1\cb^2
-\lam_2\sb^2-\lamtil\ctwob-i\Im(\lam_5 e^{2i\xi})\right]\nonumber \\
&&\qquad
+\cb\cthreeb\Re(\lam_6 e^{i\xi})+\sb\sthreeb\Re(\lam_7 e^{i\xi})
+i\cb^2\Im(\lam_6 e^{i\xi})+i\sb^2\Im(\lam_7 e^{i\xi})\,,
\label{Lam6def}      \\
\!\!\!\!\!\!\!\!\!\!\!\!\Lamg e^{i\xi}&=&
-\half s_{2\beta}\left[\lam_1\sb^2-\lam_2\cb^2+
\lamtil c_{2\beta}+i\Im(\lam_5 e^{2i\xi})\right]\nonumber \\
&&\qquad
+\sb\sthreeb\Re(\lam_6 e^{i\xi})+\cb\cthreeb\Re(\lam_7e^{i\xi})
+i\sb^2\Im(\lam_6 e^{i\xi})+i\cb^2\Im(\lam_7e^{i\xi})\,,
\label{Lam7def}
\eeqa
where
\beq \label{lamtildef}
\lamtil\equiv\lam_3+\lam_4+\Re(\lam_5 e^{2i\xi})\,.
\eeq

\end{appendix}


\end{document}